\documentclass[11pt]{svjour3} \usepackage{fullpage} 

\usepackage{xspace}

\usepackage{graphicx}

\usepackage{amssymb}

\renewcommand{\paragraph}[1]{\smallskip\vspace{2pt}\par{\em #1}~}

\newcommand{\primalOf}[1]{{{#1}}}
\newcommand{\dualOf}[1]{{\hat{#1}}}

\newcommand{\MM}{M}
\newcommand{\JJ}{J}
\newcommand{\xp}{\primalOf x}
\newcommand{\yp}{\primalOf y}
\newcommand{\pp}{\primalOf p}
\newcommand{\wwp}{\primalOf u}
\newcommand{\Ip}{\calI}

\newcommand{\xd}{\dualOf x}
\newcommand{\yd}{\dualOf y}
\newcommand{\pd}{\dualOf p}
\newcommand{\wwd}{\dualOf u}
\newcommand{\Jd}{\calJ}

\newcommand{\dd}{\delta}

\newenvironment{Proof}{\begin{proof}}{{} ~\hfill\hfill\qed~\end{proof}}

\newcommand{\algfont}{}
\newcommand{\tab}{\hspace*{0.13in}}

\newcounter{myline}

\newenvironment{alg}{
  \medskip
  \par
  \algfont
  \centering
    \begin{tabular}{|@{$\!$}|l|}\hline
      \begin{minipage}{0.96\linewidth}\raggedright
          \smallskip
          \begin{list}{\arabic{myline}.}{
            \usecounter{myline}
            \setlength{\listparindent}{0in}
            \setlength{\topsep}{0in}
            \setlength{\itemsep}{.013in}
            \setlength{\parsep}{.013in}
            \setlength{\rightmargin}{0in}
            \setlength{\itemindent}{0in}
            \setlength{\labelsep}{0.065in}
            \setlength{\leftmargin}{0.2in}
          }
          }{
        \end{list}
        \smallskip
      \end{minipage}\\\hline
    \end{tabular}
    \par
    \noindent
}

\newcommand{\A}{\item}

\newcommand{\Ahead}[1]{\item[]\hspace*{-\leftmargin}{\textrm{#1}}}

\newcommand{\algbeg}{%
  \addtolength{\labelsep}{0.13in}
  \addtolength{\itemindent}{0.13in}
  \addtolength{\listparindent}{0.13in}
}
\newcommand{\algend}{%
  \addtolength{\labelsep}{-0.13in}
  \addtolength{\itemindent}{-0.13in}
  \addtolength{\listparindent}{-0.13in}
}

\newcommand{\eps}{\varepsilon}
\newcommand{\tran}{^{\scriptscriptstyle \sf T\hspace*{-0.05em}}}

\makeatletter
\newcommand{\mymathfnnolimits}[1]{\mathop {\operator@font\sf #1}\nolimits }
\newcommand{\mymathfn}[1]{\mathop {\operator@font\sf #1}}
\makeatother

\newcommand{\prob}[1]{{\sc #1}\xspace}
\renewcommand{\prob}[1]{{#1}\xspace}

\newcommand{\E}{{\rm E}}
\newcommand{\opt}{\mbox{{\sc opt}}}

\newcommand{\giv}{\,|\,}

\newcommand{\calI}{{\cal I}}
\newcommand{\calJ}{{\cal J}}

\newcommand{\R}{{\sf R\hspace*{-1.5ex}\rule{0.15ex}{1.6ex}\hspace*{1.2ex}}}
\newcommand{\Rp}{\R_{\scriptscriptstyle +}}

\newcommand{\text}[1]{\mbox{#1}}
\newcommand{\rows}{r}
\newcommand{\columns}{c}
\newcommand{\inputsize}{n}

\begin{document}
\journalname{}
\def\makeheadbox{}

\title{A Nearly Linear-Time PTAS for Explicit \\Fractional Packing and Covering Linear Programs}
\titlerunning{Nearly Linear-Time PTAS for Packing and Covering LPs}
\author
{
	Christos Koufogiannakis \and Neal E.\ Young
}
\institute
        {Department~of Computer Science and Engineering, University~of California, Riverside.
          The first author would like to thank the Greek State Scholarship Foundation (IKY).
          The second author's research was partially supported by
          NSF grants 0626912, 0729071, and 1117954.
}

\date{}




\maketitle

\begin{abstract}
We give an approximation algorithm for fractional
packing and covering linear programs (linear programs with non-negative coefficients).
Given a constraint matrix with $\inputsize$ non-zeros, $\rows$ rows, and $\columns$ columns,
the algorithm (with high probability) computes feasible primal and dual solutions
whose costs are within a factor of $1+\eps$ of  $\opt$ (the optimal cost)
in time $O((\rows+\columns)\log(\inputsize)/\eps^2 + \inputsize)$.\footnote
{Accepted to Algorithmica, 2013.  The conference version of this paper was ``Beating Simplex for fractional packing and covering linear programs'' \cite{Koufogiannakis2007Beating}.
}
\end{abstract}

\section{Introduction}\label{sec:intro}

A {\em packing} problem is a linear program of the form $\max\{a\cdot x : \MM \xp \le b, \xp \in P\}$, where the entries of the constraint matrix $\MM$ are non-negative and $P$ is a convex polytope admitting some form of optimization oracle.  
A {\em covering} problem is of the form $\min\{a\cdot \xd : \MM \xd \ge b, \xd \in P\}$.  

This paper focuses on {\em explicitly given} packing and covering problems, that is, $\max\{a\cdot \xp : \MM \xp \le b, \xp \ge 0\}$ and $\min\{a\cdot \xd : \MM \xd \ge b, \xd \ge 0\}$, where the polytope $P$ is just the positive orthant.  
Explicitly given packing and covering are important special cases of linear programming,
including, for example, fractional set cover, multicommodity flow problems with given paths, two-player zero-sum matrix games with non-negative payoffs, and variants of these problems.

The paper gives a $(1+\eps)$-approximation algorithm ---
that is, an algorithm that returns feasible primal and dual solutions whose costs are within a given factor $1+\eps$ of $\opt$.
With high probability, it runs in time
$O((\rows+\columns)\log(\inputsize)/\eps^2 + \inputsize)$, where $\inputsize$ -- the input size -- is the number of non-zero entries in the constraint matrix and $\rows+\columns$ is the number of rows plus columns (i.e., constraints plus variables).  

For dense instances, $r+c$ can be as small as $O(\sqrt\inputsize)$.
For moderately dense instances -- as long as $r+c = o(n/\log n)$ --
the $1/\eps^2$ factor multiplies a sub-linear term.
Generally, the time is linear in the input size $n$ as long as $\eps \ge \Omega(\sqrt{(\rows+\columns)\log(\inputsize)/\inputsize})$.

\subsection{Related work}
The algorithm is a Lagrangian-relaxation (a.k.a.~price-directed decomposition, multiplicative weights) algorithm.
Broadly, these algorithms work by replacing a set of hard constraints by a sum of smooth penalties, one per constraint, and then iteratively augmenting a solution while trading off the increase in the objective against the increase in the sum of penalties.
Here the penalties are exponential in the constraint violation,
and, in each iteration, only the first-order (linear) approximation 
is used to estimate the change in the sum of penalties.

Such algorithms, which can provide useful alternatives to interior-point and Simplex methods,  have a long history and a large literature.
Bienstock gives an implementation-oriented, operations-research perspective \cite{Bienstock00Potential}.  
Arora {\em et al.}\ discuss them from a computer-science perspective, highlighting connections to other fields such as learning theory \cite{arora2012multiplicative}.
An overview by Todd places them in the context of general linear programming \cite{todd2002mfl}.

The running times of algorithms of this type increase as the approximation parameter $\eps$ gets small.   For algorithms that rely on linear approximation of the penalty changes in each iteration, the running times grow at least quadratically in $1/\eps$ (times a polynomial in the other parameters).  For explicitly given packing and covering, the fastest previous such algorithm that we know of runs in time $O((\rows+\columns)\bar{\columns}\log(\inputsize)/\eps^2)$, where $\bar{\columns}$ is the maximum number of columns in which any variable appears \cite{Young01Sequential}.  
That algorithm applies to {\em mixed} packing and covering --- a more general problem.
Using some of the techniques in this paper, one can improve that algorithm to run in time $O(\inputsize\log(\inputsize)/\eps^2)$ (an unpublished result),
which is slower than the algorithm here for dense problems.

Technically, the starting point for the work here is a remarkable algorithm by Grigoriadis and Khachiyan for the following special case of packing and covering \cite{Grigoriadis95Sublinear}.
The input is a two-player zero-sum matrix game with payoffs in $[-1,1]$.
The output is a pair of mixed strategies that guarantee an expected payoff within an {\em additive} $\eps$ of optimal.   (Note that achieving additive error $\eps$ is, however, easier than achieving multiplicative error $1+\eps$.)
The algorithm computes the desired output 
in $O((\rows+\columns)\log(\inputsize)/\eps^2)$ time.
This is remarkable in that, for dense matrices, it is {\em sub-linear} in the input size $n = \Theta(r c)$.\footnote{
The problem studied here, packing and covering, can be reduced to Grigoriadis and Khachiyan's problem.  This reduction leads to an $O((\rows+\columns)\log(\inputsize)(U\,\opt)^2/\eps^2)$-time algorithm to find a $(1+\eps)$-approximate packing/covering solution,
where $U\doteq\max_{ij} \MM_{ij}/(b_i a_j)$.
A pre-processing step \cite[\S2.1]{Luby93Parallel} can bound $U$,
leading to a running time bound of $O((\rows+\columns)\log(\inputsize)\min(r,c)^4/\eps^4)$.}
(For a machine-learning algorithm closely related to Grigoriadis and Khachiyan's result,
see \cite{clarkson2010sublinear,clarkson2012machine}.)

We also use the idea of {\em non-uniform increments}
from algorithms by Garg and K\"onemann
\cite{Garg98Faster,Konemann98Fast,garg2007faster}.

\paragraph{Dependence on $\eps$.}
Building on work by Nesterov (e.g.,~\cite{nesterov2005smooth,nesterov2009unconstrained}),
recent algorithms for packing and covering problems
have reduced the dependence on $1/\eps$ from quadratic to linear, 
at the expense of increased dependence on other parameters.
Roughly, these algorithms better approximate the change in the penalty function in each iteration, leading to fewer iterations but more time per iteration
(although not to the same extent as interior-point algorithms).
For example, Bienstock and Iyengar give an algorithm for concurrent multicommodity flow that solves $O^*(\eps^{-1} k^{1.5}|V|^{0.5})$ shortest-path problems, where $k$ is the number of commodities and $|V|$ is the number of vertices \cite{Bienstock04Solving}.  
Chudak and Eleuterio continue this direction --- for example, they give an algorithm for fractional set cover running in worst-case time $O^*(\columns^{1.5}(\rows+\columns)/\eps + \columns^2\rows)$ \cite{chudak2005ias}.

\paragraph{Comparison to Simplex and Interior-Point methods.}
Currently, the most commonly used algorithms for solving linear programs in practice are Simplex and interior-point methods.  Regarding Simplex algorithms, commercial implementations algorithms use many carefully tuned heuristics (e.g.~pre-solvers and heuristics for maintaining sparsity and numerical stability), enabling them to quickly solve many practical problems with millions of non-zeros to optimality.  But, as is well known, 
their worst-case running times are exponential.
Also, for both Simplex and interior-point methods,
running times can vary widely depending on the structure of the underlying problem.
(A detailed analysis of Simplex and interior-point running times is outside the scope of this paper.)
These issues make rigorous comparison between the various algorithms difficult.

Still, here is a meta-argument that may allow some meaningful comparison. Focus on ``square'' constraint matrices, where $r=\Theta(c)$.   Note that at a minimum, any Simplex implementation must identify a non-trivial basic feasible solution.  Likewise, interior-point algorithms require (in each iteration) a Cholesky decomposition or other matrix factorization.  Thus, essentially, both methods require implicitly (at least) solving an $r\times r$ system of linear equations.  Solving such a system is a relatively well-understood problem, both in theory and in practice, and (barring special structure) takes $\Omega(r^3)$ time, or $\Omega(r^{2.8})$ time using Strassen's algorithm.  Thus, on ``square'' instances, Simplex and interior-point algorithms should have running times growing at least with $\Omega(r^{2.8})$ (and probably more).
This reasoning applies even if Simplex or interior-point methods are terminated early so as to find {\em approximately} optimal solutions.  

In comparison, on ``square'' matrices, the algorithm in this paper takes time $O(n + r\log(r)/\eps^2)$ where $n=O(r^2)$ or less.  
If the meta-argument holds, 
then, for applications where $(1+\eps)$-approximate solutions suffice
for some fixed and moderate $\eps$ (say, $\eps\approx 1\%$),
for very large instances (say, $r\ge 10^4$),
the algorithm here should be orders of magnitude faster than Simplex
or interior-point algorithms.

This conclusion is consistent with experiments reported here, in which the running times of Simplex and interior-point algorithms on large random instances exceed $\Omega(r^{2.8})$.  Concretely, with $\eps=1\%$, the algorithm here is faster when $r$ is on the order of $10^3$, with a super-linear (in $r$) speed-up for larger $r$.

\subsection{Technical roadmap}
Broadly, the running times of iterative optimization algorithms are determined by (1) the number of iterations and (2) the time per iteration.  Various algorithms trade off these two factors in different ways.  The technical approach taken here is to accept a high number of iterations --- $(r+c)\log(n)/\eps^2$, a typical bound for an algorithm of this class
(see e.g.~\cite{Klein99Number} for further discussion) ---
and to focus on implementing each iteration as quickly as possible
(ideally in constant amortized time).

\paragraph{Coupling.}
Grigoriadis and Khachiyan's sub-linear time algorithm
uses an unusual technique of {\em coupling} primal and dual algorithms
that is critical to the algorithm here.
As a starting point, to explain coupling, consider the following ``slow'' coupling algorithm.
(Throughout, assume without loss of generality by scaling that $a_j = b_i = 1$ for all $i, j$.)
The algorithm starts with all-zero primal and dual solutions, $\xp$ and $\xd$, respectively.
In each iteration, it increases one coordinate $\xp_j$ of the primal solution $\xp$ by 1, 
and increases one coordinate $\xd_i$ of the dual solution $\xd$ by 1.
The index $j$ of the primal variable to increment is chosen randomly from a distribution $\pd$ that depends on the current {\em dual} solution.
Likewise, the index $i$ of the dual variable to increment is chosen randomly from a distribution $\pp$ that depends on the current {\em primal} solution.
The distribution $\pd$ is concentrated on the indices of dual constraints $M\tran \xd$ that are ``most violated'' by $\xd$.
Likewise, the distribution $\pp$ is concentrated on the indices of primal constraints $M \xp$ that are ``most violated'' by $\xp$.
Specifically, $\pp_i$ is proportional to $(1+\eps)^{M_i \xp}$,
while $\pd_j$ is proportional to $(1-\eps)^{M\tran_j \xd}$.\footnote{
The algorithm can be interpreted as a form of fictitious play of a two-player zero-sum game, where in each round each player plays from a distribution concentrated around the best response to the aggregate of the opponent's historical plays.  In contrast, in many other fictitious-play algorithms, one or both of the player plays a {\em deterministic} pure best-response to the opponent's historical average.}

Lemma~\ref{lemma:slowalg} in the next section proves that this algorithm
achieves the desired approximation guarantee.
Here, broadly, is why coupling helps reduce the time per iteration
in comparison to the standard approach.
The standard approach is to increment the primal variable corresponding 
to a dual constraint that is ``most violated'' by $\pp$ ---
that is, to increment $\xp_{j'}$ where $j'$ (approximately) minimizes $M\tran_{j'}\pp$
(for $\pp$ defined as above).
This requires at a minimum {\em maintaining} the vector $M\tran\pp$.
Recall that $\pp_i$ is a function of $M_i \xp$.
Thus, a change in one primal variable $\xp_{j'}$ 
changes many entries in the vector $\pp$,
but {\em even more} entries in $M\tran\pp$.
(In the $r\times c$ bipartite graph $G=([r],[c],E)$ where $E=\{(i,j) : M_{ij} \neq 0\}$,
the neighbors of $j'$ change in $\pp$,
while all {\em neighbors of those neighbors} change in $M\tran\pp$.)
Thus, maintaining $M\tran\pp$ is costly.
In comparison, to implement coupling, it is enough to maintain the vectors $\pp$ and $\pd$.
The further product $M\tran\pp$ is not needed (nor is $M\pd$).
This is the basic reason why coupling helps reduce the time per iteration.

\paragraph{Non-uniform increments.}
The next main technique, used to make more progress per iteration,
is Garg and K\"onemann's {\em non-uniform increments} \cite{Garg98Faster,Konemann98Fast,garg2007faster}.
Instead of incrementing the primal and dual variables by a uniform amount each time (as described above), the algorithm increments the chosen primal and dual variables $\xp_{j'}$ and $\xd_{i'}$ by an amount $\delta_{i'j'}$ chosen small enough so that the left-hand side (LHS) of 
each constraint (each $M_i\xp$ or $M_j\tran\xd$) increases by at most 1 (so that the analysis still holds), but {\em large enough} so that the LHS of {\em at least one} such constraint increases by at least $1/4$.
This is small enough to allow the same correctness proof to go through, 
but is large enough to guarantee a small number of iterations.
The number of iterations is bounded by (roughly) the following argument:
each iteration increases the LHS of some constraint by $1/4$,
but, during the course of the algorithm, no LHS ever exceeds $N \approx \log(n)/\eps^2$.
(The particular $N$ is chosen with foresight so that the relative error works out to $1+\eps$.)
Thus, the number of iterations is $O((r+c)N) = O((r+c)\log(n)/\eps^2)$.

\paragraph{Using slowly changing estimates of $M\xp$ and $M\tran\xd$.}
In fact, we will achieve this bound not just for the number of iterations, 
but also for the total work done (outside of pre- and post-processing).
The key to this is the third main technique.
Most of the work done by the algorithm as described so far
would be in maintaining the vectors $M\xp$ and $M\tran\xd$
and the distributions $\pp$ and $\pd$ (which are functions of $M\xp$ and $M\tran\xd$).
This would require lots of time in the worst case, because, even with non-uniform increments, there can still be many {\em small} changes in elements of $M\xp$ and $M\tran\xd$.
To work around this, instead of maintaining $M\xp$ and $M\tran\xd$ exactly,
the algorithm maintains more slowly changing {\em estimates} for them (vectors $\yp$ and $\yd$, respectively), using random sampling.
The algorithm maintains $\yp \approx M\xp$ as follows.
When the algorithm increases a primal variable $\xp_{j'}$ during an iteration,
this increases some elements in the vector $M\xp$
(specifically, the elements $M_i\xp$ where $M_{ij}>0$).
For each such element $M_i\xp$, if the element increases by, say, $\delta \le 1$,
then the algorithm increases the corresponding $\yp_i$ not by $\delta$, but by 1, but {\em only with probability $\delta$}.
This maintains not only $E[\yp_i] = M_i\xp$, but also, with high probability, $\yp_i \approx M_i\xp$.
Further, the algorithm only does work for a $\yp_i$ (e.g.~updating $\pp_i$)
when $\yp_i$ increases (by 1).
The algorithm maintains the estimate vector $\yd \approx M\tran\xd$ similarly,
and defines the sampling distributions $\pp$ and $\pd$ as functions 
of $\yp$ and $\yd$ instead of $M\xp$ and $M\tran\xd$.
In this way {\em each unit of work}  done by the algorithm
can be charged to an increase in $|M \xp| + |M\tran \xd|$ 
(or more precisely, an increase in $|\yp| + |\yd|$, which never exceeds $(r+c)N$).
(Throughout the paper, $|v|$ denotes the 1-norm of any vector $v$.)

Section~\ref{sec:slow_alg} gives the formal intuition underlying coupling by describing and formally analyzing the first (simpler, slower) coupling algorithm described above.
Section~\ref{sec:correctness} describes the full (main) algorithm and its correctness proof.
Section~\ref{sec:time} gives remaining implementation details and bounds the run time.  
Section~\ref{sec:experiments} presents basic experimental results, including a comparison with the GLPK Simplex algorithm.

\subsection{Preliminaries}
For the rest of the paper, 
assume the primal and dual problems are of the following restricted forms, respectively:
~$\max \{ |\xp| : \MM \xp \le \mathbf 1, \xp \ge  0 \}$,
~$\min \{ |\xd| : \MM\tran \xd \ge \mathbf 1, \xd \ge 0 \}$.
That is, assume $a_j = b_i = 1$ for each $i,j$.
This is without loss of generality by the transformation $\MM'_{ij}=\MM_{ij}/(b_ia_j)$.
Recall that $|v|$ denotes the 1-norm of any vector $v$.

\section{Slow algorithm (coupling)}\label{sec:slow_alg}\label{sec:alg}
To illustrate the coupling technique,
in this section we analyze the first (simpler but slower) algorithm described in the roadmap
in the introduction,
a variant of Grigoriadis and Khachiyan's algorithm \cite{Grigoriadis95Sublinear}.
We show that it returns a $(1-2\eps)$-approximate primal-dual pair with high probability.

We do not analyze the running time, which can be large.
In the following section, we describe how to modify this algorithm
(using non-uniform increments
and the random sampling trick described in the previous roadmap)
to obtain the full algorithm with a good time bound.

For just this section, assume that each $\MM_{ij}\in [0,1]$.
(Assume as always that $b_i = a_j = 1$ for all $i, j$;
recall that $|v|$ denotes the 1-norm of $v$.)
Here is the algorithm:

\noindent
\begin{alg}
\Ahead{\normalsize{\bf slow-alg}$(\MM\in[0,1]^{\rows\times \columns}, \eps)$ }

\A Vectors $\xp,\xd \leftarrow \mathbf 0$;
 scalar $N=\lceil 2\ln(\rows\columns)/\eps^2\rceil$.

\A {Repeat} until $\max_i \MM_i \xp \ge N$:

\A\tab Let $\pp_i \doteq (1+\eps)^{\MM_i \xp}$ (for all $i$)
and $\pd_j \doteq (1-\eps)^{\MM_j\tran \xd}$ (for all $j$).

\A \tab Choose random indices $j'$ and $i'$ respectively
\\\tab~~ from probability distributions $\pd/|\pd|$ and $\pp/|\pp|$.

\A \tab Increase $\xp_{j'}$ and $\xd_{i'}$ each by 1.

\A Let $(\xp^\star,\xd^\star) \doteq (\xp/\max_i \MM_i\xp, \xd/\min_j \MM_j\tran\xd)$.  \label{slow:scaling}

\A Return $(\xp^\star,\xd^\star)$.
\end{alg}

\smallskip
The scaling of $\xp$ and $\xd$ in line~\ref{slow:scaling} ensures feasibility of the final primal solution $\xp^\star$ and the final dual solution $\xd^\star$.
(Recall the assumption that $b_i = a_j = 1$ for all $i, j$.)
The final primal solution cost and final dual solution costs are, respectively $|\xp^\star| = |\xp|/\max_i M_i\xp$ and $|\xd^\star| = |\xd|/ \min_jM\tran_j\xd$.
Since the algorithm keeps the 1-norms $|\xp|$ and $|\xd|$ of the intermediate primal and dual solutions equal, the final primal and dual costs will be within a factor of $1-2\eps$ of each other as long as $\min_jM\tran_j\xd \ge (1-2\eps)\max_i M_i\xp$.
If this event happens, then by weak duality implies that each solution is a $(1-2\eps)$-approximation of its respective optimum.

To prove that the event
$\min_jM\tran_j\xd \ge (1-2\eps)\max_i M_i\xp$
happens with high probability,
we show that  $|\pp|\cdot|\pd|$ (the product of the 1-norms of $\pp$ and $\pd$,
as defined in the algorithm) is a Lyapunov function --- that is, the product  is non-increasing in expectation with each iteration.
Thus, its expected final value is at most its initial value $rc$,
and with high probability, its final value is at most, say, $(rc)^2$.
If that happens, then by careful inspection of $\pp$ and $\pd$,
it must be that $(1-\eps)\max_i M_i \xp \le \min_j M\tran_j \xd + \eps N$,
which (with the termination condition $\max_i M_i x \ge N$) implies the desired event.\footnote
{It may be instructive to compare this algorithm to the more standard algorithm.
In fact there are two standard algorithms related to this one: a primal algorithm and a dual algorithm.
In each iteration, the primal algorithm would choose $j'$ to minimize $M\tran_{j'}\pp$
and increments $\xp_{j'}$.
Separately and simultaneously, the dual algorithm would choose $i'$ to maximize $(M\pd)_{i'}$,
then increments $\xd_{i'}$.
(Note that the primal algorithm and the dual algorithm are independent,
and in fact either can be run without the other.)
To prove the approximation ratio for the primal algorithm, one would bound the increase in $|\pp|$ relative to the increase in the primal objective $|\xp|$.
To prove the approximation ratio for the dual algorithm, one would bound the decrease in $|\pd|$ relative to the increase in the dual objective $|\xd|$.
In this view, the coupled algorithm can be obtained by taking these two independent primal and dual algorithms
and randomly coupling their choices of $i'$ and $j'$.
The analysis of the coupled algorithm uses as a penalty function $|\pp||\pd|$,
the product of the respective penalty functions $|\pp|,|\pd|$ of the two underlying algorithms.
}

\begin{lemma}\label{lemma:slowalg}
The slow algorithm returns a $(1-2\eps)$-approximate primal-dual pair (feasible primal and dual solutions $\xp^\star$ and $\xd^\star$ such that $|\xp^\star| \ge (1-2\eps)|\xd^\star|$) with probability at least $1-1/(\rows\columns)$.
\end{lemma}
\begin{Proof}
In a given iteration, let $\pp$ and $\pd$ denote the vectors at the start of the iteration.  
Let $\pp'$ and $\pd'$ denote the vectors at the end of the iteration.  
Let $\Delta \xp$ denote the vector whose $j$th entry is the increase in $\xp_j$ during the iteration
(or if $z$ is a scalar, $\Delta z$ denotes the increase in $z$).  
Then, using that each $\Delta\MM_i\xp = \MM_{ij'} \in [0,1]$,
\[
|\pp'|
~=~
\sum_i \pp_i(1 + \eps)^{\MM_i \Delta \xp}
~\le~
\sum_i \pp_i(1 + \eps \MM_i \Delta \xp)
~=~
|\pp|\Big[1+\eps \frac{\pp\tran}{|\pp|} \MM \Delta\xp\Big].
\]
Likewise, for the dual,
$|\pd'|
\le
|\pd|[1-\eps (\pd/|\pd|) \tran \MM\tran \Delta\xd
]$.

\smallskip
Multiplying these bounds on $|p'|$ and $|\hat p'|$
and using that $(1+a)(1-b) = 1+a-b-ab \le 1+a-b$ for $a,b\ge 0$ gives
\[|\pp'||\pd'| 
~\le~
|\pp||\pd| \Big[~ 1 
\,+\, \eps \frac{\pp}{|\pp|} \tran \MM \Delta\xp
\,-\, \eps \Delta\xd\tran \MM \frac{\pd}{|\pd|}
~\Big].
\]

The inequality above is what motivates the ``coupling'' of primal and dual increments.
The algorithm chooses the random increments to $\xp$ and $\xd$
precisely so that $\E[\Delta \xp] = \pd/|\pd|$
and $\E[\Delta \xd] = \pp/|\pp|$.
Taking expectations of both sides of the inequality above,
and plugging these equations into the two terms on the right-hand side,
the two terms exactly cancel,
giving \(\E[|\pp'||\pd'|] \le |\pp||\pd|\).
Thus, the particular random choice of increments to $\xp$ and $\xd$
makes the quantity $|\pp|\,|\pd|$ non-increasing in expectation
with each iteration.

This and Wald's equation (Lemma~\ref{lemma:walds}, or equivalently a standard optional stopping theorem for supermartingales) imply that the expectation of $|\pp||\pd|$ at termination is at most its initial value $\rows\columns$.
So, by the Markov bound, the probability 
that $|\pp||\pd| \ge (\rows\columns)^2$ is at most $1/\rows\columns$.
Thus, with probability at least $1-1/\rows\columns$,
at termination $|\pp||\pd| \le (\rows\columns)^2$.

Assume this happens.
Note that  $(\rows\columns)^2 \le \exp({\eps^2 N})$,
so $|\pp||\pd| \le (\rows\columns)^2$ implies
\((1+\eps)^{\max_i \MM_i \xp} (1-\eps)^{\min_j \MM_j\tran \xd} \le |\pp||\pd| \le \exp({\eps^2 N}). \)
Taking logs, and using the inequalities $1/\ln(1/(1-\eps)) \le 1/\eps$
and $\ln(1+\eps)/\ln(1/(1-\eps))\ge 1-\eps$, gives
\((1-\eps) \max_i \MM_i \xp \le \min_j \MM_j\tran \xd + \eps N. \)

By the termination condition $\max_i \MM_i\xp \ge N$, so the above inequality implies
\((1-2\eps) \max_i \MM_i \xp \le \min_j \MM_j\tran \xd. \)

This and  $|\xp| = |\xd|$ (and weak duality) imply the approximation guarantee
for the primal-dual pair $(\xp^\star, \xd^\star)$ returned by the algorithm.
\end{Proof}

\section{Full algorithm}
This section describes the full algorithm 
and gives a proof of its approximation guarantee.
In addition to the coupling idea explained in the previous section,
for speed
the full algorithm uses non-uniform increments
and estimates of $M\xp$ and $M\tran\xd$ as described in the introduction.
Next we describe some more details of those techniques.
After that we give the algorithm in detail
(although some implementation details that are not crucial to the approximation guarantee
are delayed to the next section).

Recall that WLOG we are assuming $a_i=b_j = 1$ for all $i,j$.
The only assumption on $M$ is $M_{ij} \ge 0$.

\paragraph{Non-uniform increments.}
In each iteration, instead of increasing the randomly chosen $\xp_{j'}$ and $\xd_{i'}$ by 1, the algorithm increases them both by an increment $\dd_{i'j'}$, chosen just so that the maximum resulting increase in any left-hand side (LHS) of any constraint  (i.e.\ $\max_i \Delta \MM_i \xp$ or $\max_j \Delta \MM\tran_j \xd$) is in $[1/4,1]$.  
The algorithm also deletes covering constraints once they become satisfied
(the set $J$ contains indices of not-yet-satisfied covering constraints,
that is $j$ such that $M\tran_j\xd < N$).

We want the analysis of the approximation ratio to continue to hold
(the analogue of Lemma~\ref{lemma:slowalg} for the slow algorithm),
even with the increments adjusted as above.
That analysis requires that the expected change in each $\xp_j$ and each $\xd_i$ should be proportional to $\pd_j$ and $\pp_i$, respectively.
Thus, we adjust the sampling distribution for the random pair $i',j'$ so that,
when we choose $i'$ and $j'$ from the distribution
and increment $\xp_{j'}$ and $\xd_{i'}$ by $\delta_{i'j'}$ as defined above,
it is the case that, for any $i$ and $j$,
$\E[\Delta \xp_j] = \alpha \pd_j/|\pd|$ and $\E[\Delta \xd_i] = \alpha \pp_i/|\pp|$ for an $\alpha>0$.
This is done by scaling the probability of choosing each given $i',j'$ pair by a factor proportional to $1/\delta_{i'j'}$.

To implement the above non-uniform increments 
and the adjusted sampling distribution,
the algorithm maintains the following data structures
as a function of the current primal and dual solutions $\xp$ and $\xd$:
a set $J$ of indices of still-active (not yet met) covering constraints (columns);
for each column $\MM\tran_j$ its maximum entry $\wwp_j = \max_i \MM_{ij}$;
and for each row $\MM_i$ a close upper bound $\wwd_i$ on its maximum active entry $\max_{j\in J} \MM_{ij}$
(specifically, the algorithm maintains $\wwd_i \in [1,2] \times \max_{j\in J} \MM_{ij}$).

Then, the algorithm takes the increment $\dd_{i'j'}$ to be $1/(\wwd_{i'}+\wwp_{j'})$.
This seemingly odd choice has two key properties:
(1) It satisfies $\dd_{i'j'} = \Theta(1/\max(\wwd_{i'},\wwp_{j'}))$,
which ensures that when $\xp_{j'}$ and $\xd_{i'}$ are increased by $\dd_{i'j'}$,
the maximum increase in any LHS (any $\MM_i\xp$, or $\MM\tran_j\xd$ with $j\in J$)
is $\Omega(1)$.
(2) It allows the algorithm to select the random pair $(i',j')$ in constant time using the following subroutine, called {\algfont random-pair} (the notation $\pp\times\wwd$ denotes the vector with $i$th entry $\pp_i\wwd_i$):


\begin{alg}
\Ahead{{\algfont\bf random-pair}$(\pp, \pd,\pp\times\wwd,\pd\times\wwp)$}

\A With probability
$|\pp\times\wwd| |\pd| / (|\pp\times\wwd| |\pd| + |\pp| |\pd\times\wwp|)$
\\ ~~~ choose random $i'$ from distribution  $\pp\times\wwd/|\pp\times\wwd|$, 
\\ ~~~ and independently choose $j'$ from $\pd/|\pd|$,

\A or, otherwise,
\\ ~~~ choose random $i'$ from distribution $\pp/|\pp|$, 
\\ ~~~ and independently choose $j'$ from $\pd\times\wwp/|\pd\times\wwp|$.

\A Return $(i',j')$.
\end{alg}
\smallskip

The key property of {\bf random-pair} is that it makes
the expected changes in $\xp$ and $\xd$ correct:
any given pair $(i,j)$ is chosen with probability proportional to
$\pp_{i}\pd_{j}/\dd_{ij}$,
which makes the expected change in any $\xp_j$ and $\xd_i$,
respectively, is proportional to $\pd_j$ and $\pp_i$.
(See Lemma~\ref{lemma:facts} below.)

\paragraph{Maintaining estimates ($\yp$ and $\yd$) of $M\xp$ and $M\tran\xd$.}
Instead of maintaining the vectors $\pp$ and $\pd$ as direct functions
of the vectors $M\xp$ and $M\tran\xd$, 
to save work,
the algorithm maintains more slowly changing {\em estimates} ($\yp$ and $\yd$)
of the vectors $\MM \xp$ and $\MM\tran \xd$,
and maintains $\pp$ and $\pd$ as functions of the estimates,
rather than as functions of $M\xp$ and $M\tran \xd$.

Specifically, the algorithm maintains $\yp$ and $\yd$ as follows.
When any $\MM_i\xp$ increases by some $\delta \in [0,1]$ in an iteration,
the algorithm increases the corresponding estimate $\yp_i$ 
by 1 with probability $\delta$.
Likewise, when any $\MM\tran_j\xd$ increases by some $\hat\delta\in [0,1]$ in an iteration,
the algorithm increases the corresponding estimate $\yd_j$ 
by 1 with probability $\hat\delta$.
Then, each $\pp_i$ is maintained as $\pp_i = (1+\eps)^{\yp_i}$ instead of $(1+\eps)^{M_i \xp}$,
and each $\pd_j$ is maintained as $\pd_j = (1-\eps)^{\yd_j}$ instead of $(1+\eps)^{M_i \xp}$.
This reduces the frequency of updates to $\pp$ and $\pd$ (and so reduces the total work),
yet maintains $\yp\approx \MM\xp$ and $\yd\approx\MM\tran\xd$ 
with high probability,
which is enough to still allow a (suitably modified) coupling argument to go through.

Each change to a $\yp_i$ or a $\yd_j$ increases the changed element by 1.
Also, no element of $\yp$ or $\yd$ gets larger than $N$ before the algorithm stops 
(or the corresponding covering constraint is deleted).
Thus, in total the elements
of $\yp$ and $\yd$ are changed at most $O((r+c)N) = O((r+c)\log(n)/\eps^2)$ times.
We implement the algorithm to do only {\em constant work} maintaining the remaining
vectors for each such change.
This allows us to bound the total time 
by $O((r+c)\log(n)/\eps^2)$ (plus $O(n)$ pre- and post-processing time).

As a step towards this goal, 
in each iteration, in order to {\em determine} the elements in $\yp$ and $\yd$ that
change, using just $O(1)$ work per changed element, the algorithm uses the following trick.
It chooses a random $\beta\in[0,1]$.
It then increments $\yp_i$ by 1 for those $i$ 
such that the increase $\MM_{ij'}\delta_{i'j'}$ in $\MM_i\xp$ is at least $\beta$.
Likewise, it increments $\yd_j$ by 1 for $j$ such that the increase 
$\MM_{i'j}\delta_{i'j'}$ in $\MM\tran_j\xd$ is at least $\beta$.
To do this efficiently, the algorithm preprocesses $\MM$,
so that within each row $\MM_i$ or column $\MM\tran_j$ of $\MM$,
the elements can be accessed in (approximately) decreasing order
in constant time per element accessed.
(This preprocessing is described in Section~\ref{sec:implementation}.)
This method of incrementing the elements of $\yp$ and $\yd$ uses constant work per changed element and increments each element with the correct probability.
(The random increments of different elements are not independent,
but this is okay because, in the end, each estimate $\yp_j$ and $\yd_i$
will be shown seperately to be correct with high probability.)

\smallskip
The detailed algorithm is shown in Fig.~\ref{fig:alg}, except for
the subroutine {\algfont random-pair} (above)
and some implementation details that are left until Section~\ref{sec:time}.  
\begin{figure*}[t]
\begin{alg}
\Ahead{{\bf solve}$(\MM\in\Rp^{\rows\times \columns}, \eps)$ 
--- {\em return a $(1-6\eps)$-approximate primal-dual pair w/ high prob.}
}

\A {Initialize} vectors $\xp,\xd,\yp,\yd \leftarrow \mathbf 0$, 
and scalar $N=\lceil 2\ln(\rows\columns)/\eps^2\rceil$.

\A Precompute $\wwp_j \doteq \max \{ \MM_{ij} : i \in [\rows]\}$ for $j\in[\columns]$.
(The max.~entry in column $M_j$.)

As $\xp$ and $\xd$ are incremented,
the alg.~maintains $\yp$ and $\yd$ 
so $\E[\yp] = \MM\xp$, $\E[\yd] = \MM\tran\xd$.

It maintains vectors $\pp$ defined by
$\pp_i \doteq (1+\eps)^{\yp_i}$ 
and, as a function of $\yd$:
\[
\begin{array}[c]{rcll}
\JJ & \doteq& \{ j \in [\columns]  : \yd_j \le N \}
& \mbox{(the active columns)}
\smallskip
\\
\wwd_i &\in& [1,2]\times\max \{ \MM_{ij} : j\in \JJ\}
& \mbox{(approximates the max.~active entry in row $i$ of $M$)}
\\
\pd_j & \doteq & 
\left\{
\begin{array}{ll}
\displaystyle (1-\eps)^{\yd_j}   & \text{if } j\in \JJ
\\ 0 & \text{otherwise.}
\end{array}
\right.
\end{array}
\]
It maintains vectors $\pp\times\wwd$ and $\pd\times\wwp$,
where $a\times b$ is a vector whose $i$th entry is $a_i b_i$.

\A {Repeat} until $\max_i \yp_i = N$ or $\min_j \yd_j = N$:

\algbeg
\A Let $(i',j')\leftarrow \mbox{\bf random-pair}(\pp, \pd,\pp\times\wwd, \pd\times\wwp)$.
\label{line:sample}

\A {Increase} $\xp_{j'}$ and $\xd_{i'}$ each 
by the same amount $\dd_{i'j'} \doteq 1/(\wwd_{i'} + \wwp_{j'})$.
\label{line:incx}

\A Update $\yp$, $\yd$, and the other vectors as follows:  

\A \tab Choose random $\beta\in[0,1]$ uniformly, and 

\A \tab~ for each $i\in[\rows]$ with $\MM_{ij'}\dd_{i'j'}\ge \beta$, 
{increase} $\yp_{i}$ by $1$
\label{line:incyp}
\A \tab~ \hfill (and multiply $\pp_{i}$ and $(\pp\times\wwd)_{i}$ by $1+\eps$);

\A \tab~ for each\, $j\,\in\, \JJ$\, with $\MM_{i'j}\dd_{i'j'}\ge \beta$,
{increase} $\yd_{j}$ by $1$
\label{line:incyd}
\A \tab~ \hfill (and multiply $\pd_{j}$ and $(\pd\times\wwp)_{j}$ by $1-\eps$).

\A For each $j$ leaving $\JJ$, update $\JJ$, $\wwd$, and $\pp\times\wwd$.

\algend

\A Let $(\xp^\star,\xd^\star) \doteq (\xp/\max_i \MM_i\xp, \xd/\min_j \MM_j\tran\xd)$.
{Return} $(\xp^\star,\xd^\star)$.
\end{alg}

\caption{The full algorithm.
$[i]$ denotes $\{1,2,\ldots,i\}$.
Implementation details are in Section~\ref{sec:implementation}.
}
\label{fig:alg}
\end{figure*}

\paragraph{Approximation guarantee.} \label{sec:correctness}
Next we state and prove the approximation guarantee for the full algorithm
in Fig.~\ref{fig:alg}.   We first prove three utility lemmas.
The first utility lemma establishes that (in expectation)
$\xp$, $\xd$, $\yp$, and $\yd$ change 
as desired in each iteration.
\newpage

\begin{lemma}\label{lemma:facts}
In each iteration, 
\begin{enumerate}
\item 
The largest change in any relevant LHS is at least 1/4:
\[\max\{\max_{i} \Delta \MM_{i} \xp, \max_{j\in \JJ} \Delta \MM\tran_{j} \xd\}
~\in~ [1/4,1].\]

\item Let $\alpha\doteq |\pp||\pd|/\sum_{ij} \pp_i\pd_j/\dd_{ij}$.
The expected changes in each $\xp_j$, $\xp_j$, $\yp_i$, $\yd_j$ satisfy

\[
\begin{array}{r@{~}c@{~}lr@{~}c@{~}c@{~}c@{~}l}
E[\Delta \xp_j] & = & \alpha \pd_j/|\pd|,
& \E[\Delta \yp_i] & = & \E[\Delta \MM_i \xp] & = & \alpha \MM\pd_i/|\pd|,
\\[6pt]
\E[\Delta \xd_i] &=& \alpha \pp_i/|\pp|,
& \E[\Delta \yd_j] &=& \E[\Delta \MM_j\tran \xd] &=& \alpha \MM\tran \pp_j/|\pp|.
\end{array}
\]
\end{enumerate}
\end{lemma}
\begin{Proof}
(i) By the choice of $\wwd$ and $\wwp$, for the $(i',j')$ chosen,
the largest change in a relevant LHS is
\begin{eqnarray*}
\delta_{i'j'}\max\big(
  \max_{i} \MM_{ij'}, 
  \max_{j\in \JJ} \MM_{i'j}\big)
&\in& [1/2,1]\,\dd_{i'j'} \max(\wwd_{i'}, \wwp_{j'})\\
&\subseteq& [1/4,1]\, \dd_{i'j'} (\wwd_{i'}+ \wwp_{j'})\\
&=& [1/4,1].
\end{eqnarray*}

\noindent
(ii) First, we verify that the probability that {\algfont random-pair} returns a given $(i,j)$ is 
$\alpha (\pp_{i}/|\pp|)(\pd_{j}/|\pd|)/\dd_{ij}$.
Here is the calculation.
By inspection of {\algfont random-pair}, the probability is proportional to
\[
|\pp\times\wwd|\,|\pd| 
\frac{\pp_{i}\wwd_{i}}{|\pp\times\wwd|}
\frac{\pd_{j}}{|\pd|}
+
|\pp|\,|\pd\times\wwp| 
\frac{\pp_{i}}{\pp}
\frac{\pd_{j}\wwp_{j}}{|\pd\times\wwp|}
\]
which by algebra simplifies to
$\pp_i\pd_j(\wwd_i+\wwp_j) = \pp_{i}\pd_{j}/\dd_{ij}$.

Hence, the probability must be $\alpha (\pp_{i}/|\pp|)(\pd_{j}/|\pd|)/\dd_{ij}$,
because the choice of $\alpha$ makes the sum over all $i$ and $j$ of the probabilities equal 1.

Next, note that part (i) of the lemma implies that in line~\ref{line:incyp} (given the chosen $i'$ and $j'$) the probability that a given $\yp_{i}$ is incremented is $\MM_{ij'}\dd_{i'j'}$, while in line~\ref{line:incyd} the probability that a given $\yd_{j}$ is incremented is $\MM_{i'j}\dd_{i'j'}$.

Now, the remaining equalities in (ii) follow by direct calculation.
For example:

\bigskip
\hfill
$\displaystyle\E[\Delta\xp_j] = \sum_{i} (\alpha \pp_{i}/|\pp|)(\pd_j/|\pd|)/\dd_{ij})\dd_{ij} = \alpha \pd_j/|\pd|$.
\hfill
\end{Proof}

The next lemma shows that (with high probability) the estimate vectors $\yp$ and $\yd$ suitably approximate $\MM\xp$ and $\MM\tran\xd$, respectively.
The proof is simply an application of an appropriate Azuma-like inequality
(tailored to deal with the random stopping time of the algorithm).

\begin{lemma} \label{lemma:bleah}~
\begin{enumerate}
\item
For any $i$, with probability at least
$1-1/(\rows\columns)^2$, at termination
$(1-\eps) \MM_i\xp \,\le\, \yp_i + \eps N$.
\item
For any $j$, with probability at least
$1-1/(\rows\columns)^2$, after the last iteration with $j\in\JJ$,
it holds that $(1-\eps) \yd_j \,\le\, \MM_j\tran \xd + \eps N$.
\end{enumerate}
\end{lemma}
\begin{Proof}
(i) By Lemma~\ref{lemma:facts}, in each iteration each $\MM_i\xp$ and $\yp_i$ increase by at most 1 and the expected increases in these two quantities are the same.
So, by the Azuma inequality for random stopping times (Lemma~\ref{lemma:chernoff}),
$\Pr[(1-\eps) \MM_i\xp \ge \yp_i + \eps N]$ 
is at most
$\exp(-\eps^2N) \le 1/(\rows\columns)^2$.
This proves (i).  

The proof for (ii) is similar, noting that, while $j\in\JJ$,
the quantity $\MM_j\tran\xd$ increases by at most 1 each iteration.
\end{Proof}

Finally, here is the main utility lemma.
Recall that the heart of the analysis of the slow algorithm
(Lemma~\ref{lemma:slowalg})
was showing that in expectation $|\pp||\pd|$ was non-increasing.
This allowed us to conclude that (with high probability at the end)
$\max_i M_i\xp$ was not much larger than $\min_j M\tran_j \xd$.
This was the key to proving the approximation ratio.

The next lemma gives the analogous argument for the full algorithm.
It shows that the quantity $|\pp||\pd|$
is non-increasing in expectation,
which, by definition of $\pp$ and $\pd$,
implies that (with high probability at the end)
$\max_i \yp_i$ is not much larger than $\min_j \yd_j$.
The proof is essentially the same as that of Lemma~\ref{lemma:slowalg},
but with some technical complications accounting for the deletion
of covering constraints.

Since (with high probability by Lemma~\ref{lemma:bleah})
the estimates $\yp$ and $\yd$ approximate $M\xp$
and $M\xd$, respectively,
this implies that (with high probability at the end)
$\max_i M_i\xp$ is not much larger than $\min_j M_j \tran\xd$.
Since the algorithm maintains $|\xp| = |\xd|$,
this is enough to prove the approximation ratio.

\begin{lemma} \label{lemma:lyapunov}
With probability at least $1-1/\rows\columns$,
when the algorithm stops,
$\max_i \yp_i  \le N$ and $\min_j \yd_j \ge (1-2\eps)N$.
\end{lemma}
\begin{Proof} 
Let $\pp'$ and $\pd'$ denote $\pp$ and $\pd$ after a given iteration,
while $\pp$ and $\pd$ denote the values before the iteration.
We claim that, given $\pp$ and $\pd$, 
$\E[|\pp'|\,|\pd'|] \le |\pp|\,|\pd|$
--- with each iteration $|\pp|\,|\pd|$ is non-increasing in expectation.
To prove it, note
\(
|\pp'| \,=\,
\sum_{i} \pp_i(1+\eps \Delta\yp_i)
\,=\,
|\pp| + \eps \pp\tran\Delta\yp
\)
and, similarly,
\(
|\pd'| \,=\,|\pd| - \eps \pd\tran\Delta\yd
\)
(recall $\Delta\yp_i,\Delta\yd_j \in \{0,1\}$).
Multiplying these two equations
and dropping a negative term gives
\[
|\pp'|\,|\pd'| ~\le~
|\pp|\,|\pd|
 + \eps |\pd|\pp\tran\Delta\yp
 - \eps |\pp|\pd\tran\Delta\yd.
\]
The claim follows by taking expectations of both sides,
then, in the right-hand side applying linearity of expectation and
substituting
 $\E[\Delta \yp] = \alpha\MM\pd/|\pd|$
and
 $\E[\Delta \yd] = \alpha\MM\tran\pp/|\pp|$
from Lemma~\ref{lemma:facts}.

By Wald's equation (Lemma~\ref{lemma:walds}),
the claim implies that $\E[|\pp|\,|\pd|]$ for $\pp$ and $\pd$ at termination
is at most its initial value $\rows\columns$.
Applying the Markov bound, with probability at least $1-1/\rows\columns$,
at termination $\max_i \pp_i \max_j \pd_j \le |\pp||\pd| \le (\rows\columns)^2 \le \exp(\eps^2 N)$.

Assume this event happens.
The index set $\JJ$ is not empty at termination, so the minimum
$\yd_j$ is achieved for $j\in\JJ$.
Substitute in the definitions of $\pp_i$ and $\pd_j$
and take log to get
$\max_i \yp_i\ln(1+\eps)  
\le \min_j \yd_j\ln(1/(1-\eps)) + \eps^2 N$.

Divide by $\ln(1/(1-\eps))$, apply
$1/\ln(1/(1-\eps))\le 1/\eps$
and also
$\ln(1+\eps)/\ln(1/(1-\eps))\ge 1-\eps$.
This gives
$(1-\eps) \max_i \yp_i 
\le \min_j \yd_j + \eps N$.

By the termination condition $\max_i\yp_i\le N$ is guaranteed,
and either  $\max_i\yp_i=N$ or $\min_j\yd_j = N$.
If $\min_j\yd_j = N$, then the event in the lemma occurs.
If not, then $\max_i\yp_i=N$, 
which (with the inequality in previous paragraph) 
implies $(1-\eps) N \le \min_j\yd_j + \eps N$,
again implying the event in the lemma.
\end{Proof}

Finally, here is the approximation guarantee (Theorem~\ref{thm:correctness}).
It follows from the three lemmas above by straightforward algebra.

\begin{theorem}\label{thm:correctness}
With probability at least $1-3/\rows\columns$, the algorithm in Fig.~\ref{fig:alg} returns feasible primal and dual solutions
$(\xp^\star,\xd^\star)$ with $|\xp^\star|/|\xd^\star| \ge 1-6\eps$.
\end{theorem}
\begin{Proof}
Recall that the algorithm returns
 $(\xp^\star,\xd^\star) \doteq (\xp/\max_i \MM_i\xp, \xd/\min_j \MM_j\tran\xd)$.
By the naive union bound,
with probability at least $1-3/\rows\columns$
(for all $i$ and $j$) the events in Lemma~\ref{lemma:bleah} occur,
and the event in Lemma~\ref{lemma:lyapunov} occurs.
Assume all of these events happen.  Then, at termination, for all $i$ and $j$,

\bigskip
\noindent
$
\begin{array}{@{}l@{}r@{~~}c@{~~}lcr@{~~}c@{~~}l}
&(1-\eps)M_i \xp & \le & \yp_i + \eps N 
& & (1-2\eps) N  & \le & \yd_j
\\
&\yp_i & \le &  N
& \raisebox{6pt}{\mbox{~~~~and~~~~}} & (1-\eps)\yd_j&\le&M\tran_j \xd + \eps N .
\\[9pt]
\lefteqn{\parbox{0.95\textwidth}{~~By algebra, using $(1-a)(1-b)\ge 1-a-b$ and $1/(1+\eps)\ge 1-\eps$, it follows for all $i$ and $j$ that}}
\\[13pt]
& \hspace*{0.5in}(1-2\eps)\MM_i\xp& \le & N
& \mbox{ and } &
(1-4\eps)N & \le & \MM_j\tran \xd. 
\end{array}
$
\bigskip

This implies $\min_j\MM_j\tran\xd / \max_i\MM_i\xp \ge 1-6\eps$.

The scaling at the end of the algorithm assures that $\xp^\star$ and $\xd^\star$ are feasible.
Since the sizes $|\xp|$ and $|\xd|$ increase by the same amount each iteration, they are equal.
Thus, the ratio of the primal and dual objectives 
is $|\xp^\star|/|\xd^\star| = \min_j\MM_j\tran\xd / \max_i\MM_i\xp \ge 1-6\eps$.
\end{Proof}

\section{Implementation details and running time}\label{sec:implementation}\label{sec:time}

This section gives remaining implementation details for the algorithm
and bounds the running time.
The remaining implementation details concern the maintenance of the vectors
$(\xp,\xd,\yp,\yd,\pp,\pd,\wwp,\wwd, \pp\times \wwd, \pd\times\wwp)$
so that each update to these vectors can be implemented in constant time
and {\algfont random-pair} can be implemented in constant time.

The matrix $\MM$ should be given in any standard sparse representation,
so that the non-zero entries can be traversed 
in time proportional to the number of non-zero entries.

\subsection{Simpler implementation}
First, here is an implementation that takes 
$O(\inputsize\log \inputsize + (\rows+\columns)\log(\inputsize)/\eps^2)$ time.
(After this we describe how to modify this implementation
to remove the $\log \inputsize$ factor from the first term.)

\begin{theorem}\label{thm:slow}
The algorithm can be implemented to return a $(1-6\eps)$-approximate
primal-dual pair for \prob{packing} and \prob{covering} in time
$O(\inputsize\log \inputsize +(\rows+\columns)\log(\inputsize)/\eps^2)$
with probability at least $1-4/rc$.
\end{theorem}
\begin{Proof}
To support {\algfont random-pair},
store each of the four vectors $\pp,\pd,\pp\times\wwd,\pd\times\wwp$ in its own random-sampling data structure \cite{matias2003dgd}
(see also \cite{hagerup1993oag}).
This data structure maintains a vector $v$;
it supports random sampling from the distribution $v/|v|$
and changing any entry of $v$ in constant time.
Then {\algfont random-pair} runs in constant time,
and each update of an entry of  $\pp$, $\pd$, $\pp\times\wwd$, or $\pd\times\wwp$
takes constant time.

Updating the estimates $\yp$ and $\yd$ in each iteration
requires, given $i'$ and $j'$, identifying which $j$ and $i$ are such that $M_{i'j}$
and $M_{ij'}$ are at least $\beta/\delta_{i'j'}$ (the corresponding elements $\yp_i$ and $\yd_j$ get increased).
To support this efficiently, at the start of the algorithm, preprocess the matrix $\MM$.
Build, for each row and column, a doubly linked list of the non-zero entries.
{\em Sort each list in descending order.} 
Cross-reference the lists so that, 
given an entry $\MM_{ij}$ in the $i$th row list, 
the corresponding entry $\MM_{ij}$ in the $j$th column list 
can be found in constant time.
The total time for preprocessing is
$O(\inputsize\log \inputsize)$.

Now implement each iteration as follows.
Let $\Ip_t$ denote the set of indices $i$ for which $\yp_i$ is incremented in line \ref{line:incyp} in iteration $t$.
From the random $\beta\in[0,1]$ and the sorted list for row $j'$, 
compute this set $\Ip_t$ by traversing the list for row $j'$ in order of decreasing $M_{ij'}$, 
collecting elements until an $i$ with $M_{ij'}<\beta/\dd_{i'j'}$ is encountered.
Then, for each $i\in\Ip_t$, 
update $\yp_i$, $\pp_{i}$, and the $i$th entry in $\pp\times\wwd$ in constant time.
Likewise, let $\Jd_t$ denote the set of indices $j$ for which $\yd_j$ is incremented in line \ref{line:incyd}.
Compute $\Jd_t$ from the sorted list for column $i'$.
For each $j\in\Jd_t$, 
update $\pd_{j}$,  and the $j$th entry in $\pd\times\wwp$.
The total time for these operations during the course of the algorithm is
$O(\sum_t 1+|\Ip_t|+|\Jd_t|)$.

For each element $j$ that leaves $\JJ$ during the iteration, update $\pd_j$.
Delete all entries in the $j$th column list from all row lists.
For each row list $i$ whose first (largest) entry is deleted, 
update the corresponding $\wwd_{i}$ 
by setting $\wwd_{i}$ to be the next (now first and maximum) entry
remaining in the row list;
also update $(\pp\times\wwd)_i$.
The total time for this during the course of the algorithm is
$O(\inputsize )$, because each $\MM_{ij}$ is deleted at most once.

This completes the implementation.

By inspection, the total time is
$O(\inputsize\log \inputsize)$ (for preprocessing, and deletion of covering constraints)
plus $O(\sum_t1+|\Ip_t|+|\Jd_t|)$ (for the work done as a result of the increments).

The first term $O(n\log n)$ above is in its final form.
The next three lemmas bound the second term (the sum).
The first lemma bounds the sum except for the ``1''.
That is, it bounds the number of times any $\yp_i$ or $\yd_j$ is incremented.
(There are $r+c$ elements, and each can be incremented at most $N$ times
during the course of the algorithm.)

\begin{lemma}\label{lemma:totaltime}
\[\sum_t |\Ip_t|+|\Jd_t| \le (\rows+\columns) N 
~=~
O((\rows+\columns)\log(\inputsize)/\eps^2).\]
\end{lemma}
\begin{Proof}
First, $\sum_t |\Ip_t| \le \rows N$ 
because each $\yp_{i}$ 
can be increased at most $N$ times 
before $\max_i\yp_i \ge N$ (causing termination).
Second, $\sum_t |\Jd_t| \le \columns N$ 
because each $\yd_{j}$ 
can be increased at most $N$ times 
before $j$ leaves $\JJ$ and ceases to be updated.
\end{Proof}

The next lemma bounds the remaining part of the second term,
which is $O(\sum_t1)$.
Given that $\sum_t|\Ip_t|+|\Jd_t| \le (\rows+\columns)N$,
it's enough to bound the number of iterations $t$
where $|\Ip_t|+|\Jd_t|=0$.
Call such an iteration {\em empty}.
(The 1's in the non-empty iterations contribute 
at most $\sum_t |\Ip_t|+|\Jd_t| \le (\rows+\columns)N$ to the sum.)

We first show that each iteration is non-empty with probability at least 1/4.
This is so because, for any $(i',j')$ pair chosen in an iteration,
for the constraint that determines the increment $\delta_{i'j'}$,
the expected increase in the corresponding $\yp_i$ or $\yd_j$ 
must be at least 1/4, and that element will be incremented
(making the iteration non-empty)
with probability at least 1/4.

\begin{lemma}\label{lemma:threequarters}
Given the state at the start of an iteration,
the probability that it is empty is at most 3/4.
\end{lemma}
\begin{Proof}
Given the $(i',j')$ chosen in the iteration, 
by (1) of Lemma~\ref{lemma:facts},
by definition of $\delta_{i'j'}$,
there is either an $i$ such that
$\MM_{ij'}\dd_{i'j'} \ge 1/4$
or a $j$ such that
$\MM_{i'j}\dd_{i'j'} \ge 1/4$.
In the former case, $i\in\Ip_t$ with probability at least $1/4$.
In the latter case, $j\in\Jd_t$ with probability at least $1/4$.
\end{Proof}

This implies that, with high probability, the number of empty iterations
does not exceed three times the number of non-empty iterations by much.
(This follows from the Azuma-like inequality.)   We have already bounded
the number of non-empty iterations, so this implies a bound (with high probability)
on the number of empty iterations.

\begin{lemma}\label{lemma:empty}
With probability at least $1-1/rc$, the number of empty iterations is $O((r+c)N)$.
\end{lemma}
\begin{Proof}
Let $E_t$ be 1 for empty iterations and 0 otherwise.
By the previous lemma and the Azuma-like inequality tailored for random stopping times (Lemma~\ref{lemma:chernoff}), for any $\delta,A\ge 0$,
\[\Pr\Big[~\textstyle (1-\delta)\sum_{t=1}^T E_t ~\ge~ 3 \sum_{t=1}^T (1-E_t) ~+~ A~\Big]
~\le~\exp(-\delta A).
\]
Taking $\delta=1/2$ and $A =2\ln(rc)$,
it follows that with probability at least $1-1/rc$,
the number of  empty iterations
is bounded by a constant times the number of non-empty iterations
plus $2\ln(rc)$.
The number of non-empty iterations is at most $(r+c)N$,
hence, with probability at least $1-1/rc$
the number of empty iterations is $O((r+c)N)$.
\end{Proof}

Finally we complete the proof of Theorem~\ref{thm:slow}, stated at the top of the section.

As discussed above, the total time is
$O(\inputsize\log \inputsize)$ (for preprocessing, and deletion of covering constraints)
plus $O(\sum_t1+|\Ip_t|+|\Jd_t|)$ (for the work done as a result of the increments).

By Lemma~\ref{lemma:totaltime},
$\sum_t |\Ip_t|+|\Jd_t| = O((r+c)\log(n)/\eps^2)$.
By Lemma~\ref{lemma:empty}, with probability $1-1/rc$,
the number of iterations $t$ such that $|\Ip_t|+|\Jd_t| = 0$
is $O((r+c)\log(n)/\eps^2)$.
Together, these imply that, with probability $1-1/rc$,
and the total time is
$O(n\log n + (r+c)\log(n)/\eps^2)$.
This and Theorem~\ref{thm:correctness} imply Theorem~\ref{thm:slow}.
\end{Proof}

\subsection{Faster implementation.}

To prove the main result,
it remains to describe how to remove the $\log \inputsize$ factor
from the $\inputsize\log \inputsize$ term in the time bound in the previous section.

The idea is that it suffices 
to {\em approximately} sort the row and column lists,
and that this can be done in linear time.

\begin{theorem}\label{thm:fast}
The algorithm can be implemented to return a $(1-7\eps)$-approximate primal-dual pair for \prob{packing} and \prob{covering} in time $O(\inputsize+(\columns+\rows)\log(\inputsize)/\eps^2)$
with probability at least $1-5/rc$.
\end{theorem}
\begin{Proof}
Modify the algorithm as follows.

First, preprocess $\MM$ as described in \cite[\S2.1]{Luby93Parallel}
so that the non-zero entries have bounded range.
Specifically, 
let $\beta=\min_j\max_i \MM_{ij}$. 
Let $\MM'_{ij} \doteq 0$ if $\MM_{ij} < \beta \eps/\columns$
and $\MM'_{ij} \doteq \min\{\beta \columns/\eps,\MM_{ij}\}$ otherwise.
As shown in \cite{Luby93Parallel},
any $(1-6\eps)$-approximate primal-dual pair for the transformed problem
will be a $(1-7\eps)$-approximate primal-dual pair for the original problem.

In the preprocessing step, instead of sorting the row and column lists,
{\em pseudo-sort} them --- sort them based on keys
$\lfloor \log_2 \MM_{ij}\rfloor$.  
These keys will be integers in the range
$\log_2(\beta)\pm \log(\columns/\eps)$.  
Use bucket sort, so that a row or column with $k$ entries
can be processed in $O(k+\log(\columns/\eps))$ time.  
The total time for pseudo-sorting the rows and columns is
$O(\inputsize +(\rows+\columns)\log(\columns/\eps))$.

Then, in the $t$th iteration, maintain the data structures as before, 
except as follows.

Compute the set $\Ip_t$ as follows.
Traverse the pseudo-sorted $j$th column
until an index $i$ with $\MM_{ij'}\dd_{i'j'} < \beta/2$ is found.
(No indices later in the list can be in $\Ip_t$.)
Take all the indices $i$ seen with $\MM_{ij'}\dd_{i'j'} \ge \beta$.
Compute the set $\Jd_t$ similarly.
Total time for this is 
$O(\sum_t 1+ |\Ip'_t| + |\Jd'_t|)$,
where $\Ip'_t$ and $\Jd'_t$
denote the sets of indices actually traversed
(so $\Ip_t\subseteq\Ip'_t$
and $\Jd_t\subseteq\Jd'_t$).

When an index $j$ leaves the set $\JJ$,
delete all entries in the $j$th column list from all row lists.
For each row list affected, set $\wwd_{i}$ 
to {\em two} times the first element remaining in the row list.
This ensures  $\wwd_{i} \in [1,2]\max_{j\in\JJ}\MM_{ij}$.

These are the only details that are changed.  

The total time is now
$O(\inputsize + (\rows+\columns)\log(\columns/\eps))$ for preprocessing and deletion of covering constraints,
plus $O(\sum_t 1+|\Ip'_t| + |\Jd'_t|)$ to implement the increments and vector updates.
To finish, the next lemma bounds the latter term.
The basic idea is that, in each iteration,
each matrix entry is at most {\em twice} as likely to be examined
as it was in the previous algorithm.
Thus, with high probability, 
each matrix element is examined at most about twice as often
as it would have been in the previous algorithm.

\begin{lemma}\label{lemma:losetwo}
With probability at least $1-2/rc$,
it happens that
\(\sum_t (1+ |\Ip'_t| + |\Jd'_t|)~=~O((r+c)N).\)
\end{lemma}
\begin{Proof}
Consider a given iteration.  
Fix $i'$ and $j'$ chosen in the iteration.
For each $i$, note that, for the random $\beta\in[0,1]$,
\begin{eqnarray*}
\Pr[ i \in \Ip'_t ] 
~\le~
\Pr[ \beta/2 \le \MM_{ij'}\dd_{i'j'} ]
&~\le~&
2\MM_{ij'}\dd_{i'j'}
\\&=&
2\Pr[ \beta \le \MM_{ij'}\dd_{i'j'} ]
~=~
 2\Pr[i\in\Ip_t].
\end{eqnarray*}
Fix an $i$.
Applying Azuma-like inequality for random stopping times (Lemma~\ref{lemma:chernoff}),
for any $\delta,A\ge 0$,
\[\textstyle
\Pr\Big[~ (1-\delta) \sum_t [i\in\Ip'_t] ~\ge~ 2\sum_t [i\in\Ip_t] ~+~ A ~\Big]
~\le~\exp(-\delta A).
\]
(Above $[i\in S]$ denotes 1 if $i\in S$ and 0 otherwise.)

Taking $\delta=1/2$ and $A=4\ln(rc)$,
with probability at least $1-(rc)^2$,
it happens that
\[\textstyle
\sum_t [i\in\Ip'_t]~\le~ 4\sum_t [i\in\Ip_t] \,+\, 8\ln(rc).\]

Likewise, for any $j$, with probability at least $1-1/(rc)^2$,
we have that $\sum_t [j\in\Jd'_t] ~\le~2\sum_t [j\in\Jd_t] \,+\, 8\ln(rc)$.

Summing the naive union bound over all $i$ and $j$,
with probability at least $1-1/rc$,
it happens that the sum
$\sum_t ( |\Ip'_t| + |\Jd'_t|)$ 
is at most
$4\sum_t (|\Ip_t| + |\Jd_t|) \,+\, 8(r+c)\ln(rc)$.

By Lemma~\ref{lemma:totaltime} the latter quantity is $O((r+c)N)$.

By Lemma~\ref{lemma:empty},
the number of empty iterations is still
$O((r+c)N)$ with probability at least $1-1/rc$.
The lemma follows by applying the naive union bound.
\end{Proof}

If the event in the lemma happens, then the total time is
$O(\inputsize + (\rows+\columns)\log(\inputsize)/\eps^2)$.
This proves Theorem~\ref{thm:fast}.
\end{Proof}

\section{Empirical Results}\label{sec:experiments}

We performed an experimental evaluation of our algorithm
and compared it against Simplex on randomly generated 0/1 input matrices.
These experiments suffer from the following limitations:
(i) the instances are relatively small,
(ii) the instances are random and thus not representative of practical applications,
(iii) the comparison is to the publicly available GLPK (GNU Linear Programming Kit),
not the industry standard CPLEX.
With those caveats, here are the findings.

The running time of our algorithm is well-predicted by the analysis, with a leading constant factor of about 12 basic operations in the big-O term in which $\eps$ occurs.

For moderately large inputs, the algorithm can be substantially faster than Simplex (GLPK -- Gnu Linear Programming Kit -- Simplex algorithm glpsol version 4.15 with default options).\footnote{%
Preliminary experiments suggest that the more sophisticated CPLEX implementation
is faster than GLPK Simplex, but, often, only by a factor of five or so.
Also, preliminary experiments on larger instances than are considered here
suggest that the running time of Simplex and interior-point methods, including CPLEX implementations on random instances grows more rapidly than estimated here.
}
{\em The empirical running times reported here for Simplex are to find a $(1\pm\eps)$-approximate solution.}  

For inputs with 2500-5000 rows and columns, the algorithm (with $\eps=0.01$) is faster than Simplex by factors ranging from tens to hundreds.
For larger instances, the speedup grows roughly linearly in $\rows\columns$.
For instances with moderately small $\eps$ 
and thousands (or more) rows and columns,
the algorithm is orders of magnitude faster than Simplex.

The test inputs had
$r,c \in [739,5000]$,
$\eps\in \{0.02,0.01,0.005\}$,
and matrix density $d\in\{1/2^k : k = 1,2,3,4,5,6\}$.
For each $(r,c,d)$ tuple 
there was a random 0/1 matrix with $r$ rows and $c$ columns,
where each entry was 1 with probability $d$.
The algorithm here was run on each such input, with each $\eps$.
The running time was compared to that taken by a Simplex solver
to find a $(1-\eps)$-approximate solution.

GLPK Simplex failed to finish due to cycling on about 10\% of the initial runs; 
those inputs are excluded from the final data.
This left 167 runs.
The complete data for the non-excluded runs is given in the tables at the end of the section.

\subsection{Empirical evaluation of this algorithm}

The running time of  the algorithm here includes
(A) time for preprocessing and initialization,
(B) time for sampling (line~\ref{line:sample}, once per iteration of the outer loop),
and (C) time for increments (lines~\ref{line:incyp} and~\ref{line:incyd}, once per iteration of the inner loops).
Theoretically the dominant terms are $O(n)$ for (A) and $O((r+c)\log(n)/\eps^2)$ for (C).
For the inputs tested here, the significant terms in practice are for (B) and (C),
with the role of (B) diminishing for larger instances.
The time (number of basic operations) is well-predicted by the expression
\begin{equation}\label{eqn:predtime}
[12(r+c) ~+~ 480 d^{-1}]\frac{\ln(rc)}{\eps^2}
\end{equation}
where $d=1/2^k$ is the density (fraction of matrix entries that are non-zero, at least $1/\min(r,c)$).

The $12(r+c)\ln(rc)/\eps^2$ term is the time spent in (C), the inner loops;
it is the most significant term in the experiments as $r$ and $c$ grow.
The less significant term $480 d^{-1}\ln(rc)/\eps^2$ is for (B), and is
proportional to the number of samples (that is, iterations of the outer loop).
Note that this term {\em decreases} as matrix density increases.
(For the implementation we focused on reducing the time for (C), not for (B).
It is probable that the constant 480 above can be reduced with a more careful implementation.)


The plot below shows the run time in seconds, divided by the predicted time
(the predicted number of basic operations~(\ref{eqn:predtime})
times the predicted time per basic operation):
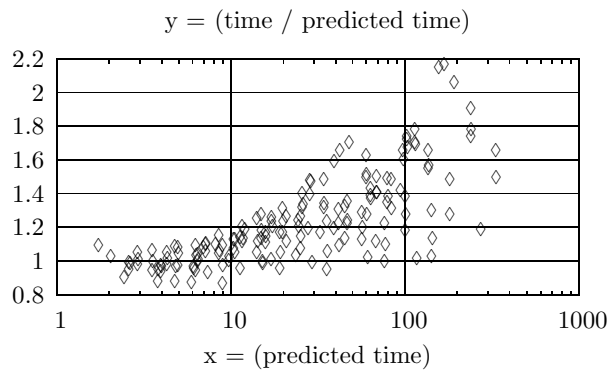
\begin{center}
\noindent\hspace*{-1em}
\scalebox{0.67}{
\setlength{\unitlength}{0.240900pt}
\ifx\plotpoint\undefined\newsavebox{\plotpoint}\fi
\sbox{\plotpoint}{\rule[-0.200pt]{0.400pt}{0.400pt}}%
\begin{picture}(1500,900)(0,0)
\font\gnuplot=cmr10 at 14pt
\gnuplot
\sbox{\plotpoint}{\rule[-0.200pt]{0.400pt}{0.400pt}}%
\put(203.0,174.0){\rule[-0.200pt]{291.489pt}{0.400pt}}
\put(203.0,174.0){\rule[-0.200pt]{4.818pt}{0.400pt}}
\put(174,174){\makebox(0,0)[r]{ 0.8}}
\put(1393.0,174.0){\rule[-0.200pt]{4.818pt}{0.400pt}}
\put(203.0,253.0){\rule[-0.200pt]{291.489pt}{0.400pt}}
\put(203.0,253.0){\rule[-0.200pt]{4.818pt}{0.400pt}}
\put(174,253){\makebox(0,0)[r]{ 1}}
\put(1393.0,253.0){\rule[-0.200pt]{4.818pt}{0.400pt}}
\put(203.0,332.0){\rule[-0.200pt]{291.489pt}{0.400pt}}
\put(203.0,332.0){\rule[-0.200pt]{4.818pt}{0.400pt}}
\put(174,332){\makebox(0,0)[r]{ 1.2}}
\put(1393.0,332.0){\rule[-0.200pt]{4.818pt}{0.400pt}}
\put(203.0,411.0){\rule[-0.200pt]{291.489pt}{0.400pt}}
\put(203.0,411.0){\rule[-0.200pt]{4.818pt}{0.400pt}}
\put(174,411){\makebox(0,0)[r]{ 1.4}}
\put(1393.0,411.0){\rule[-0.200pt]{4.818pt}{0.400pt}}
\put(203.0,489.0){\rule[-0.200pt]{291.489pt}{0.400pt}}
\put(203.0,489.0){\rule[-0.200pt]{4.818pt}{0.400pt}}
\put(174,489){\makebox(0,0)[r]{ 1.6}}
\put(1393.0,489.0){\rule[-0.200pt]{4.818pt}{0.400pt}}
\put(203.0,568.0){\rule[-0.200pt]{291.489pt}{0.400pt}}
\put(203.0,568.0){\rule[-0.200pt]{4.818pt}{0.400pt}}
\put(174,568){\makebox(0,0)[r]{ 1.8}}
\put(1393.0,568.0){\rule[-0.200pt]{4.818pt}{0.400pt}}
\put(203.0,647.0){\rule[-0.200pt]{291.489pt}{0.400pt}}
\put(203.0,647.0){\rule[-0.200pt]{4.818pt}{0.400pt}}
\put(174,647){\makebox(0,0)[r]{ 2}}
\put(1393.0,647.0){\rule[-0.200pt]{4.818pt}{0.400pt}}
\put(203.0,726.0){\rule[-0.200pt]{291.489pt}{0.400pt}}
\put(203.0,726.0){\rule[-0.200pt]{4.818pt}{0.400pt}}
\put(174,726){\makebox(0,0)[r]{ 2.2}}
\put(1393.0,726.0){\rule[-0.200pt]{4.818pt}{0.400pt}}
\put(203.0,174.0){\rule[-0.200pt]{0.400pt}{132.977pt}}
\put(203.0,174.0){\rule[-0.200pt]{0.400pt}{4.818pt}}
\put(203,116){\makebox(0,0){ 1}}
\put(203.0,706.0){\rule[-0.200pt]{0.400pt}{4.818pt}}
\put(324.0,174.0){\rule[-0.200pt]{0.400pt}{2.409pt}}
\put(324.0,716.0){\rule[-0.200pt]{0.400pt}{2.409pt}}
\put(395.0,174.0){\rule[-0.200pt]{0.400pt}{2.409pt}}
\put(395.0,716.0){\rule[-0.200pt]{0.400pt}{2.409pt}}
\put(446.0,174.0){\rule[-0.200pt]{0.400pt}{2.409pt}}
\put(446.0,716.0){\rule[-0.200pt]{0.400pt}{2.409pt}}
\put(485.0,174.0){\rule[-0.200pt]{0.400pt}{2.409pt}}
\put(485.0,716.0){\rule[-0.200pt]{0.400pt}{2.409pt}}
\put(517.0,174.0){\rule[-0.200pt]{0.400pt}{2.409pt}}
\put(517.0,716.0){\rule[-0.200pt]{0.400pt}{2.409pt}}
\put(544.0,174.0){\rule[-0.200pt]{0.400pt}{2.409pt}}
\put(544.0,716.0){\rule[-0.200pt]{0.400pt}{2.409pt}}
\put(567.0,174.0){\rule[-0.200pt]{0.400pt}{2.409pt}}
\put(567.0,716.0){\rule[-0.200pt]{0.400pt}{2.409pt}}
\put(588.0,174.0){\rule[-0.200pt]{0.400pt}{2.409pt}}
\put(588.0,716.0){\rule[-0.200pt]{0.400pt}{2.409pt}}
\put(606.0,174.0){\rule[-0.200pt]{0.400pt}{132.977pt}}
\put(606.0,174.0){\rule[-0.200pt]{0.400pt}{4.818pt}}
\put(606,116){\makebox(0,0){ 10}}
\put(606.0,706.0){\rule[-0.200pt]{0.400pt}{4.818pt}}
\put(728.0,174.0){\rule[-0.200pt]{0.400pt}{2.409pt}}
\put(728.0,716.0){\rule[-0.200pt]{0.400pt}{2.409pt}}
\put(799.0,174.0){\rule[-0.200pt]{0.400pt}{2.409pt}}
\put(799.0,716.0){\rule[-0.200pt]{0.400pt}{2.409pt}}
\put(849.0,174.0){\rule[-0.200pt]{0.400pt}{2.409pt}}
\put(849.0,716.0){\rule[-0.200pt]{0.400pt}{2.409pt}}
\put(888.0,174.0){\rule[-0.200pt]{0.400pt}{2.409pt}}
\put(888.0,716.0){\rule[-0.200pt]{0.400pt}{2.409pt}}
\put(920.0,174.0){\rule[-0.200pt]{0.400pt}{2.409pt}}
\put(920.0,716.0){\rule[-0.200pt]{0.400pt}{2.409pt}}
\put(947.0,174.0){\rule[-0.200pt]{0.400pt}{2.409pt}}
\put(947.0,716.0){\rule[-0.200pt]{0.400pt}{2.409pt}}
\put(971.0,174.0){\rule[-0.200pt]{0.400pt}{2.409pt}}
\put(971.0,716.0){\rule[-0.200pt]{0.400pt}{2.409pt}}
\put(991.0,174.0){\rule[-0.200pt]{0.400pt}{2.409pt}}
\put(991.0,716.0){\rule[-0.200pt]{0.400pt}{2.409pt}}
\put(1010.0,174.0){\rule[-0.200pt]{0.400pt}{132.977pt}}
\put(1010.0,174.0){\rule[-0.200pt]{0.400pt}{4.818pt}}
\put(1010,116){\makebox(0,0){ 100}}
\put(1010.0,706.0){\rule[-0.200pt]{0.400pt}{4.818pt}}
\put(1131.0,174.0){\rule[-0.200pt]{0.400pt}{2.409pt}}
\put(1131.0,716.0){\rule[-0.200pt]{0.400pt}{2.409pt}}
\put(1202.0,174.0){\rule[-0.200pt]{0.400pt}{2.409pt}}
\put(1202.0,716.0){\rule[-0.200pt]{0.400pt}{2.409pt}}
\put(1252.0,174.0){\rule[-0.200pt]{0.400pt}{2.409pt}}
\put(1252.0,716.0){\rule[-0.200pt]{0.400pt}{2.409pt}}
\put(1292.0,174.0){\rule[-0.200pt]{0.400pt}{2.409pt}}
\put(1292.0,716.0){\rule[-0.200pt]{0.400pt}{2.409pt}}
\put(1324.0,174.0){\rule[-0.200pt]{0.400pt}{2.409pt}}
\put(1324.0,716.0){\rule[-0.200pt]{0.400pt}{2.409pt}}
\put(1351.0,174.0){\rule[-0.200pt]{0.400pt}{2.409pt}}
\put(1351.0,716.0){\rule[-0.200pt]{0.400pt}{2.409pt}}
\put(1374.0,174.0){\rule[-0.200pt]{0.400pt}{2.409pt}}
\put(1374.0,716.0){\rule[-0.200pt]{0.400pt}{2.409pt}}
\put(1395.0,174.0){\rule[-0.200pt]{0.400pt}{2.409pt}}
\put(1395.0,716.0){\rule[-0.200pt]{0.400pt}{2.409pt}}
\put(1413.0,174.0){\rule[-0.200pt]{0.400pt}{132.977pt}}
\put(1413.0,174.0){\rule[-0.200pt]{0.400pt}{4.818pt}}
\put(1413,116){\makebox(0,0){ 1000}}
\put(1413.0,706.0){\rule[-0.200pt]{0.400pt}{4.818pt}}
\put(203.0,174.0){\rule[-0.200pt]{0.400pt}{132.977pt}}
\put(203.0,174.0){\rule[-0.200pt]{291.489pt}{0.400pt}}
\put(1413.0,174.0){\rule[-0.200pt]{0.400pt}{132.977pt}}
\put(203.0,726.0){\rule[-0.200pt]{291.489pt}{0.400pt}}
\put(808,29){\makebox(0,0){x = (predicted time)}}
\put(808,813){\makebox(0,0){y = (time / predicted time)}}
\put(300,293){\raisebox{-.8pt}{\makebox(0,0){$\Diamond$}}}
\put(543,290){\raisebox{-.8pt}{\makebox(0,0){$\Diamond$}}}
\put(786,333){\raisebox{-.8pt}{\makebox(0,0){$\Diamond$}}}
\put(438,238){\raisebox{-.8pt}{\makebox(0,0){$\Diamond$}}}
\put(681,251){\raisebox{-.8pt}{\makebox(0,0){$\Diamond$}}}
\put(924,265){\raisebox{-.8pt}{\makebox(0,0){$\Diamond$}}}
\put(373,252){\raisebox{-.8pt}{\makebox(0,0){$\Diamond$}}}
\put(616,278){\raisebox{-.8pt}{\makebox(0,0){$\Diamond$}}}
\put(858,295){\raisebox{-.8pt}{\makebox(0,0){$\Diamond$}}}
\put(328,267){\raisebox{-.8pt}{\makebox(0,0){$\Diamond$}}}
\put(571,297){\raisebox{-.8pt}{\makebox(0,0){$\Diamond$}}}
\put(814,325){\raisebox{-.8pt}{\makebox(0,0){$\Diamond$}}}
\put(390,249){\raisebox{-.8pt}{\makebox(0,0){$\Diamond$}}}
\put(633,304){\raisebox{-.8pt}{\makebox(0,0){$\Diamond$}}}
\put(876,345){\raisebox{-.8pt}{\makebox(0,0){$\Diamond$}}}
\put(370,238){\raisebox{-.8pt}{\makebox(0,0){$\Diamond$}}}
\put(613,287){\raisebox{-.8pt}{\makebox(0,0){$\Diamond$}}}
\put(856,345){\raisebox{-.8pt}{\makebox(0,0){$\Diamond$}}}
\put(477,209){\raisebox{-.8pt}{\makebox(0,0){$\Diamond$}}}
\put(720,239){\raisebox{-.8pt}{\makebox(0,0){$\Diamond$}}}
\put(963,256){\raisebox{-.8pt}{\makebox(0,0){$\Diamond$}}}
\put(424,236){\raisebox{-.8pt}{\makebox(0,0){$\Diamond$}}}
\put(667,277){\raisebox{-.8pt}{\makebox(0,0){$\Diamond$}}}
\put(910,308){\raisebox{-.8pt}{\makebox(0,0){$\Diamond$}}}
\put(390,277){\raisebox{-.8pt}{\makebox(0,0){$\Diamond$}}}
\put(633,338){\raisebox{-.8pt}{\makebox(0,0){$\Diamond$}}}
\put(876,396){\raisebox{-.8pt}{\makebox(0,0){$\Diamond$}}}
\put(370,253){\raisebox{-.8pt}{\makebox(0,0){$\Diamond$}}}
\put(613,307){\raisebox{-.8pt}{\makebox(0,0){$\Diamond$}}}
\put(856,379){\raisebox{-.8pt}{\makebox(0,0){$\Diamond$}}}
\put(477,292){\raisebox{-.8pt}{\makebox(0,0){$\Diamond$}}}
\put(720,325){\raisebox{-.8pt}{\makebox(0,0){$\Diamond$}}}
\put(963,345){\raisebox{-.8pt}{\makebox(0,0){$\Diamond$}}}
\put(359,220){\raisebox{-.8pt}{\makebox(0,0){$\Diamond$}}}
\put(602,259){\raisebox{-.8pt}{\makebox(0,0){$\Diamond$}}}
\put(845,333){\raisebox{-.8pt}{\makebox(0,0){$\Diamond$}}}
\put(424,282){\raisebox{-.8pt}{\makebox(0,0){$\Diamond$}}}
\put(667,356){\raisebox{-.8pt}{\makebox(0,0){$\Diamond$}}}
\put(910,372){\raisebox{-.8pt}{\makebox(0,0){$\Diamond$}}}
\put(392,262){\raisebox{-.8pt}{\makebox(0,0){$\Diamond$}}}
\put(635,310){\raisebox{-.8pt}{\makebox(0,0){$\Diamond$}}}
\put(877,353){\raisebox{-.8pt}{\makebox(0,0){$\Diamond$}}}
\put(553,230){\raisebox{-.8pt}{\makebox(0,0){$\Diamond$}}}
\put(795,254){\raisebox{-.8pt}{\makebox(0,0){$\Diamond$}}}
\put(1038,264){\raisebox{-.8pt}{\makebox(0,0){$\Diamond$}}}
\put(479,245){\raisebox{-.8pt}{\makebox(0,0){$\Diamond$}}}
\put(722,278){\raisebox{-.8pt}{\makebox(0,0){$\Diamond$}}}
\put(965,297){\raisebox{-.8pt}{\makebox(0,0){$\Diamond$}}}
\put(426,257){\raisebox{-.8pt}{\makebox(0,0){$\Diamond$}}}
\put(669,304){\raisebox{-.8pt}{\makebox(0,0){$\Diamond$}}}
\put(912,335){\raisebox{-.8pt}{\makebox(0,0){$\Diamond$}}}
\put(437,210){\raisebox{-.8pt}{\makebox(0,0){$\Diamond$}}}
\put(680,257){\raisebox{-.8pt}{\makebox(0,0){$\Diamond$}}}
\put(922,367){\raisebox{-.8pt}{\makebox(0,0){$\Diamond$}}}
\put(484,289){\raisebox{-.8pt}{\makebox(0,0){$\Diamond$}}}
\put(726,380){\raisebox{-.8pt}{\makebox(0,0){$\Diamond$}}}
\put(969,450){\raisebox{-.8pt}{\makebox(0,0){$\Diamond$}}}
\put(458,278){\raisebox{-.8pt}{\makebox(0,0){$\Diamond$}}}
\put(701,359){\raisebox{-.8pt}{\makebox(0,0){$\Diamond$}}}
\put(944,455){\raisebox{-.8pt}{\makebox(0,0){$\Diamond$}}}
\put(587,295){\raisebox{-.8pt}{\makebox(0,0){$\Diamond$}}}
\put(829,351){\raisebox{-.8pt}{\makebox(0,0){$\Diamond$}}}
\put(1072,376){\raisebox{-.8pt}{\makebox(0,0){$\Diamond$}}}
\put(444,247){\raisebox{-.8pt}{\makebox(0,0){$\Diamond$}}}
\put(687,324){\raisebox{-.8pt}{\makebox(0,0){$\Diamond$}}}
\put(930,426){\raisebox{-.8pt}{\makebox(0,0){$\Diamond$}}}
\put(525,294){\raisebox{-.8pt}{\makebox(0,0){$\Diamond$}}}
\put(768,361){\raisebox{-.8pt}{\makebox(0,0){$\Diamond$}}}
\put(1011,409){\raisebox{-.8pt}{\makebox(0,0){$\Diamond$}}}
\put(484,248){\raisebox{-.8pt}{\makebox(0,0){$\Diamond$}}}
\put(726,325){\raisebox{-.8pt}{\makebox(0,0){$\Diamond$}}}
\put(969,394){\raisebox{-.8pt}{\makebox(0,0){$\Diamond$}}}
\put(458,249){\raisebox{-.8pt}{\makebox(0,0){$\Diamond$}}}
\put(701,350){\raisebox{-.8pt}{\makebox(0,0){$\Diamond$}}}
\put(944,418){\raisebox{-.8pt}{\makebox(0,0){$\Diamond$}}}
\put(587,206){\raisebox{-.8pt}{\makebox(0,0){$\Diamond$}}}
\put(829,238){\raisebox{-.8pt}{\makebox(0,0){$\Diamond$}}}
\put(1072,269){\raisebox{-.8pt}{\makebox(0,0){$\Diamond$}}}
\put(444,232){\raisebox{-.8pt}{\makebox(0,0){$\Diamond$}}}
\put(687,300){\raisebox{-.8pt}{\makebox(0,0){$\Diamond$}}}
\put(930,411){\raisebox{-.8pt}{\makebox(0,0){$\Diamond$}}}
\put(525,235){\raisebox{-.8pt}{\makebox(0,0){$\Diamond$}}}
\put(768,287){\raisebox{-.8pt}{\makebox(0,0){$\Diamond$}}}
\put(1011,330){\raisebox{-.8pt}{\makebox(0,0){$\Diamond$}}}
\put(485,276){\raisebox{-.8pt}{\makebox(0,0){$\Diamond$}}}
\put(728,351){\raisebox{-.8pt}{\makebox(0,0){$\Diamond$}}}
\put(971,408){\raisebox{-.8pt}{\makebox(0,0){$\Diamond$}}}
\put(460,264){\raisebox{-.8pt}{\makebox(0,0){$\Diamond$}}}
\put(703,338){\raisebox{-.8pt}{\makebox(0,0){$\Diamond$}}}
\put(945,418){\raisebox{-.8pt}{\makebox(0,0){$\Diamond$}}}
\put(588,246){\raisebox{-.8pt}{\makebox(0,0){$\Diamond$}}}
\put(831,279){\raisebox{-.8pt}{\makebox(0,0){$\Diamond$}}}
\put(1074,311){\raisebox{-.8pt}{\makebox(0,0){$\Diamond$}}}
\put(446,245){\raisebox{-.8pt}{\makebox(0,0){$\Diamond$}}}
\put(688,314){\raisebox{-.8pt}{\makebox(0,0){$\Diamond$}}}
\put(931,403){\raisebox{-.8pt}{\makebox(0,0){$\Diamond$}}}
\put(526,270){\raisebox{-.8pt}{\makebox(0,0){$\Diamond$}}}
\put(769,316){\raisebox{-.8pt}{\makebox(0,0){$\Diamond$}}}
\put(1012,367){\raisebox{-.8pt}{\makebox(0,0){$\Diamond$}}}
\put(520,244){\raisebox{-.8pt}{\makebox(0,0){$\Diamond$}}}
\put(763,353){\raisebox{-.8pt}{\makebox(0,0){$\Diamond$}}}
\put(1005,516){\raisebox{-.8pt}{\makebox(0,0){$\Diamond$}}}
\put(579,318){\raisebox{-.8pt}{\makebox(0,0){$\Diamond$}}}
\put(822,449){\raisebox{-.8pt}{\makebox(0,0){$\Diamond$}}}
\put(1064,515){\raisebox{-.8pt}{\makebox(0,0){$\Diamond$}}}
\put(515,208){\raisebox{-.8pt}{\makebox(0,0){$\Diamond$}}}
\put(757,270){\raisebox{-.8pt}{\makebox(0,0){$\Diamond$}}}
\put(1000,422){\raisebox{-.8pt}{\makebox(0,0){$\Diamond$}}}
\put(548,296){\raisebox{-.8pt}{\makebox(0,0){$\Diamond$}}}
\put(790,445){\raisebox{-.8pt}{\makebox(0,0){$\Diamond$}}}
\put(1033,564){\raisebox{-.8pt}{\makebox(0,0){$\Diamond$}}}
\put(529,272){\raisebox{-.8pt}{\makebox(0,0){$\Diamond$}}}
\put(772,415){\raisebox{-.8pt}{\makebox(0,0){$\Diamond$}}}
\put(1015,549){\raisebox{-.8pt}{\makebox(0,0){$\Diamond$}}}
\put(628,312){\raisebox{-.8pt}{\makebox(0,0){$\Diamond$}}}
\put(871,385){\raisebox{-.8pt}{\makebox(0,0){$\Diamond$}}}
\put(1114,449){\raisebox{-.8pt}{\makebox(0,0){$\Diamond$}}}
\put(579,274){\raisebox{-.8pt}{\makebox(0,0){$\Diamond$}}}
\put(822,382){\raisebox{-.8pt}{\makebox(0,0){$\Diamond$}}}
\put(1064,473){\raisebox{-.8pt}{\makebox(0,0){$\Diamond$}}}
\put(548,280){\raisebox{-.8pt}{\makebox(0,0){$\Diamond$}}}
\put(790,413){\raisebox{-.8pt}{\makebox(0,0){$\Diamond$}}}
\put(1033,534){\raisebox{-.8pt}{\makebox(0,0){$\Diamond$}}}
\put(529,256){\raisebox{-.8pt}{\makebox(0,0){$\Diamond$}}}
\put(772,383){\raisebox{-.8pt}{\makebox(0,0){$\Diamond$}}}
\put(1015,541){\raisebox{-.8pt}{\makebox(0,0){$\Diamond$}}}
\put(628,241){\raisebox{-.8pt}{\makebox(0,0){$\Diamond$}}}
\put(871,311){\raisebox{-.8pt}{\makebox(0,0){$\Diamond$}}}
\put(1114,366){\raisebox{-.8pt}{\makebox(0,0){$\Diamond$}}}
\put(521,240){\raisebox{-.8pt}{\makebox(0,0){$\Diamond$}}}
\put(764,338){\raisebox{-.8pt}{\makebox(0,0){$\Diamond$}}}
\put(1007,496){\raisebox{-.8pt}{\makebox(0,0){$\Diamond$}}}
\put(580,283){\raisebox{-.8pt}{\makebox(0,0){$\Diamond$}}}
\put(823,391){\raisebox{-.8pt}{\makebox(0,0){$\Diamond$}}}
\put(1066,482){\raisebox{-.8pt}{\makebox(0,0){$\Diamond$}}}
\put(549,298){\raisebox{-.8pt}{\makebox(0,0){$\Diamond$}}}
\put(792,443){\raisebox{-.8pt}{\makebox(0,0){$\Diamond$}}}
\put(1035,529){\raisebox{-.8pt}{\makebox(0,0){$\Diamond$}}}
\put(700,261){\raisebox{-.8pt}{\makebox(0,0){$\Diamond$}}}
\put(943,303){\raisebox{-.8pt}{\makebox(0,0){$\Diamond$}}}
\put(1186,331){\raisebox{-.8pt}{\makebox(0,0){$\Diamond$}}}
\put(531,262){\raisebox{-.8pt}{\makebox(0,0){$\Diamond$}}}
\put(774,390){\raisebox{-.8pt}{\makebox(0,0){$\Diamond$}}}
\put(1016,522){\raisebox{-.8pt}{\makebox(0,0){$\Diamond$}}}
\put(677,366){\raisebox{-.8pt}{\makebox(0,0){$\Diamond$}}}
\put(920,504){\raisebox{-.8pt}{\makebox(0,0){$\Diamond$}}}
\put(1163,614){\raisebox{-.8pt}{\makebox(0,0){$\Diamond$}}}
\put(735,362){\raisebox{-.8pt}{\makebox(0,0){$\Diamond$}}}
\put(978,448){\raisebox{-.8pt}{\makebox(0,0){$\Diamond$}}}
\put(1221,516){\raisebox{-.8pt}{\makebox(0,0){$\Diamond$}}}
\put(603,268){\raisebox{-.8pt}{\makebox(0,0){$\Diamond$}}}
\put(846,490){\raisebox{-.8pt}{\makebox(0,0){$\Diamond$}}}
\put(1089,709){\raisebox{-.8pt}{\makebox(0,0){$\Diamond$}}}
\put(677,310){\raisebox{-.8pt}{\makebox(0,0){$\Diamond$}}}
\put(920,452){\raisebox{-.8pt}{\makebox(0,0){$\Diamond$}}}
\put(1163,548){\raisebox{-.8pt}{\makebox(0,0){$\Diamond$}}}
\put(639,332){\raisebox{-.8pt}{\makebox(0,0){$\Diamond$}}}
\put(881,535){\raisebox{-.8pt}{\makebox(0,0){$\Diamond$}}}
\put(1124,674){\raisebox{-.8pt}{\makebox(0,0){$\Diamond$}}}
\put(616,309){\raisebox{-.8pt}{\makebox(0,0){$\Diamond$}}}
\put(859,516){\raisebox{-.8pt}{\makebox(0,0){$\Diamond$}}}
\put(1101,716){\raisebox{-.8pt}{\makebox(0,0){$\Diamond$}}}
\put(736,304){\raisebox{-.8pt}{\makebox(0,0){$\Diamond$}}}
\put(979,380){\raisebox{-.8pt}{\makebox(0,0){$\Diamond$}}}
\put(1222,453){\raisebox{-.8pt}{\makebox(0,0){$\Diamond$}}}
\put(678,330){\raisebox{-.8pt}{\makebox(0,0){$\Diamond$}}}
\put(921,459){\raisebox{-.8pt}{\makebox(0,0){$\Diamond$}}}
\put(1164,566){\raisebox{-.8pt}{\makebox(0,0){$\Diamond$}}}
\put(203.0,174.0){\rule[-0.200pt]{0.400pt}{132.977pt}}
\put(203.0,174.0){\rule[-0.200pt]{291.489pt}{0.400pt}}
\put(1413.0,174.0){\rule[-0.200pt]{0.400pt}{132.977pt}}
\put(203.0,726.0){\rule[-0.200pt]{291.489pt}{0.400pt}}
\end{picture}

}
\end{center}

The time exceeds the predicted time by up to a factor of two for large instances.

To understand this further, consider the next two plots.
The plot on the left plots the actual the number of basic operations
(obtained by instrumenting the code),
divided by the estimate~(\ref{eqn:predtime}).
The plot on the right plots the average time per operation.

{
\centering

\noindent
\renewcommand{\Diamond}{+}

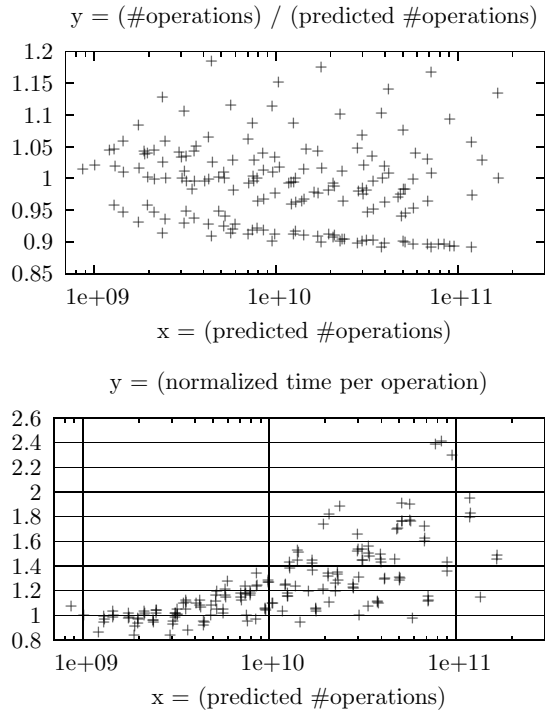
\scalebox{0.63}{
\setlength{\unitlength}{0.240900pt}
\ifx\plotpoint\undefined\newsavebox{\plotpoint}\fi
\sbox{\plotpoint}{\rule[-0.200pt]{0.400pt}{0.400pt}}%
\begin{picture}(1500,900)(0,0)
\font\gnuplot=cmr10 at 14pt
\gnuplot
\sbox{\plotpoint}{\rule[-0.200pt]{0.400pt}{0.400pt}}%
\put(232.0,174.0){\rule[-0.200pt]{4.818pt}{0.400pt}}
\put(203,174){\makebox(0,0)[r]{ 0.85}}
\put(1393.0,174.0){\rule[-0.200pt]{4.818pt}{0.400pt}}
\put(232.0,253.0){\rule[-0.200pt]{4.818pt}{0.400pt}}
\put(203,253){\makebox(0,0)[r]{ 0.9}}
\put(1393.0,253.0){\rule[-0.200pt]{4.818pt}{0.400pt}}
\put(232.0,332.0){\rule[-0.200pt]{4.818pt}{0.400pt}}
\put(203,332){\makebox(0,0)[r]{ 0.95}}
\put(1393.0,332.0){\rule[-0.200pt]{4.818pt}{0.400pt}}
\put(232.0,411.0){\rule[-0.200pt]{4.818pt}{0.400pt}}
\put(203,411){\makebox(0,0)[r]{ 1}}
\put(1393.0,411.0){\rule[-0.200pt]{4.818pt}{0.400pt}}
\put(232.0,489.0){\rule[-0.200pt]{4.818pt}{0.400pt}}
\put(203,489){\makebox(0,0)[r]{ 1.05}}
\put(1393.0,489.0){\rule[-0.200pt]{4.818pt}{0.400pt}}
\put(232.0,568.0){\rule[-0.200pt]{4.818pt}{0.400pt}}
\put(203,568){\makebox(0,0)[r]{ 1.1}}
\put(1393.0,568.0){\rule[-0.200pt]{4.818pt}{0.400pt}}
\put(232.0,647.0){\rule[-0.200pt]{4.818pt}{0.400pt}}
\put(203,647){\makebox(0,0)[r]{ 1.15}}
\put(1393.0,647.0){\rule[-0.200pt]{4.818pt}{0.400pt}}
\put(232.0,726.0){\rule[-0.200pt]{4.818pt}{0.400pt}}
\put(203,726){\makebox(0,0)[r]{ 1.2}}
\put(1393.0,726.0){\rule[-0.200pt]{4.818pt}{0.400pt}}
\put(232.0,174.0){\rule[-0.200pt]{0.400pt}{2.409pt}}
\put(232.0,716.0){\rule[-0.200pt]{0.400pt}{2.409pt}}
\put(258.0,174.0){\rule[-0.200pt]{0.400pt}{2.409pt}}
\put(258.0,716.0){\rule[-0.200pt]{0.400pt}{2.409pt}}
\put(281.0,174.0){\rule[-0.200pt]{0.400pt}{2.409pt}}
\put(281.0,716.0){\rule[-0.200pt]{0.400pt}{2.409pt}}
\put(302.0,174.0){\rule[-0.200pt]{0.400pt}{4.818pt}}
\put(302,116){\makebox(0,0){ 1e+09}}
\put(302.0,706.0){\rule[-0.200pt]{0.400pt}{4.818pt}}
\put(437.0,174.0){\rule[-0.200pt]{0.400pt}{2.409pt}}
\put(437.0,716.0){\rule[-0.200pt]{0.400pt}{2.409pt}}
\put(516.0,174.0){\rule[-0.200pt]{0.400pt}{2.409pt}}
\put(516.0,716.0){\rule[-0.200pt]{0.400pt}{2.409pt}}
\put(572.0,174.0){\rule[-0.200pt]{0.400pt}{2.409pt}}
\put(572.0,716.0){\rule[-0.200pt]{0.400pt}{2.409pt}}
\put(615.0,174.0){\rule[-0.200pt]{0.400pt}{2.409pt}}
\put(615.0,716.0){\rule[-0.200pt]{0.400pt}{2.409pt}}
\put(651.0,174.0){\rule[-0.200pt]{0.400pt}{2.409pt}}
\put(651.0,716.0){\rule[-0.200pt]{0.400pt}{2.409pt}}
\put(681.0,174.0){\rule[-0.200pt]{0.400pt}{2.409pt}}
\put(681.0,716.0){\rule[-0.200pt]{0.400pt}{2.409pt}}
\put(707.0,174.0){\rule[-0.200pt]{0.400pt}{2.409pt}}
\put(707.0,716.0){\rule[-0.200pt]{0.400pt}{2.409pt}}
\put(730.0,174.0){\rule[-0.200pt]{0.400pt}{2.409pt}}
\put(730.0,716.0){\rule[-0.200pt]{0.400pt}{2.409pt}}
\put(750.0,174.0){\rule[-0.200pt]{0.400pt}{4.818pt}}
\put(750,116){\makebox(0,0){ 1e+10}}
\put(750.0,706.0){\rule[-0.200pt]{0.400pt}{4.818pt}}
\put(885.0,174.0){\rule[-0.200pt]{0.400pt}{2.409pt}}
\put(885.0,716.0){\rule[-0.200pt]{0.400pt}{2.409pt}}
\put(964.0,174.0){\rule[-0.200pt]{0.400pt}{2.409pt}}
\put(964.0,716.0){\rule[-0.200pt]{0.400pt}{2.409pt}}
\put(1020.0,174.0){\rule[-0.200pt]{0.400pt}{2.409pt}}
\put(1020.0,716.0){\rule[-0.200pt]{0.400pt}{2.409pt}}
\put(1064.0,174.0){\rule[-0.200pt]{0.400pt}{2.409pt}}
\put(1064.0,716.0){\rule[-0.200pt]{0.400pt}{2.409pt}}
\put(1099.0,174.0){\rule[-0.200pt]{0.400pt}{2.409pt}}
\put(1099.0,716.0){\rule[-0.200pt]{0.400pt}{2.409pt}}
\put(1129.0,174.0){\rule[-0.200pt]{0.400pt}{2.409pt}}
\put(1129.0,716.0){\rule[-0.200pt]{0.400pt}{2.409pt}}
\put(1155.0,174.0){\rule[-0.200pt]{0.400pt}{2.409pt}}
\put(1155.0,716.0){\rule[-0.200pt]{0.400pt}{2.409pt}}
\put(1178.0,174.0){\rule[-0.200pt]{0.400pt}{2.409pt}}
\put(1178.0,716.0){\rule[-0.200pt]{0.400pt}{2.409pt}}
\put(1199.0,174.0){\rule[-0.200pt]{0.400pt}{4.818pt}}
\put(1199,116){\makebox(0,0){ 1e+11}}
\put(1199.0,706.0){\rule[-0.200pt]{0.400pt}{4.818pt}}
\put(1334.0,174.0){\rule[-0.200pt]{0.400pt}{2.409pt}}
\put(1334.0,716.0){\rule[-0.200pt]{0.400pt}{2.409pt}}
\put(1413.0,174.0){\rule[-0.200pt]{0.400pt}{2.409pt}}
\put(1413.0,716.0){\rule[-0.200pt]{0.400pt}{2.409pt}}
\put(232.0,174.0){\rule[-0.200pt]{0.400pt}{132.977pt}}
\put(232.0,174.0){\rule[-0.200pt]{284.503pt}{0.400pt}}
\put(1413.0,174.0){\rule[-0.200pt]{0.400pt}{132.977pt}}
\put(232.0,726.0){\rule[-0.200pt]{284.503pt}{0.400pt}}
\put(822,29){\makebox(0,0){x = (predicted \#operations)}}
\put(822,813){\makebox(0,0){y = (\#operations) / (predicted \#operations)}}
\put(275,438){\raisebox{-.8pt}{\makebox(0,0){$\Diamond$}}}
\put(545,387){\raisebox{-.8pt}{\makebox(0,0){$\Diamond$}}}
\put(815,359){\raisebox{-.8pt}{\makebox(0,0){$\Diamond$}}}
\put(428,475){\raisebox{-.8pt}{\makebox(0,0){$\Diamond$}}}
\put(698,460){\raisebox{-.8pt}{\makebox(0,0){$\Diamond$}}}
\put(968,447){\raisebox{-.8pt}{\makebox(0,0){$\Diamond$}}}
\put(355,446){\raisebox{-.8pt}{\makebox(0,0){$\Diamond$}}}
\put(625,415){\raisebox{-.8pt}{\makebox(0,0){$\Diamond$}}}
\put(896,396){\raisebox{-.8pt}{\makebox(0,0){$\Diamond$}}}
\put(305,447){\raisebox{-.8pt}{\makebox(0,0){$\Diamond$}}}
\put(576,407){\raisebox{-.8pt}{\makebox(0,0){$\Diamond$}}}
\put(846,380){\raisebox{-.8pt}{\makebox(0,0){$\Diamond$}}}
\put(375,331){\raisebox{-.8pt}{\makebox(0,0){$\Diamond$}}}
\put(645,288){\raisebox{-.8pt}{\makebox(0,0){$\Diamond$}}}
\put(915,263){\raisebox{-.8pt}{\makebox(0,0){$\Diamond$}}}
\put(353,348){\raisebox{-.8pt}{\makebox(0,0){$\Diamond$}}}
\put(623,296){\raisebox{-.8pt}{\makebox(0,0){$\Diamond$}}}
\put(893,269){\raisebox{-.8pt}{\makebox(0,0){$\Diamond$}}}
\put(472,279){\raisebox{-.8pt}{\makebox(0,0){$\Diamond$}}}
\put(742,258){\raisebox{-.8pt}{\makebox(0,0){$\Diamond$}}}
\put(1012,244){\raisebox{-.8pt}{\makebox(0,0){$\Diamond$}}}
\put(413,307){\raisebox{-.8pt}{\makebox(0,0){$\Diamond$}}}
\put(683,275){\raisebox{-.8pt}{\makebox(0,0){$\Diamond$}}}
\put(953,254){\raisebox{-.8pt}{\makebox(0,0){$\Diamond$}}}
\put(375,508){\raisebox{-.8pt}{\makebox(0,0){$\Diamond$}}}
\put(645,459){\raisebox{-.8pt}{\makebox(0,0){$\Diamond$}}}
\put(915,433){\raisebox{-.8pt}{\makebox(0,0){$\Diamond$}}}
\put(353,487){\raisebox{-.8pt}{\makebox(0,0){$\Diamond$}}}
\put(623,431){\raisebox{-.8pt}{\makebox(0,0){$\Diamond$}}}
\put(893,402){\raisebox{-.8pt}{\makebox(0,0){$\Diamond$}}}
\put(472,617){\raisebox{-.8pt}{\makebox(0,0){$\Diamond$}}}
\put(742,593){\raisebox{-.8pt}{\makebox(0,0){$\Diamond$}}}
\put(1012,577){\raisebox{-.8pt}{\makebox(0,0){$\Diamond$}}}
\put(341,484){\raisebox{-.8pt}{\makebox(0,0){$\Diamond$}}}
\put(611,417){\raisebox{-.8pt}{\makebox(0,0){$\Diamond$}}}
\put(881,386){\raisebox{-.8pt}{\makebox(0,0){$\Diamond$}}}
\put(413,547){\raisebox{-.8pt}{\makebox(0,0){$\Diamond$}}}
\put(683,511){\raisebox{-.8pt}{\makebox(0,0){$\Diamond$}}}
\put(953,490){\raisebox{-.8pt}{\makebox(0,0){$\Diamond$}}}
\put(376,430){\raisebox{-.8pt}{\makebox(0,0){$\Diamond$}}}
\put(647,385){\raisebox{-.8pt}{\makebox(0,0){$\Diamond$}}}
\put(917,359){\raisebox{-.8pt}{\makebox(0,0){$\Diamond$}}}
\put(555,495){\raisebox{-.8pt}{\makebox(0,0){$\Diamond$}}}
\put(826,487){\raisebox{-.8pt}{\makebox(0,0){$\Diamond$}}}
\put(1096,477){\raisebox{-.8pt}{\makebox(0,0){$\Diamond$}}}
\put(473,454){\raisebox{-.8pt}{\makebox(0,0){$\Diamond$}}}
\put(744,430){\raisebox{-.8pt}{\makebox(0,0){$\Diamond$}}}
\put(1014,414){\raisebox{-.8pt}{\makebox(0,0){$\Diamond$}}}
\put(414,439){\raisebox{-.8pt}{\makebox(0,0){$\Diamond$}}}
\put(685,404){\raisebox{-.8pt}{\makebox(0,0){$\Diamond$}}}
\put(955,382){\raisebox{-.8pt}{\makebox(0,0){$\Diamond$}}}
\put(427,482){\raisebox{-.8pt}{\makebox(0,0){$\Diamond$}}}
\put(697,408){\raisebox{-.8pt}{\makebox(0,0){$\Diamond$}}}
\put(967,375){\raisebox{-.8pt}{\makebox(0,0){$\Diamond$}}}
\put(478,508){\raisebox{-.8pt}{\makebox(0,0){$\Diamond$}}}
\put(749,467){\raisebox{-.8pt}{\makebox(0,0){$\Diamond$}}}
\put(1019,444){\raisebox{-.8pt}{\makebox(0,0){$\Diamond$}}}
\put(451,486){\raisebox{-.8pt}{\makebox(0,0){$\Diamond$}}}
\put(721,435){\raisebox{-.8pt}{\makebox(0,0){$\Diamond$}}}
\put(991,408){\raisebox{-.8pt}{\makebox(0,0){$\Diamond$}}}
\put(593,705){\raisebox{-.8pt}{\makebox(0,0){$\Diamond$}}}
\put(863,691){\raisebox{-.8pt}{\makebox(0,0){$\Diamond$}}}
\put(1134,679){\raisebox{-.8pt}{\makebox(0,0){$\Diamond$}}}
\put(435,477){\raisebox{-.8pt}{\makebox(0,0){$\Diamond$}}}
\put(705,416){\raisebox{-.8pt}{\makebox(0,0){$\Diamond$}}}
\put(975,386){\raisebox{-.8pt}{\makebox(0,0){$\Diamond$}}}
\put(525,581){\raisebox{-.8pt}{\makebox(0,0){$\Diamond$}}}
\put(795,553){\raisebox{-.8pt}{\makebox(0,0){$\Diamond$}}}
\put(1065,535){\raisebox{-.8pt}{\makebox(0,0){$\Diamond$}}}
\put(478,313){\raisebox{-.8pt}{\makebox(0,0){$\Diamond$}}}
\put(749,275){\raisebox{-.8pt}{\makebox(0,0){$\Diamond$}}}
\put(1019,254){\raisebox{-.8pt}{\makebox(0,0){$\Diamond$}}}
\put(451,331){\raisebox{-.8pt}{\makebox(0,0){$\Diamond$}}}
\put(721,285){\raisebox{-.8pt}{\makebox(0,0){$\Diamond$}}}
\put(991,260){\raisebox{-.8pt}{\makebox(0,0){$\Diamond$}}}
\put(593,272){\raisebox{-.8pt}{\makebox(0,0){$\Diamond$}}}
\put(863,255){\raisebox{-.8pt}{\makebox(0,0){$\Diamond$}}}
\put(1134,244){\raisebox{-.8pt}{\makebox(0,0){$\Diamond$}}}
\put(435,347){\raisebox{-.8pt}{\makebox(0,0){$\Diamond$}}}
\put(705,290){\raisebox{-.8pt}{\makebox(0,0){$\Diamond$}}}
\put(975,261){\raisebox{-.8pt}{\makebox(0,0){$\Diamond$}}}
\put(525,303){\raisebox{-.8pt}{\makebox(0,0){$\Diamond$}}}
\put(795,275){\raisebox{-.8pt}{\makebox(0,0){$\Diamond$}}}
\put(1065,258){\raisebox{-.8pt}{\makebox(0,0){$\Diamond$}}}
\put(480,415){\raisebox{-.8pt}{\makebox(0,0){$\Diamond$}}}
\put(750,377){\raisebox{-.8pt}{\makebox(0,0){$\Diamond$}}}
\put(1020,355){\raisebox{-.8pt}{\makebox(0,0){$\Diamond$}}}
\put(452,413){\raisebox{-.8pt}{\makebox(0,0){$\Diamond$}}}
\put(722,364){\raisebox{-.8pt}{\makebox(0,0){$\Diamond$}}}
\put(992,338){\raisebox{-.8pt}{\makebox(0,0){$\Diamond$}}}
\put(595,454){\raisebox{-.8pt}{\makebox(0,0){$\Diamond$}}}
\put(865,439){\raisebox{-.8pt}{\makebox(0,0){$\Diamond$}}}
\put(1135,427){\raisebox{-.8pt}{\makebox(0,0){$\Diamond$}}}
\put(436,417){\raisebox{-.8pt}{\makebox(0,0){$\Diamond$}}}
\put(707,358){\raisebox{-.8pt}{\makebox(0,0){$\Diamond$}}}
\put(977,330){\raisebox{-.8pt}{\makebox(0,0){$\Diamond$}}}
\put(526,433){\raisebox{-.8pt}{\makebox(0,0){$\Diamond$}}}
\put(796,405){\raisebox{-.8pt}{\makebox(0,0){$\Diamond$}}}
\put(1066,387){\raisebox{-.8pt}{\makebox(0,0){$\Diamond$}}}
\put(519,468){\raisebox{-.8pt}{\makebox(0,0){$\Diamond$}}}
\put(789,405){\raisebox{-.8pt}{\makebox(0,0){$\Diamond$}}}
\put(1059,375){\raisebox{-.8pt}{\makebox(0,0){$\Diamond$}}}
\put(585,517){\raisebox{-.8pt}{\makebox(0,0){$\Diamond$}}}
\put(855,482){\raisebox{-.8pt}{\makebox(0,0){$\Diamond$}}}
\put(1125,463){\raisebox{-.8pt}{\makebox(0,0){$\Diamond$}}}
\put(513,480){\raisebox{-.8pt}{\makebox(0,0){$\Diamond$}}}
\put(783,402){\raisebox{-.8pt}{\makebox(0,0){$\Diamond$}}}
\put(1053,368){\raisebox{-.8pt}{\makebox(0,0){$\Diamond$}}}
\put(550,483){\raisebox{-.8pt}{\makebox(0,0){$\Diamond$}}}
\put(820,438){\raisebox{-.8pt}{\makebox(0,0){$\Diamond$}}}
\put(1090,413){\raisebox{-.8pt}{\makebox(0,0){$\Diamond$}}}
\put(530,471){\raisebox{-.8pt}{\makebox(0,0){$\Diamond$}}}
\put(800,416){\raisebox{-.8pt}{\makebox(0,0){$\Diamond$}}}
\put(1070,388){\raisebox{-.8pt}{\makebox(0,0){$\Diamond$}}}
\put(640,597){\raisebox{-.8pt}{\makebox(0,0){$\Diamond$}}}
\put(910,575){\raisebox{-.8pt}{\makebox(0,0){$\Diamond$}}}
\put(1180,561){\raisebox{-.8pt}{\makebox(0,0){$\Diamond$}}}
\put(585,302){\raisebox{-.8pt}{\makebox(0,0){$\Diamond$}}}
\put(855,270){\raisebox{-.8pt}{\makebox(0,0){$\Diamond$}}}
\put(1125,252){\raisebox{-.8pt}{\makebox(0,0){$\Diamond$}}}
\put(550,315){\raisebox{-.8pt}{\makebox(0,0){$\Diamond$}}}
\put(820,273){\raisebox{-.8pt}{\makebox(0,0){$\Diamond$}}}
\put(1090,250){\raisebox{-.8pt}{\makebox(0,0){$\Diamond$}}}
\put(530,332){\raisebox{-.8pt}{\makebox(0,0){$\Diamond$}}}
\put(800,283){\raisebox{-.8pt}{\makebox(0,0){$\Diamond$}}}
\put(1070,257){\raisebox{-.8pt}{\makebox(0,0){$\Diamond$}}}
\put(640,278){\raisebox{-.8pt}{\makebox(0,0){$\Diamond$}}}
\put(910,259){\raisebox{-.8pt}{\makebox(0,0){$\Diamond$}}}
\put(1180,246){\raisebox{-.8pt}{\makebox(0,0){$\Diamond$}}}
\put(520,416){\raisebox{-.8pt}{\makebox(0,0){$\Diamond$}}}
\put(790,351){\raisebox{-.8pt}{\makebox(0,0){$\Diamond$}}}
\put(1061,321){\raisebox{-.8pt}{\makebox(0,0){$\Diamond$}}}
\put(586,412){\raisebox{-.8pt}{\makebox(0,0){$\Diamond$}}}
\put(856,378){\raisebox{-.8pt}{\makebox(0,0){$\Diamond$}}}
\put(1126,359){\raisebox{-.8pt}{\makebox(0,0){$\Diamond$}}}
\put(551,408){\raisebox{-.8pt}{\makebox(0,0){$\Diamond$}}}
\put(821,364){\raisebox{-.8pt}{\makebox(0,0){$\Diamond$}}}
\put(1091,340){\raisebox{-.8pt}{\makebox(0,0){$\Diamond$}}}
\put(719,478){\raisebox{-.8pt}{\makebox(0,0){$\Diamond$}}}
\put(989,470){\raisebox{-.8pt}{\makebox(0,0){$\Diamond$}}}
\put(1260,461){\raisebox{-.8pt}{\makebox(0,0){$\Diamond$}}}
\put(531,409){\raisebox{-.8pt}{\makebox(0,0){$\Diamond$}}}
\put(801,355){\raisebox{-.8pt}{\makebox(0,0){$\Diamond$}}}
\put(1071,329){\raisebox{-.8pt}{\makebox(0,0){$\Diamond$}}}
\put(694,551){\raisebox{-.8pt}{\makebox(0,0){$\Diamond$}}}
\put(964,522){\raisebox{-.8pt}{\makebox(0,0){$\Diamond$}}}
\put(1234,505){\raisebox{-.8pt}{\makebox(0,0){$\Diamond$}}}
\put(758,653){\raisebox{-.8pt}{\makebox(0,0){$\Diamond$}}}
\put(1029,637){\raisebox{-.8pt}{\makebox(0,0){$\Diamond$}}}
\put(1299,626){\raisebox{-.8pt}{\makebox(0,0){$\Diamond$}}}
\put(611,333){\raisebox{-.8pt}{\makebox(0,0){$\Diamond$}}}
\put(882,277){\raisebox{-.8pt}{\makebox(0,0){$\Diamond$}}}
\put(1152,251){\raisebox{-.8pt}{\makebox(0,0){$\Diamond$}}}
\put(694,287){\raisebox{-.8pt}{\makebox(0,0){$\Diamond$}}}
\put(964,259){\raisebox{-.8pt}{\makebox(0,0){$\Diamond$}}}
\put(1234,243){\raisebox{-.8pt}{\makebox(0,0){$\Diamond$}}}
\put(651,302){\raisebox{-.8pt}{\makebox(0,0){$\Diamond$}}}
\put(921,264){\raisebox{-.8pt}{\makebox(0,0){$\Diamond$}}}
\put(1191,245){\raisebox{-.8pt}{\makebox(0,0){$\Diamond$}}}
\put(626,321){\raisebox{-.8pt}{\makebox(0,0){$\Diamond$}}}
\put(896,274){\raisebox{-.8pt}{\makebox(0,0){$\Diamond$}}}
\put(1166,250){\raisebox{-.8pt}{\makebox(0,0){$\Diamond$}}}
\put(760,443){\raisebox{-.8pt}{\makebox(0,0){$\Diamond$}}}
\put(1030,427){\raisebox{-.8pt}{\makebox(0,0){$\Diamond$}}}
\put(1300,416){\raisebox{-.8pt}{\makebox(0,0){$\Diamond$}}}
\put(695,420){\raisebox{-.8pt}{\makebox(0,0){$\Diamond$}}}
\put(965,391){\raisebox{-.8pt}{\makebox(0,0){$\Diamond$}}}
\put(1235,374){\raisebox{-.8pt}{\makebox(0,0){$\Diamond$}}}
\put(232.0,174.0){\rule[-0.200pt]{0.400pt}{132.977pt}}
\put(232.0,174.0){\rule[-0.200pt]{284.503pt}{0.400pt}}
\put(1413.0,174.0){\rule[-0.200pt]{0.400pt}{132.977pt}}
\put(232.0,726.0){\rule[-0.200pt]{284.503pt}{0.400pt}}
\end{picture}
}

\scalebox{0.63}{
\setlength{\unitlength}{0.240900pt}
\ifx\plotpoint\undefined\newsavebox{\plotpoint}\fi
\sbox{\plotpoint}{\rule[-0.200pt]{0.400pt}{0.400pt}}%
\begin{picture}(1500,900)(0,0)
\font\gnuplot=cmr10 at 14pt
\gnuplot
\sbox{\plotpoint}{\rule[-0.200pt]{0.400pt}{0.400pt}}%
\put(203.0,174.0){\rule[-0.200pt]{291.489pt}{0.400pt}}
\put(203.0,174.0){\rule[-0.200pt]{4.818pt}{0.400pt}}
\put(174,174){\makebox(0,0)[r]{ 0.8}}
\put(1393.0,174.0){\rule[-0.200pt]{4.818pt}{0.400pt}}
\put(203.0,235.0){\rule[-0.200pt]{291.489pt}{0.400pt}}
\put(203.0,235.0){\rule[-0.200pt]{4.818pt}{0.400pt}}
\put(174,235){\makebox(0,0)[r]{ 1}}
\put(1393.0,235.0){\rule[-0.200pt]{4.818pt}{0.400pt}}
\put(203.0,297.0){\rule[-0.200pt]{291.489pt}{0.400pt}}
\put(203.0,297.0){\rule[-0.200pt]{4.818pt}{0.400pt}}
\put(174,297){\makebox(0,0)[r]{ 1.2}}
\put(1393.0,297.0){\rule[-0.200pt]{4.818pt}{0.400pt}}
\put(203.0,358.0){\rule[-0.200pt]{291.489pt}{0.400pt}}
\put(203.0,358.0){\rule[-0.200pt]{4.818pt}{0.400pt}}
\put(174,358){\makebox(0,0)[r]{ 1.4}}
\put(1393.0,358.0){\rule[-0.200pt]{4.818pt}{0.400pt}}
\put(203.0,419.0){\rule[-0.200pt]{291.489pt}{0.400pt}}
\put(203.0,419.0){\rule[-0.200pt]{4.818pt}{0.400pt}}
\put(174,419){\makebox(0,0)[r]{ 1.6}}
\put(1393.0,419.0){\rule[-0.200pt]{4.818pt}{0.400pt}}
\put(203.0,481.0){\rule[-0.200pt]{291.489pt}{0.400pt}}
\put(203.0,481.0){\rule[-0.200pt]{4.818pt}{0.400pt}}
\put(174,481){\makebox(0,0)[r]{ 1.8}}
\put(1393.0,481.0){\rule[-0.200pt]{4.818pt}{0.400pt}}
\put(203.0,542.0){\rule[-0.200pt]{291.489pt}{0.400pt}}
\put(203.0,542.0){\rule[-0.200pt]{4.818pt}{0.400pt}}
\put(174,542){\makebox(0,0)[r]{ 2}}
\put(1393.0,542.0){\rule[-0.200pt]{4.818pt}{0.400pt}}
\put(203.0,603.0){\rule[-0.200pt]{291.489pt}{0.400pt}}
\put(203.0,603.0){\rule[-0.200pt]{4.818pt}{0.400pt}}
\put(174,603){\makebox(0,0)[r]{ 2.2}}
\put(1393.0,603.0){\rule[-0.200pt]{4.818pt}{0.400pt}}
\put(203.0,665.0){\rule[-0.200pt]{291.489pt}{0.400pt}}
\put(203.0,665.0){\rule[-0.200pt]{4.818pt}{0.400pt}}
\put(174,665){\makebox(0,0)[r]{ 2.4}}
\put(1393.0,665.0){\rule[-0.200pt]{4.818pt}{0.400pt}}
\put(203.0,726.0){\rule[-0.200pt]{291.489pt}{0.400pt}}
\put(203.0,726.0){\rule[-0.200pt]{4.818pt}{0.400pt}}
\put(174,726){\makebox(0,0)[r]{ 2.6}}
\put(1393.0,726.0){\rule[-0.200pt]{4.818pt}{0.400pt}}
\put(203.0,174.0){\rule[-0.200pt]{0.400pt}{2.409pt}}
\put(203.0,716.0){\rule[-0.200pt]{0.400pt}{2.409pt}}
\put(230.0,174.0){\rule[-0.200pt]{0.400pt}{2.409pt}}
\put(230.0,716.0){\rule[-0.200pt]{0.400pt}{2.409pt}}
\put(253.0,174.0){\rule[-0.200pt]{0.400pt}{2.409pt}}
\put(253.0,716.0){\rule[-0.200pt]{0.400pt}{2.409pt}}
\put(274.0,174.0){\rule[-0.200pt]{0.400pt}{132.977pt}}
\put(274.0,174.0){\rule[-0.200pt]{0.400pt}{4.818pt}}
\put(274,116){\makebox(0,0){ 1e+09}}
\put(274.0,706.0){\rule[-0.200pt]{0.400pt}{4.818pt}}
\put(413.0,174.0){\rule[-0.200pt]{0.400pt}{2.409pt}}
\put(413.0,716.0){\rule[-0.200pt]{0.400pt}{2.409pt}}
\put(494.0,174.0){\rule[-0.200pt]{0.400pt}{2.409pt}}
\put(494.0,716.0){\rule[-0.200pt]{0.400pt}{2.409pt}}
\put(551.0,174.0){\rule[-0.200pt]{0.400pt}{2.409pt}}
\put(551.0,716.0){\rule[-0.200pt]{0.400pt}{2.409pt}}
\put(596.0,174.0){\rule[-0.200pt]{0.400pt}{2.409pt}}
\put(596.0,716.0){\rule[-0.200pt]{0.400pt}{2.409pt}}
\put(632.0,174.0){\rule[-0.200pt]{0.400pt}{2.409pt}}
\put(632.0,716.0){\rule[-0.200pt]{0.400pt}{2.409pt}}
\put(663.0,174.0){\rule[-0.200pt]{0.400pt}{2.409pt}}
\put(663.0,716.0){\rule[-0.200pt]{0.400pt}{2.409pt}}
\put(689.0,174.0){\rule[-0.200pt]{0.400pt}{2.409pt}}
\put(689.0,716.0){\rule[-0.200pt]{0.400pt}{2.409pt}}
\put(713.0,174.0){\rule[-0.200pt]{0.400pt}{2.409pt}}
\put(713.0,716.0){\rule[-0.200pt]{0.400pt}{2.409pt}}
\put(734.0,174.0){\rule[-0.200pt]{0.400pt}{132.977pt}}
\put(734.0,174.0){\rule[-0.200pt]{0.400pt}{4.818pt}}
\put(734,116){\makebox(0,0){ 1e+10}}
\put(734.0,706.0){\rule[-0.200pt]{0.400pt}{4.818pt}}
\put(872.0,174.0){\rule[-0.200pt]{0.400pt}{2.409pt}}
\put(872.0,716.0){\rule[-0.200pt]{0.400pt}{2.409pt}}
\put(953.0,174.0){\rule[-0.200pt]{0.400pt}{2.409pt}}
\put(953.0,716.0){\rule[-0.200pt]{0.400pt}{2.409pt}}
\put(1011.0,174.0){\rule[-0.200pt]{0.400pt}{2.409pt}}
\put(1011.0,716.0){\rule[-0.200pt]{0.400pt}{2.409pt}}
\put(1055.0,174.0){\rule[-0.200pt]{0.400pt}{2.409pt}}
\put(1055.0,716.0){\rule[-0.200pt]{0.400pt}{2.409pt}}
\put(1092.0,174.0){\rule[-0.200pt]{0.400pt}{2.409pt}}
\put(1092.0,716.0){\rule[-0.200pt]{0.400pt}{2.409pt}}
\put(1122.0,174.0){\rule[-0.200pt]{0.400pt}{2.409pt}}
\put(1122.0,716.0){\rule[-0.200pt]{0.400pt}{2.409pt}}
\put(1149.0,174.0){\rule[-0.200pt]{0.400pt}{2.409pt}}
\put(1149.0,716.0){\rule[-0.200pt]{0.400pt}{2.409pt}}
\put(1173.0,174.0){\rule[-0.200pt]{0.400pt}{2.409pt}}
\put(1173.0,716.0){\rule[-0.200pt]{0.400pt}{2.409pt}}
\put(1194.0,174.0){\rule[-0.200pt]{0.400pt}{132.977pt}}
\put(1194.0,174.0){\rule[-0.200pt]{0.400pt}{4.818pt}}
\put(1194,116){\makebox(0,0){ 1e+11}}
\put(1194.0,706.0){\rule[-0.200pt]{0.400pt}{4.818pt}}
\put(1332.0,174.0){\rule[-0.200pt]{0.400pt}{2.409pt}}
\put(1332.0,716.0){\rule[-0.200pt]{0.400pt}{2.409pt}}
\put(1413.0,174.0){\rule[-0.200pt]{0.400pt}{2.409pt}}
\put(1413.0,716.0){\rule[-0.200pt]{0.400pt}{2.409pt}}
\put(203.0,174.0){\rule[-0.200pt]{0.400pt}{132.977pt}}
\put(203.0,174.0){\rule[-0.200pt]{291.489pt}{0.400pt}}
\put(1413.0,174.0){\rule[-0.200pt]{0.400pt}{132.977pt}}
\put(203.0,726.0){\rule[-0.200pt]{291.489pt}{0.400pt}}
\put(808,29){\makebox(0,0){x = (predicted \#operations)}}
\put(808,813){\makebox(0,0){y = (normalized time per operation)}}
\put(247,261){\raisebox{-.8pt}{\makebox(0,0){$\Diamond$}}}
\put(524,270){\raisebox{-.8pt}{\makebox(0,0){$\Diamond$}}}
\put(800,310){\raisebox{-.8pt}{\makebox(0,0){$\Diamond$}}}
\put(404,213){\raisebox{-.8pt}{\makebox(0,0){$\Diamond$}}}
\put(681,224){\raisebox{-.8pt}{\makebox(0,0){$\Diamond$}}}
\put(957,238){\raisebox{-.8pt}{\makebox(0,0){$\Diamond$}}}
\put(329,228){\raisebox{-.8pt}{\makebox(0,0){$\Diamond$}}}
\put(606,254){\raisebox{-.8pt}{\makebox(0,0){$\Diamond$}}}
\put(883,271){\raisebox{-.8pt}{\makebox(0,0){$\Diamond$}}}
\put(278,239){\raisebox{-.8pt}{\makebox(0,0){$\Diamond$}}}
\put(555,270){\raisebox{-.8pt}{\makebox(0,0){$\Diamond$}}}
\put(832,299){\raisebox{-.8pt}{\makebox(0,0){$\Diamond$}}}
\put(349,248){\raisebox{-.8pt}{\makebox(0,0){$\Diamond$}}}
\put(626,304){\raisebox{-.8pt}{\makebox(0,0){$\Diamond$}}}
\put(903,347){\raisebox{-.8pt}{\makebox(0,0){$\Diamond$}}}
\put(327,236){\raisebox{-.8pt}{\makebox(0,0){$\Diamond$}}}
\put(603,288){\raisebox{-.8pt}{\makebox(0,0){$\Diamond$}}}
\put(880,344){\raisebox{-.8pt}{\makebox(0,0){$\Diamond$}}}
\put(449,226){\raisebox{-.8pt}{\makebox(0,0){$\Diamond$}}}
\put(725,257){\raisebox{-.8pt}{\makebox(0,0){$\Diamond$}}}
\put(1002,274){\raisebox{-.8pt}{\makebox(0,0){$\Diamond$}}}
\put(388,243){\raisebox{-.8pt}{\makebox(0,0){$\Diamond$}}}
\put(665,285){\raisebox{-.8pt}{\makebox(0,0){$\Diamond$}}}
\put(942,317){\raisebox{-.8pt}{\makebox(0,0){$\Diamond$}}}
\put(349,235){\raisebox{-.8pt}{\makebox(0,0){$\Diamond$}}}
\put(626,290){\raisebox{-.8pt}{\makebox(0,0){$\Diamond$}}}
\put(903,341){\raisebox{-.8pt}{\makebox(0,0){$\Diamond$}}}
\put(327,221){\raisebox{-.8pt}{\makebox(0,0){$\Diamond$}}}
\put(603,273){\raisebox{-.8pt}{\makebox(0,0){$\Diamond$}}}
\put(880,336){\raisebox{-.8pt}{\makebox(0,0){$\Diamond$}}}
\put(449,227){\raisebox{-.8pt}{\makebox(0,0){$\Diamond$}}}
\put(725,254){\raisebox{-.8pt}{\makebox(0,0){$\Diamond$}}}
\put(1002,271){\raisebox{-.8pt}{\makebox(0,0){$\Diamond$}}}
\put(314,197){\raisebox{-.8pt}{\makebox(0,0){$\Diamond$}}}
\put(591,239){\raisebox{-.8pt}{\makebox(0,0){$\Diamond$}}}
\put(868,304){\raisebox{-.8pt}{\makebox(0,0){$\Diamond$}}}
\put(388,232){\raisebox{-.8pt}{\makebox(0,0){$\Diamond$}}}
\put(665,293){\raisebox{-.8pt}{\makebox(0,0){$\Diamond$}}}
\put(942,309){\raisebox{-.8pt}{\makebox(0,0){$\Diamond$}}}
\put(351,239){\raisebox{-.8pt}{\makebox(0,0){$\Diamond$}}}
\put(628,285){\raisebox{-.8pt}{\makebox(0,0){$\Diamond$}}}
\put(905,326){\raisebox{-.8pt}{\makebox(0,0){$\Diamond$}}}
\put(534,203){\raisebox{-.8pt}{\makebox(0,0){$\Diamond$}}}
\put(811,222){\raisebox{-.8pt}{\makebox(0,0){$\Diamond$}}}
\put(1088,231){\raisebox{-.8pt}{\makebox(0,0){$\Diamond$}}}
\put(450,221){\raisebox{-.8pt}{\makebox(0,0){$\Diamond$}}}
\put(727,251){\raisebox{-.8pt}{\makebox(0,0){$\Diamond$}}}
\put(1004,269){\raisebox{-.8pt}{\makebox(0,0){$\Diamond$}}}
\put(390,233){\raisebox{-.8pt}{\makebox(0,0){$\Diamond$}}}
\put(667,277){\raisebox{-.8pt}{\makebox(0,0){$\Diamond$}}}
\put(943,306){\raisebox{-.8pt}{\makebox(0,0){$\Diamond$}}}
\put(402,190){\raisebox{-.8pt}{\makebox(0,0){$\Diamond$}}}
\put(679,239){\raisebox{-.8pt}{\makebox(0,0){$\Diamond$}}}
\put(956,333){\raisebox{-.8pt}{\makebox(0,0){$\Diamond$}}}
\put(456,244){\raisebox{-.8pt}{\makebox(0,0){$\Diamond$}}}
\put(732,320){\raisebox{-.8pt}{\makebox(0,0){$\Diamond$}}}
\put(1009,379){\raisebox{-.8pt}{\makebox(0,0){$\Diamond$}}}
\put(427,240){\raisebox{-.8pt}{\makebox(0,0){$\Diamond$}}}
\put(704,312){\raisebox{-.8pt}{\makebox(0,0){$\Diamond$}}}
\put(980,394){\raisebox{-.8pt}{\makebox(0,0){$\Diamond$}}}
\put(573,215){\raisebox{-.8pt}{\makebox(0,0){$\Diamond$}}}
\put(850,254){\raisebox{-.8pt}{\makebox(0,0){$\Diamond$}}}
\put(1127,273){\raisebox{-.8pt}{\makebox(0,0){$\Diamond$}}}
\put(411,219){\raisebox{-.8pt}{\makebox(0,0){$\Diamond$}}}
\put(688,289){\raisebox{-.8pt}{\makebox(0,0){$\Diamond$}}}
\put(964,377){\raisebox{-.8pt}{\makebox(0,0){$\Diamond$}}}
\put(503,234){\raisebox{-.8pt}{\makebox(0,0){$\Diamond$}}}
\put(780,287){\raisebox{-.8pt}{\makebox(0,0){$\Diamond$}}}
\put(1056,325){\raisebox{-.8pt}{\makebox(0,0){$\Diamond$}}}
\put(456,251){\raisebox{-.8pt}{\makebox(0,0){$\Diamond$}}}
\put(732,326){\raisebox{-.8pt}{\makebox(0,0){$\Diamond$}}}
\put(1009,391){\raisebox{-.8pt}{\makebox(0,0){$\Diamond$}}}
\put(427,248){\raisebox{-.8pt}{\makebox(0,0){$\Diamond$}}}
\put(704,344){\raisebox{-.8pt}{\makebox(0,0){$\Diamond$}}}
\put(980,410){\raisebox{-.8pt}{\makebox(0,0){$\Diamond$}}}
\put(573,225){\raisebox{-.8pt}{\makebox(0,0){$\Diamond$}}}
\put(850,256){\raisebox{-.8pt}{\makebox(0,0){$\Diamond$}}}
\put(1127,286){\raisebox{-.8pt}{\makebox(0,0){$\Diamond$}}}
\put(411,231){\raisebox{-.8pt}{\makebox(0,0){$\Diamond$}}}
\put(688,300){\raisebox{-.8pt}{\makebox(0,0){$\Diamond$}}}
\put(964,403){\raisebox{-.8pt}{\makebox(0,0){$\Diamond$}}}
\put(503,243){\raisebox{-.8pt}{\makebox(0,0){$\Diamond$}}}
\put(780,293){\raisebox{-.8pt}{\makebox(0,0){$\Diamond$}}}
\put(1056,334){\raisebox{-.8pt}{\makebox(0,0){$\Diamond$}}}
\put(457,252){\raisebox{-.8pt}{\makebox(0,0){$\Diamond$}}}
\put(734,320){\raisebox{-.8pt}{\makebox(0,0){$\Diamond$}}}
\put(1011,372){\raisebox{-.8pt}{\makebox(0,0){$\Diamond$}}}
\put(428,244){\raisebox{-.8pt}{\makebox(0,0){$\Diamond$}}}
\put(705,313){\raisebox{-.8pt}{\makebox(0,0){$\Diamond$}}}
\put(982,385){\raisebox{-.8pt}{\makebox(0,0){$\Diamond$}}}
\put(575,222){\raisebox{-.8pt}{\makebox(0,0){$\Diamond$}}}
\put(851,250){\raisebox{-.8pt}{\makebox(0,0){$\Diamond$}}}
\put(1128,277){\raisebox{-.8pt}{\makebox(0,0){$\Diamond$}}}
\put(412,228){\raisebox{-.8pt}{\makebox(0,0){$\Diamond$}}}
\put(689,295){\raisebox{-.8pt}{\makebox(0,0){$\Diamond$}}}
\put(966,375){\raisebox{-.8pt}{\makebox(0,0){$\Diamond$}}}
\put(504,244){\raisebox{-.8pt}{\makebox(0,0){$\Diamond$}}}
\put(781,286){\raisebox{-.8pt}{\makebox(0,0){$\Diamond$}}}
\put(1058,330){\raisebox{-.8pt}{\makebox(0,0){$\Diamond$}}}
\put(497,218){\raisebox{-.8pt}{\makebox(0,0){$\Diamond$}}}
\put(774,314){\raisebox{-.8pt}{\makebox(0,0){$\Diamond$}}}
\put(1050,452){\raisebox{-.8pt}{\makebox(0,0){$\Diamond$}}}
\put(564,264){\raisebox{-.8pt}{\makebox(0,0){$\Diamond$}}}
\put(841,368){\raisebox{-.8pt}{\makebox(0,0){$\Diamond$}}}
\put(1118,423){\raisebox{-.8pt}{\makebox(0,0){$\Diamond$}}}
\put(491,189){\raisebox{-.8pt}{\makebox(0,0){$\Diamond$}}}
\put(768,250){\raisebox{-.8pt}{\makebox(0,0){$\Diamond$}}}
\put(1045,379){\raisebox{-.8pt}{\makebox(0,0){$\Diamond$}}}
\put(529,254){\raisebox{-.8pt}{\makebox(0,0){$\Diamond$}}}
\put(805,377){\raisebox{-.8pt}{\makebox(0,0){$\Diamond$}}}
\put(1082,476){\raisebox{-.8pt}{\makebox(0,0){$\Diamond$}}}
\put(508,238){\raisebox{-.8pt}{\makebox(0,0){$\Diamond$}}}
\put(785,360){\raisebox{-.8pt}{\makebox(0,0){$\Diamond$}}}
\put(1062,473){\raisebox{-.8pt}{\makebox(0,0){$\Diamond$}}}
\put(621,244){\raisebox{-.8pt}{\makebox(0,0){$\Diamond$}}}
\put(898,299){\raisebox{-.8pt}{\makebox(0,0){$\Diamond$}}}
\put(1174,348){\raisebox{-.8pt}{\makebox(0,0){$\Diamond$}}}
\put(564,276){\raisebox{-.8pt}{\makebox(0,0){$\Diamond$}}}
\put(841,375){\raisebox{-.8pt}{\makebox(0,0){$\Diamond$}}}
\put(1118,460){\raisebox{-.8pt}{\makebox(0,0){$\Diamond$}}}
\put(529,277){\raisebox{-.8pt}{\makebox(0,0){$\Diamond$}}}
\put(805,401){\raisebox{-.8pt}{\makebox(0,0){$\Diamond$}}}
\put(1082,514){\raisebox{-.8pt}{\makebox(0,0){$\Diamond$}}}
\put(508,254){\raisebox{-.8pt}{\makebox(0,0){$\Diamond$}}}
\put(785,372){\raisebox{-.8pt}{\makebox(0,0){$\Diamond$}}}
\put(1062,517){\raisebox{-.8pt}{\makebox(0,0){$\Diamond$}}}
\put(621,253){\raisebox{-.8pt}{\makebox(0,0){$\Diamond$}}}
\put(898,318){\raisebox{-.8pt}{\makebox(0,0){$\Diamond$}}}
\put(1174,370){\raisebox{-.8pt}{\makebox(0,0){$\Diamond$}}}
\put(498,224){\raisebox{-.8pt}{\makebox(0,0){$\Diamond$}}}
\put(775,316){\raisebox{-.8pt}{\makebox(0,0){$\Diamond$}}}
\put(1052,455){\raisebox{-.8pt}{\makebox(0,0){$\Diamond$}}}
\put(566,258){\raisebox{-.8pt}{\makebox(0,0){$\Diamond$}}}
\put(842,351){\raisebox{-.8pt}{\makebox(0,0){$\Diamond$}}}
\put(1119,430){\raisebox{-.8pt}{\makebox(0,0){$\Diamond$}}}
\put(530,271){\raisebox{-.8pt}{\makebox(0,0){$\Diamond$}}}
\put(807,397){\raisebox{-.8pt}{\makebox(0,0){$\Diamond$}}}
\put(1084,474){\raisebox{-.8pt}{\makebox(0,0){$\Diamond$}}}
\put(702,229){\raisebox{-.8pt}{\makebox(0,0){$\Diamond$}}}
\put(979,262){\raisebox{-.8pt}{\makebox(0,0){$\Diamond$}}}
\put(1256,285){\raisebox{-.8pt}{\makebox(0,0){$\Diamond$}}}
\put(509,242){\raisebox{-.8pt}{\makebox(0,0){$\Diamond$}}}
\put(786,357){\raisebox{-.8pt}{\makebox(0,0){$\Diamond$}}}
\put(1063,473){\raisebox{-.8pt}{\makebox(0,0){$\Diamond$}}}
\put(676,291){\raisebox{-.8pt}{\makebox(0,0){$\Diamond$}}}
\put(953,398){\raisebox{-.8pt}{\makebox(0,0){$\Diamond$}}}
\put(1230,483){\raisebox{-.8pt}{\makebox(0,0){$\Diamond$}}}
\put(742,268){\raisebox{-.8pt}{\makebox(0,0){$\Diamond$}}}
\put(1019,329){\raisebox{-.8pt}{\makebox(0,0){$\Diamond$}}}
\put(1296,379){\raisebox{-.8pt}{\makebox(0,0){$\Diamond$}}}
\put(592,263){\raisebox{-.8pt}{\makebox(0,0){$\Diamond$}}}
\put(869,465){\raisebox{-.8pt}{\makebox(0,0){$\Diamond$}}}
\put(1145,665){\raisebox{-.8pt}{\makebox(0,0){$\Diamond$}}}
\put(676,310){\raisebox{-.8pt}{\makebox(0,0){$\Diamond$}}}
\put(953,440){\raisebox{-.8pt}{\makebox(0,0){$\Diamond$}}}
\put(1230,529){\raisebox{-.8pt}{\makebox(0,0){$\Diamond$}}}
\put(632,324){\raisebox{-.8pt}{\makebox(0,0){$\Diamond$}}}
\put(909,509){\raisebox{-.8pt}{\makebox(0,0){$\Diamond$}}}
\put(1186,637){\raisebox{-.8pt}{\makebox(0,0){$\Diamond$}}}
\put(606,300){\raisebox{-.8pt}{\makebox(0,0){$\Diamond$}}}
\put(883,489){\raisebox{-.8pt}{\makebox(0,0){$\Diamond$}}}
\put(1160,671){\raisebox{-.8pt}{\makebox(0,0){$\Diamond$}}}
\put(744,268){\raisebox{-.8pt}{\makebox(0,0){$\Diamond$}}}
\put(1020,330){\raisebox{-.8pt}{\makebox(0,0){$\Diamond$}}}
\put(1297,389){\raisebox{-.8pt}{\makebox(0,0){$\Diamond$}}}
\put(677,293){\raisebox{-.8pt}{\makebox(0,0){$\Diamond$}}}
\put(954,402){\raisebox{-.8pt}{\makebox(0,0){$\Diamond$}}}
\put(1231,492){\raisebox{-.8pt}{\makebox(0,0){$\Diamond$}}}
\put(203.0,174.0){\rule[-0.200pt]{0.400pt}{132.977pt}}
\put(203.0,174.0){\rule[-0.200pt]{291.489pt}{0.400pt}}
\put(1413.0,174.0){\rule[-0.200pt]{0.400pt}{132.977pt}}
\put(203.0,726.0){\rule[-0.200pt]{291.489pt}{0.400pt}}
\end{picture}
}

}

The conclusion seems to be that
the number of basic operations is as predicted,
but, unexpectedly,
the {\em time per basic operation}
is larger (by as much as a factor of two) for large inputs.
We observed this effect on a number of different machines.  
We don't know why.
Perhaps caching or memory allocation issues could be the culprit.

\subsection{Empirical evaluation of Simplex}

We estimate the time for Simplex to find a near-optimal approximation to be
at least $5\min(r,c)rc$ basic operations.
This estimate comes from assuming that at least $\Omega(\min(r,c))$ pivot steps are required (because this many variables will be non-zero in the final solution), and each pivot step will take $\Omega(rc)$ time.  (This holds even for sparse matrices due to rapid fill-in.)
The leading constant 5 comes from experimental evaluation.
This estimate seems conservative, and indeed GLPK Simplex often exceeded it.

Here's a plot of the actual time for Simplex to find a ($1-\eps$)-approximate solution
(for each test input),
divided by this estimate
($5\min(r,c)rc$ times the estimated time per operation).

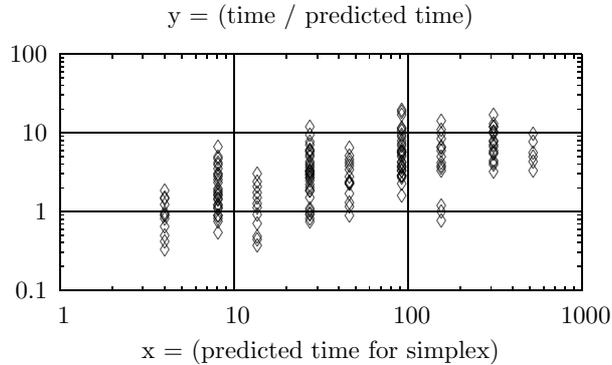
\begin{center}
\noindent\hspace*{-1em}
\scalebox{0.67}{
\setlength{\unitlength}{0.240900pt}
\ifx\plotpoint\undefined\newsavebox{\plotpoint}\fi
\sbox{\plotpoint}{\rule[-0.200pt]{0.400pt}{0.400pt}}%
\begin{picture}(1500,900)(0,0)
\font\gnuplot=cmr10 at 14pt
\gnuplot
\sbox{\plotpoint}{\rule[-0.200pt]{0.400pt}{0.400pt}}%
\put(203.0,174.0){\rule[-0.200pt]{291.489pt}{0.400pt}}
\put(203.0,174.0){\rule[-0.200pt]{4.818pt}{0.400pt}}
\put(174,174){\makebox(0,0)[r]{ 0.1}}
\put(1393.0,174.0){\rule[-0.200pt]{4.818pt}{0.400pt}}
\put(203.0,229.0){\rule[-0.200pt]{2.409pt}{0.400pt}}
\put(1403.0,229.0){\rule[-0.200pt]{2.409pt}{0.400pt}}
\put(203.0,262.0){\rule[-0.200pt]{2.409pt}{0.400pt}}
\put(1403.0,262.0){\rule[-0.200pt]{2.409pt}{0.400pt}}
\put(203.0,285.0){\rule[-0.200pt]{2.409pt}{0.400pt}}
\put(1403.0,285.0){\rule[-0.200pt]{2.409pt}{0.400pt}}
\put(203.0,303.0){\rule[-0.200pt]{2.409pt}{0.400pt}}
\put(1403.0,303.0){\rule[-0.200pt]{2.409pt}{0.400pt}}
\put(203.0,317.0){\rule[-0.200pt]{2.409pt}{0.400pt}}
\put(1403.0,317.0){\rule[-0.200pt]{2.409pt}{0.400pt}}
\put(203.0,329.0){\rule[-0.200pt]{2.409pt}{0.400pt}}
\put(1403.0,329.0){\rule[-0.200pt]{2.409pt}{0.400pt}}
\put(203.0,340.0){\rule[-0.200pt]{2.409pt}{0.400pt}}
\put(1403.0,340.0){\rule[-0.200pt]{2.409pt}{0.400pt}}
\put(203.0,350.0){\rule[-0.200pt]{2.409pt}{0.400pt}}
\put(1403.0,350.0){\rule[-0.200pt]{2.409pt}{0.400pt}}
\put(203.0,358.0){\rule[-0.200pt]{291.489pt}{0.400pt}}
\put(203.0,358.0){\rule[-0.200pt]{4.818pt}{0.400pt}}
\put(174,358){\makebox(0,0)[r]{ 1}}
\put(1393.0,358.0){\rule[-0.200pt]{4.818pt}{0.400pt}}
\put(203.0,413.0){\rule[-0.200pt]{2.409pt}{0.400pt}}
\put(1403.0,413.0){\rule[-0.200pt]{2.409pt}{0.400pt}}
\put(203.0,446.0){\rule[-0.200pt]{2.409pt}{0.400pt}}
\put(1403.0,446.0){\rule[-0.200pt]{2.409pt}{0.400pt}}
\put(203.0,469.0){\rule[-0.200pt]{2.409pt}{0.400pt}}
\put(1403.0,469.0){\rule[-0.200pt]{2.409pt}{0.400pt}}
\put(203.0,487.0){\rule[-0.200pt]{2.409pt}{0.400pt}}
\put(1403.0,487.0){\rule[-0.200pt]{2.409pt}{0.400pt}}
\put(203.0,501.0){\rule[-0.200pt]{2.409pt}{0.400pt}}
\put(1403.0,501.0){\rule[-0.200pt]{2.409pt}{0.400pt}}
\put(203.0,513.0){\rule[-0.200pt]{2.409pt}{0.400pt}}
\put(1403.0,513.0){\rule[-0.200pt]{2.409pt}{0.400pt}}
\put(203.0,524.0){\rule[-0.200pt]{2.409pt}{0.400pt}}
\put(1403.0,524.0){\rule[-0.200pt]{2.409pt}{0.400pt}}
\put(203.0,534.0){\rule[-0.200pt]{2.409pt}{0.400pt}}
\put(1403.0,534.0){\rule[-0.200pt]{2.409pt}{0.400pt}}
\put(203.0,542.0){\rule[-0.200pt]{291.489pt}{0.400pt}}
\put(203.0,542.0){\rule[-0.200pt]{4.818pt}{0.400pt}}
\put(174,542){\makebox(0,0)[r]{ 10}}
\put(1393.0,542.0){\rule[-0.200pt]{4.818pt}{0.400pt}}
\put(203.0,597.0){\rule[-0.200pt]{2.409pt}{0.400pt}}
\put(1403.0,597.0){\rule[-0.200pt]{2.409pt}{0.400pt}}
\put(203.0,630.0){\rule[-0.200pt]{2.409pt}{0.400pt}}
\put(1403.0,630.0){\rule[-0.200pt]{2.409pt}{0.400pt}}
\put(203.0,653.0){\rule[-0.200pt]{2.409pt}{0.400pt}}
\put(1403.0,653.0){\rule[-0.200pt]{2.409pt}{0.400pt}}
\put(203.0,671.0){\rule[-0.200pt]{2.409pt}{0.400pt}}
\put(1403.0,671.0){\rule[-0.200pt]{2.409pt}{0.400pt}}
\put(203.0,685.0){\rule[-0.200pt]{2.409pt}{0.400pt}}
\put(1403.0,685.0){\rule[-0.200pt]{2.409pt}{0.400pt}}
\put(203.0,697.0){\rule[-0.200pt]{2.409pt}{0.400pt}}
\put(1403.0,697.0){\rule[-0.200pt]{2.409pt}{0.400pt}}
\put(203.0,708.0){\rule[-0.200pt]{2.409pt}{0.400pt}}
\put(1403.0,708.0){\rule[-0.200pt]{2.409pt}{0.400pt}}
\put(203.0,718.0){\rule[-0.200pt]{2.409pt}{0.400pt}}
\put(1403.0,718.0){\rule[-0.200pt]{2.409pt}{0.400pt}}
\put(203.0,726.0){\rule[-0.200pt]{291.489pt}{0.400pt}}
\put(203.0,726.0){\rule[-0.200pt]{4.818pt}{0.400pt}}
\put(174,726){\makebox(0,0)[r]{ 100}}
\put(1393.0,726.0){\rule[-0.200pt]{4.818pt}{0.400pt}}
\put(203.0,174.0){\rule[-0.200pt]{0.400pt}{132.977pt}}
\put(203.0,174.0){\rule[-0.200pt]{0.400pt}{4.818pt}}
\put(203,116){\makebox(0,0){ 1}}
\put(203.0,706.0){\rule[-0.200pt]{0.400pt}{4.818pt}}
\put(324.0,174.0){\rule[-0.200pt]{0.400pt}{2.409pt}}
\put(324.0,716.0){\rule[-0.200pt]{0.400pt}{2.409pt}}
\put(395.0,174.0){\rule[-0.200pt]{0.400pt}{2.409pt}}
\put(395.0,716.0){\rule[-0.200pt]{0.400pt}{2.409pt}}
\put(446.0,174.0){\rule[-0.200pt]{0.400pt}{2.409pt}}
\put(446.0,716.0){\rule[-0.200pt]{0.400pt}{2.409pt}}
\put(485.0,174.0){\rule[-0.200pt]{0.400pt}{2.409pt}}
\put(485.0,716.0){\rule[-0.200pt]{0.400pt}{2.409pt}}
\put(517.0,174.0){\rule[-0.200pt]{0.400pt}{2.409pt}}
\put(517.0,716.0){\rule[-0.200pt]{0.400pt}{2.409pt}}
\put(544.0,174.0){\rule[-0.200pt]{0.400pt}{2.409pt}}
\put(544.0,716.0){\rule[-0.200pt]{0.400pt}{2.409pt}}
\put(567.0,174.0){\rule[-0.200pt]{0.400pt}{2.409pt}}
\put(567.0,716.0){\rule[-0.200pt]{0.400pt}{2.409pt}}
\put(588.0,174.0){\rule[-0.200pt]{0.400pt}{2.409pt}}
\put(588.0,716.0){\rule[-0.200pt]{0.400pt}{2.409pt}}
\put(606.0,174.0){\rule[-0.200pt]{0.400pt}{132.977pt}}
\put(606.0,174.0){\rule[-0.200pt]{0.400pt}{4.818pt}}
\put(606,116){\makebox(0,0){ 10}}
\put(606.0,706.0){\rule[-0.200pt]{0.400pt}{4.818pt}}
\put(728.0,174.0){\rule[-0.200pt]{0.400pt}{2.409pt}}
\put(728.0,716.0){\rule[-0.200pt]{0.400pt}{2.409pt}}
\put(799.0,174.0){\rule[-0.200pt]{0.400pt}{2.409pt}}
\put(799.0,716.0){\rule[-0.200pt]{0.400pt}{2.409pt}}
\put(849.0,174.0){\rule[-0.200pt]{0.400pt}{2.409pt}}
\put(849.0,716.0){\rule[-0.200pt]{0.400pt}{2.409pt}}
\put(888.0,174.0){\rule[-0.200pt]{0.400pt}{2.409pt}}
\put(888.0,716.0){\rule[-0.200pt]{0.400pt}{2.409pt}}
\put(920.0,174.0){\rule[-0.200pt]{0.400pt}{2.409pt}}
\put(920.0,716.0){\rule[-0.200pt]{0.400pt}{2.409pt}}
\put(947.0,174.0){\rule[-0.200pt]{0.400pt}{2.409pt}}
\put(947.0,716.0){\rule[-0.200pt]{0.400pt}{2.409pt}}
\put(971.0,174.0){\rule[-0.200pt]{0.400pt}{2.409pt}}
\put(971.0,716.0){\rule[-0.200pt]{0.400pt}{2.409pt}}
\put(991.0,174.0){\rule[-0.200pt]{0.400pt}{2.409pt}}
\put(991.0,716.0){\rule[-0.200pt]{0.400pt}{2.409pt}}
\put(1010.0,174.0){\rule[-0.200pt]{0.400pt}{132.977pt}}
\put(1010.0,174.0){\rule[-0.200pt]{0.400pt}{4.818pt}}
\put(1010,116){\makebox(0,0){ 100}}
\put(1010.0,706.0){\rule[-0.200pt]{0.400pt}{4.818pt}}
\put(1131.0,174.0){\rule[-0.200pt]{0.400pt}{2.409pt}}
\put(1131.0,716.0){\rule[-0.200pt]{0.400pt}{2.409pt}}
\put(1202.0,174.0){\rule[-0.200pt]{0.400pt}{2.409pt}}
\put(1202.0,716.0){\rule[-0.200pt]{0.400pt}{2.409pt}}
\put(1252.0,174.0){\rule[-0.200pt]{0.400pt}{2.409pt}}
\put(1252.0,716.0){\rule[-0.200pt]{0.400pt}{2.409pt}}
\put(1292.0,174.0){\rule[-0.200pt]{0.400pt}{2.409pt}}
\put(1292.0,716.0){\rule[-0.200pt]{0.400pt}{2.409pt}}
\put(1324.0,174.0){\rule[-0.200pt]{0.400pt}{2.409pt}}
\put(1324.0,716.0){\rule[-0.200pt]{0.400pt}{2.409pt}}
\put(1351.0,174.0){\rule[-0.200pt]{0.400pt}{2.409pt}}
\put(1351.0,716.0){\rule[-0.200pt]{0.400pt}{2.409pt}}
\put(1374.0,174.0){\rule[-0.200pt]{0.400pt}{2.409pt}}
\put(1374.0,716.0){\rule[-0.200pt]{0.400pt}{2.409pt}}
\put(1395.0,174.0){\rule[-0.200pt]{0.400pt}{2.409pt}}
\put(1395.0,716.0){\rule[-0.200pt]{0.400pt}{2.409pt}}
\put(1413.0,174.0){\rule[-0.200pt]{0.400pt}{132.977pt}}
\put(1413.0,174.0){\rule[-0.200pt]{0.400pt}{4.818pt}}
\put(1413,116){\makebox(0,0){ 1000}}
\put(1413.0,706.0){\rule[-0.200pt]{0.400pt}{4.818pt}}
\put(203.0,174.0){\rule[-0.200pt]{0.400pt}{132.977pt}}
\put(203.0,174.0){\rule[-0.200pt]{291.489pt}{0.400pt}}
\put(1413.0,174.0){\rule[-0.200pt]{0.400pt}{132.977pt}}
\put(203.0,726.0){\rule[-0.200pt]{291.489pt}{0.400pt}}
\put(808,29){\makebox(0,0){x = (predicted time for simplex)}}
\put(808,813){\makebox(0,0){y = (time / predicted time)}}
\put(447,351){\raisebox{-.8pt}{\makebox(0,0){$\Diamond$}}}
\put(447,391){\raisebox{-.8pt}{\makebox(0,0){$\Diamond$}}}
\put(447,409){\raisebox{-.8pt}{\makebox(0,0){$\Diamond$}}}
\put(447,273){\raisebox{-.8pt}{\makebox(0,0){$\Diamond$}}}
\put(447,290){\raisebox{-.8pt}{\makebox(0,0){$\Diamond$}}}
\put(447,304){\raisebox{-.8pt}{\makebox(0,0){$\Diamond$}}}
\put(447,326){\raisebox{-.8pt}{\makebox(0,0){$\Diamond$}}}
\put(447,344){\raisebox{-.8pt}{\makebox(0,0){$\Diamond$}}}
\put(447,358){\raisebox{-.8pt}{\makebox(0,0){$\Diamond$}}}
\put(447,353){\raisebox{-.8pt}{\makebox(0,0){$\Diamond$}}}
\put(447,378){\raisebox{-.8pt}{\makebox(0,0){$\Diamond$}}}
\put(447,393){\raisebox{-.8pt}{\makebox(0,0){$\Diamond$}}}
\put(570,371){\raisebox{-.8pt}{\makebox(0,0){$\Diamond$}}}
\put(570,400){\raisebox{-.8pt}{\makebox(0,0){$\Diamond$}}}
\put(570,415){\raisebox{-.8pt}{\makebox(0,0){$\Diamond$}}}
\put(570,446){\raisebox{-.8pt}{\makebox(0,0){$\Diamond$}}}
\put(570,470){\raisebox{-.8pt}{\makebox(0,0){$\Diamond$}}}
\put(570,484){\raisebox{-.8pt}{\makebox(0,0){$\Diamond$}}}
\put(570,313){\raisebox{-.8pt}{\makebox(0,0){$\Diamond$}}}
\put(570,338){\raisebox{-.8pt}{\makebox(0,0){$\Diamond$}}}
\put(570,351){\raisebox{-.8pt}{\makebox(0,0){$\Diamond$}}}
\put(570,344){\raisebox{-.8pt}{\makebox(0,0){$\Diamond$}}}
\put(570,375){\raisebox{-.8pt}{\makebox(0,0){$\Diamond$}}}
\put(570,392){\raisebox{-.8pt}{\makebox(0,0){$\Diamond$}}}
\put(570,406){\raisebox{-.8pt}{\makebox(0,0){$\Diamond$}}}
\put(570,435){\raisebox{-.8pt}{\makebox(0,0){$\Diamond$}}}
\put(570,462){\raisebox{-.8pt}{\makebox(0,0){$\Diamond$}}}
\put(570,399){\raisebox{-.8pt}{\makebox(0,0){$\Diamond$}}}
\put(570,450){\raisebox{-.8pt}{\makebox(0,0){$\Diamond$}}}
\put(570,474){\raisebox{-.8pt}{\makebox(0,0){$\Diamond$}}}
\put(570,352){\raisebox{-.8pt}{\makebox(0,0){$\Diamond$}}}
\put(570,368){\raisebox{-.8pt}{\makebox(0,0){$\Diamond$}}}
\put(570,378){\raisebox{-.8pt}{\makebox(0,0){$\Diamond$}}}
\put(570,442){\raisebox{-.8pt}{\makebox(0,0){$\Diamond$}}}
\put(570,489){\raisebox{-.8pt}{\makebox(0,0){$\Diamond$}}}
\put(570,512){\raisebox{-.8pt}{\makebox(0,0){$\Diamond$}}}
\put(570,390){\raisebox{-.8pt}{\makebox(0,0){$\Diamond$}}}
\put(570,411){\raisebox{-.8pt}{\makebox(0,0){$\Diamond$}}}
\put(570,428){\raisebox{-.8pt}{\makebox(0,0){$\Diamond$}}}
\put(661,393){\raisebox{-.8pt}{\makebox(0,0){$\Diamond$}}}
\put(661,430){\raisebox{-.8pt}{\makebox(0,0){$\Diamond$}}}
\put(661,450){\raisebox{-.8pt}{\makebox(0,0){$\Diamond$}}}
\put(661,281){\raisebox{-.8pt}{\makebox(0,0){$\Diamond$}}}
\put(661,295){\raisebox{-.8pt}{\makebox(0,0){$\Diamond$}}}
\put(661,302){\raisebox{-.8pt}{\makebox(0,0){$\Diamond$}}}
\put(661,332){\raisebox{-.8pt}{\makebox(0,0){$\Diamond$}}}
\put(661,354){\raisebox{-.8pt}{\makebox(0,0){$\Diamond$}}}
\put(661,370){\raisebox{-.8pt}{\makebox(0,0){$\Diamond$}}}
\put(661,379){\raisebox{-.8pt}{\makebox(0,0){$\Diamond$}}}
\put(661,406){\raisebox{-.8pt}{\makebox(0,0){$\Diamond$}}}
\put(661,420){\raisebox{-.8pt}{\makebox(0,0){$\Diamond$}}}
\put(783,457){\raisebox{-.8pt}{\makebox(0,0){$\Diamond$}}}
\put(783,516){\raisebox{-.8pt}{\makebox(0,0){$\Diamond$}}}
\put(783,560){\raisebox{-.8pt}{\makebox(0,0){$\Diamond$}}}
\put(783,457){\raisebox{-.8pt}{\makebox(0,0){$\Diamond$}}}
\put(783,478){\raisebox{-.8pt}{\makebox(0,0){$\Diamond$}}}
\put(783,493){\raisebox{-.8pt}{\makebox(0,0){$\Diamond$}}}
\put(783,469){\raisebox{-.8pt}{\makebox(0,0){$\Diamond$}}}
\put(783,501){\raisebox{-.8pt}{\makebox(0,0){$\Diamond$}}}
\put(783,525){\raisebox{-.8pt}{\makebox(0,0){$\Diamond$}}}
\put(783,345){\raisebox{-.8pt}{\makebox(0,0){$\Diamond$}}}
\put(783,356){\raisebox{-.8pt}{\makebox(0,0){$\Diamond$}}}
\put(783,364){\raisebox{-.8pt}{\makebox(0,0){$\Diamond$}}}
\put(783,447){\raisebox{-.8pt}{\makebox(0,0){$\Diamond$}}}
\put(783,503){\raisebox{-.8pt}{\makebox(0,0){$\Diamond$}}}
\put(783,541){\raisebox{-.8pt}{\makebox(0,0){$\Diamond$}}}
\put(783,425){\raisebox{-.8pt}{\makebox(0,0){$\Diamond$}}}
\put(783,441){\raisebox{-.8pt}{\makebox(0,0){$\Diamond$}}}
\put(783,451){\raisebox{-.8pt}{\makebox(0,0){$\Diamond$}}}
\put(783,411){\raisebox{-.8pt}{\makebox(0,0){$\Diamond$}}}
\put(783,439){\raisebox{-.8pt}{\makebox(0,0){$\Diamond$}}}
\put(783,454){\raisebox{-.8pt}{\makebox(0,0){$\Diamond$}}}
\put(783,411){\raisebox{-.8pt}{\makebox(0,0){$\Diamond$}}}
\put(783,448){\raisebox{-.8pt}{\makebox(0,0){$\Diamond$}}}
\put(783,473){\raisebox{-.8pt}{\makebox(0,0){$\Diamond$}}}
\put(783,337){\raisebox{-.8pt}{\makebox(0,0){$\Diamond$}}}
\put(783,355){\raisebox{-.8pt}{\makebox(0,0){$\Diamond$}}}
\put(783,368){\raisebox{-.8pt}{\makebox(0,0){$\Diamond$}}}
\put(783,463){\raisebox{-.8pt}{\makebox(0,0){$\Diamond$}}}
\put(783,488){\raisebox{-.8pt}{\makebox(0,0){$\Diamond$}}}
\put(783,505){\raisebox{-.8pt}{\makebox(0,0){$\Diamond$}}}
\put(783,393){\raisebox{-.8pt}{\makebox(0,0){$\Diamond$}}}
\put(783,418){\raisebox{-.8pt}{\makebox(0,0){$\Diamond$}}}
\put(783,434){\raisebox{-.8pt}{\makebox(0,0){$\Diamond$}}}
\put(875,432){\raisebox{-.8pt}{\makebox(0,0){$\Diamond$}}}
\put(875,459){\raisebox{-.8pt}{\makebox(0,0){$\Diamond$}}}
\put(875,476){\raisebox{-.8pt}{\makebox(0,0){$\Diamond$}}}
\put(875,429){\raisebox{-.8pt}{\makebox(0,0){$\Diamond$}}}
\put(875,469){\raisebox{-.8pt}{\makebox(0,0){$\Diamond$}}}
\put(875,491){\raisebox{-.8pt}{\makebox(0,0){$\Diamond$}}}
\put(875,351){\raisebox{-.8pt}{\makebox(0,0){$\Diamond$}}}
\put(875,372){\raisebox{-.8pt}{\makebox(0,0){$\Diamond$}}}
\put(875,383){\raisebox{-.8pt}{\makebox(0,0){$\Diamond$}}}
\put(875,427){\raisebox{-.8pt}{\makebox(0,0){$\Diamond$}}}
\put(875,483){\raisebox{-.8pt}{\makebox(0,0){$\Diamond$}}}
\put(875,511){\raisebox{-.8pt}{\makebox(0,0){$\Diamond$}}}
\put(875,404){\raisebox{-.8pt}{\makebox(0,0){$\Diamond$}}}
\put(875,427){\raisebox{-.8pt}{\makebox(0,0){$\Diamond$}}}
\put(875,443){\raisebox{-.8pt}{\makebox(0,0){$\Diamond$}}}
\put(996,465){\raisebox{-.8pt}{\makebox(0,0){$\Diamond$}}}
\put(996,536){\raisebox{-.8pt}{\makebox(0,0){$\Diamond$}}}
\put(996,586){\raisebox{-.8pt}{\makebox(0,0){$\Diamond$}}}
\put(996,494){\raisebox{-.8pt}{\makebox(0,0){$\Diamond$}}}
\put(996,514){\raisebox{-.8pt}{\makebox(0,0){$\Diamond$}}}
\put(996,527){\raisebox{-.8pt}{\makebox(0,0){$\Diamond$}}}
\put(996,464){\raisebox{-.8pt}{\makebox(0,0){$\Diamond$}}}
\put(996,549){\raisebox{-.8pt}{\makebox(0,0){$\Diamond$}}}
\put(996,599){\raisebox{-.8pt}{\makebox(0,0){$\Diamond$}}}
\put(996,505){\raisebox{-.8pt}{\makebox(0,0){$\Diamond$}}}
\put(996,540){\raisebox{-.8pt}{\makebox(0,0){$\Diamond$}}}
\put(996,555){\raisebox{-.8pt}{\makebox(0,0){$\Diamond$}}}
\put(996,498){\raisebox{-.8pt}{\makebox(0,0){$\Diamond$}}}
\put(996,556){\raisebox{-.8pt}{\makebox(0,0){$\Diamond$}}}
\put(996,593){\raisebox{-.8pt}{\makebox(0,0){$\Diamond$}}}
\put(996,439){\raisebox{-.8pt}{\makebox(0,0){$\Diamond$}}}
\put(996,457){\raisebox{-.8pt}{\makebox(0,0){$\Diamond$}}}
\put(996,467){\raisebox{-.8pt}{\makebox(0,0){$\Diamond$}}}
\put(996,445){\raisebox{-.8pt}{\makebox(0,0){$\Diamond$}}}
\put(996,474){\raisebox{-.8pt}{\makebox(0,0){$\Diamond$}}}
\put(996,490){\raisebox{-.8pt}{\makebox(0,0){$\Diamond$}}}
\put(996,445){\raisebox{-.8pt}{\makebox(0,0){$\Diamond$}}}
\put(996,478){\raisebox{-.8pt}{\makebox(0,0){$\Diamond$}}}
\put(996,501){\raisebox{-.8pt}{\makebox(0,0){$\Diamond$}}}
\put(996,456){\raisebox{-.8pt}{\makebox(0,0){$\Diamond$}}}
\put(996,500){\raisebox{-.8pt}{\makebox(0,0){$\Diamond$}}}
\put(996,520){\raisebox{-.8pt}{\makebox(0,0){$\Diamond$}}}
\put(996,399){\raisebox{-.8pt}{\makebox(0,0){$\Diamond$}}}
\put(996,427){\raisebox{-.8pt}{\makebox(0,0){$\Diamond$}}}
\put(996,445){\raisebox{-.8pt}{\makebox(0,0){$\Diamond$}}}
\put(1088,456){\raisebox{-.8pt}{\makebox(0,0){$\Diamond$}}}
\put(1088,542){\raisebox{-.8pt}{\makebox(0,0){$\Diamond$}}}
\put(1088,572){\raisebox{-.8pt}{\makebox(0,0){$\Diamond$}}}
\put(1088,463){\raisebox{-.8pt}{\makebox(0,0){$\Diamond$}}}
\put(1088,488){\raisebox{-.8pt}{\makebox(0,0){$\Diamond$}}}
\put(1088,503){\raisebox{-.8pt}{\makebox(0,0){$\Diamond$}}}
\put(1088,474){\raisebox{-.8pt}{\makebox(0,0){$\Diamond$}}}
\put(1088,511){\raisebox{-.8pt}{\makebox(0,0){$\Diamond$}}}
\put(1088,532){\raisebox{-.8pt}{\makebox(0,0){$\Diamond$}}}
\put(1088,341){\raisebox{-.8pt}{\makebox(0,0){$\Diamond$}}}
\put(1088,361){\raisebox{-.8pt}{\makebox(0,0){$\Diamond$}}}
\put(1088,374){\raisebox{-.8pt}{\makebox(0,0){$\Diamond$}}}
\put(1088,468){\raisebox{-.8pt}{\makebox(0,0){$\Diamond$}}}
\put(1088,513){\raisebox{-.8pt}{\makebox(0,0){$\Diamond$}}}
\put(1088,552){\raisebox{-.8pt}{\makebox(0,0){$\Diamond$}}}
\put(1209,525){\raisebox{-.8pt}{\makebox(0,0){$\Diamond$}}}
\put(1209,547){\raisebox{-.8pt}{\makebox(0,0){$\Diamond$}}}
\put(1209,558){\raisebox{-.8pt}{\makebox(0,0){$\Diamond$}}}
\put(1209,454){\raisebox{-.8pt}{\makebox(0,0){$\Diamond$}}}
\put(1209,470){\raisebox{-.8pt}{\makebox(0,0){$\Diamond$}}}
\put(1209,479){\raisebox{-.8pt}{\makebox(0,0){$\Diamond$}}}
\put(1209,513){\raisebox{-.8pt}{\makebox(0,0){$\Diamond$}}}
\put(1209,558){\raisebox{-.8pt}{\makebox(0,0){$\Diamond$}}}
\put(1209,588){\raisebox{-.8pt}{\makebox(0,0){$\Diamond$}}}
\put(1209,475){\raisebox{-.8pt}{\makebox(0,0){$\Diamond$}}}
\put(1209,499){\raisebox{-.8pt}{\makebox(0,0){$\Diamond$}}}
\put(1209,517){\raisebox{-.8pt}{\makebox(0,0){$\Diamond$}}}
\put(1209,496){\raisebox{-.8pt}{\makebox(0,0){$\Diamond$}}}
\put(1209,527){\raisebox{-.8pt}{\makebox(0,0){$\Diamond$}}}
\put(1209,547){\raisebox{-.8pt}{\makebox(0,0){$\Diamond$}}}
\put(1209,503){\raisebox{-.8pt}{\makebox(0,0){$\Diamond$}}}
\put(1209,543){\raisebox{-.8pt}{\makebox(0,0){$\Diamond$}}}
\put(1209,566){\raisebox{-.8pt}{\makebox(0,0){$\Diamond$}}}
\put(1301,457){\raisebox{-.8pt}{\makebox(0,0){$\Diamond$}}}
\put(1301,477){\raisebox{-.8pt}{\makebox(0,0){$\Diamond$}}}
\put(1301,489){\raisebox{-.8pt}{\makebox(0,0){$\Diamond$}}}
\put(1301,499){\raisebox{-.8pt}{\makebox(0,0){$\Diamond$}}}
\put(1301,525){\raisebox{-.8pt}{\makebox(0,0){$\Diamond$}}}
\put(1301,542){\raisebox{-.8pt}{\makebox(0,0){$\Diamond$}}}
\put(203.0,174.0){\rule[-0.200pt]{0.400pt}{132.977pt}}
\put(203.0,174.0){\rule[-0.200pt]{291.489pt}{0.400pt}}
\put(1413.0,174.0){\rule[-0.200pt]{0.400pt}{132.977pt}}
\put(203.0,726.0){\rule[-0.200pt]{291.489pt}{0.400pt}}
\end{picture}

}
\end{center}

Simplex generally took at least the estimated time,
and sometimes up to a factor of ten longer.
(Note also that this experimental data excludes
about 10\% of the runs, in which GLPK Simplex 
failed to terminate due to basis cycling.)

\subsection{Speed-up of this algorithm versus Simplex.}
Combining the above estimates, a conservative estimate of the speed-up factor
in using the algorithm here instead of Simplex
(that is, the time for Simplex divided by the time for the algorithm here) is 
\begin{equation}\label{eqn:predspeedup}
\frac{\textstyle
5\min(r,c)rc
}{\textstyle
[12(r+c) ~+~ 480 d^{-1}]\,\ln(rc)/\eps^2.
}
\end{equation}

The plot below plots the actual measured speed-up divided by the conservative estimate~(\ref{eqn:predspeedup}), as a function of the estimated running time of the algorithm here.

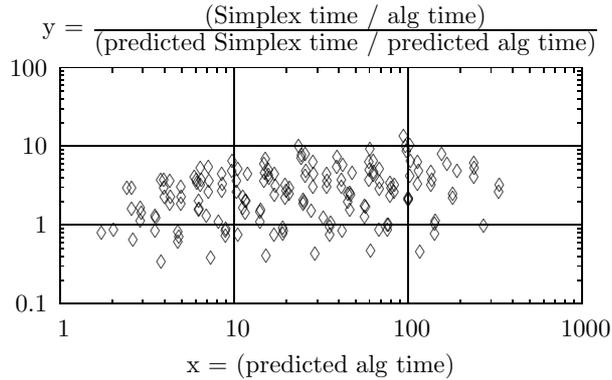
\begin{center}
\noindent\hspace*{-1em}
\scalebox{0.67}{
\setlength{\unitlength}{0.240900pt}
\ifx\plotpoint\undefined\newsavebox{\plotpoint}\fi
\sbox{\plotpoint}{\rule[-0.200pt]{0.400pt}{0.400pt}}%
\begin{picture}(1500,900)(0,0)
\font\gnuplot=cmr10 at 14pt
\gnuplot
\sbox{\plotpoint}{\rule[-0.200pt]{0.400pt}{0.400pt}}%
\put(203.0,174.0){\rule[-0.200pt]{291.489pt}{0.400pt}}
\put(203.0,174.0){\rule[-0.200pt]{4.818pt}{0.400pt}}
\put(174,174){\makebox(0,0)[r]{ 0.1}}
\put(1393.0,174.0){\rule[-0.200pt]{4.818pt}{0.400pt}}
\put(203.0,229.0){\rule[-0.200pt]{2.409pt}{0.400pt}}
\put(1403.0,229.0){\rule[-0.200pt]{2.409pt}{0.400pt}}
\put(203.0,262.0){\rule[-0.200pt]{2.409pt}{0.400pt}}
\put(1403.0,262.0){\rule[-0.200pt]{2.409pt}{0.400pt}}
\put(203.0,285.0){\rule[-0.200pt]{2.409pt}{0.400pt}}
\put(1403.0,285.0){\rule[-0.200pt]{2.409pt}{0.400pt}}
\put(203.0,303.0){\rule[-0.200pt]{2.409pt}{0.400pt}}
\put(1403.0,303.0){\rule[-0.200pt]{2.409pt}{0.400pt}}
\put(203.0,317.0){\rule[-0.200pt]{2.409pt}{0.400pt}}
\put(1403.0,317.0){\rule[-0.200pt]{2.409pt}{0.400pt}}
\put(203.0,329.0){\rule[-0.200pt]{2.409pt}{0.400pt}}
\put(1403.0,329.0){\rule[-0.200pt]{2.409pt}{0.400pt}}
\put(203.0,340.0){\rule[-0.200pt]{2.409pt}{0.400pt}}
\put(1403.0,340.0){\rule[-0.200pt]{2.409pt}{0.400pt}}
\put(203.0,350.0){\rule[-0.200pt]{2.409pt}{0.400pt}}
\put(1403.0,350.0){\rule[-0.200pt]{2.409pt}{0.400pt}}
\put(203.0,358.0){\rule[-0.200pt]{291.489pt}{0.400pt}}
\put(203.0,358.0){\rule[-0.200pt]{4.818pt}{0.400pt}}
\put(174,358){\makebox(0,0)[r]{ 1}}
\put(1393.0,358.0){\rule[-0.200pt]{4.818pt}{0.400pt}}
\put(203.0,413.0){\rule[-0.200pt]{2.409pt}{0.400pt}}
\put(1403.0,413.0){\rule[-0.200pt]{2.409pt}{0.400pt}}
\put(203.0,446.0){\rule[-0.200pt]{2.409pt}{0.400pt}}
\put(1403.0,446.0){\rule[-0.200pt]{2.409pt}{0.400pt}}
\put(203.0,469.0){\rule[-0.200pt]{2.409pt}{0.400pt}}
\put(1403.0,469.0){\rule[-0.200pt]{2.409pt}{0.400pt}}
\put(203.0,487.0){\rule[-0.200pt]{2.409pt}{0.400pt}}
\put(1403.0,487.0){\rule[-0.200pt]{2.409pt}{0.400pt}}
\put(203.0,501.0){\rule[-0.200pt]{2.409pt}{0.400pt}}
\put(1403.0,501.0){\rule[-0.200pt]{2.409pt}{0.400pt}}
\put(203.0,513.0){\rule[-0.200pt]{2.409pt}{0.400pt}}
\put(1403.0,513.0){\rule[-0.200pt]{2.409pt}{0.400pt}}
\put(203.0,524.0){\rule[-0.200pt]{2.409pt}{0.400pt}}
\put(1403.0,524.0){\rule[-0.200pt]{2.409pt}{0.400pt}}
\put(203.0,534.0){\rule[-0.200pt]{2.409pt}{0.400pt}}
\put(1403.0,534.0){\rule[-0.200pt]{2.409pt}{0.400pt}}
\put(203.0,542.0){\rule[-0.200pt]{291.489pt}{0.400pt}}
\put(203.0,542.0){\rule[-0.200pt]{4.818pt}{0.400pt}}
\put(174,542){\makebox(0,0)[r]{ 10}}
\put(1393.0,542.0){\rule[-0.200pt]{4.818pt}{0.400pt}}
\put(203.0,597.0){\rule[-0.200pt]{2.409pt}{0.400pt}}
\put(1403.0,597.0){\rule[-0.200pt]{2.409pt}{0.400pt}}
\put(203.0,630.0){\rule[-0.200pt]{2.409pt}{0.400pt}}
\put(1403.0,630.0){\rule[-0.200pt]{2.409pt}{0.400pt}}
\put(203.0,653.0){\rule[-0.200pt]{2.409pt}{0.400pt}}
\put(1403.0,653.0){\rule[-0.200pt]{2.409pt}{0.400pt}}
\put(203.0,671.0){\rule[-0.200pt]{2.409pt}{0.400pt}}
\put(1403.0,671.0){\rule[-0.200pt]{2.409pt}{0.400pt}}
\put(203.0,685.0){\rule[-0.200pt]{2.409pt}{0.400pt}}
\put(1403.0,685.0){\rule[-0.200pt]{2.409pt}{0.400pt}}
\put(203.0,697.0){\rule[-0.200pt]{2.409pt}{0.400pt}}
\put(1403.0,697.0){\rule[-0.200pt]{2.409pt}{0.400pt}}
\put(203.0,708.0){\rule[-0.200pt]{2.409pt}{0.400pt}}
\put(1403.0,708.0){\rule[-0.200pt]{2.409pt}{0.400pt}}
\put(203.0,718.0){\rule[-0.200pt]{2.409pt}{0.400pt}}
\put(1403.0,718.0){\rule[-0.200pt]{2.409pt}{0.400pt}}
\put(203.0,726.0){\rule[-0.200pt]{291.489pt}{0.400pt}}
\put(203.0,726.0){\rule[-0.200pt]{4.818pt}{0.400pt}}
\put(174,726){\makebox(0,0)[r]{ 100}}
\put(1393.0,726.0){\rule[-0.200pt]{4.818pt}{0.400pt}}
\put(203.0,174.0){\rule[-0.200pt]{0.400pt}{132.977pt}}
\put(203.0,174.0){\rule[-0.200pt]{0.400pt}{4.818pt}}
\put(203,116){\makebox(0,0){ 1}}
\put(203.0,706.0){\rule[-0.200pt]{0.400pt}{4.818pt}}
\put(324.0,174.0){\rule[-0.200pt]{0.400pt}{2.409pt}}
\put(324.0,716.0){\rule[-0.200pt]{0.400pt}{2.409pt}}
\put(395.0,174.0){\rule[-0.200pt]{0.400pt}{2.409pt}}
\put(395.0,716.0){\rule[-0.200pt]{0.400pt}{2.409pt}}
\put(446.0,174.0){\rule[-0.200pt]{0.400pt}{2.409pt}}
\put(446.0,716.0){\rule[-0.200pt]{0.400pt}{2.409pt}}
\put(485.0,174.0){\rule[-0.200pt]{0.400pt}{2.409pt}}
\put(485.0,716.0){\rule[-0.200pt]{0.400pt}{2.409pt}}
\put(517.0,174.0){\rule[-0.200pt]{0.400pt}{2.409pt}}
\put(517.0,716.0){\rule[-0.200pt]{0.400pt}{2.409pt}}
\put(544.0,174.0){\rule[-0.200pt]{0.400pt}{2.409pt}}
\put(544.0,716.0){\rule[-0.200pt]{0.400pt}{2.409pt}}
\put(567.0,174.0){\rule[-0.200pt]{0.400pt}{2.409pt}}
\put(567.0,716.0){\rule[-0.200pt]{0.400pt}{2.409pt}}
\put(588.0,174.0){\rule[-0.200pt]{0.400pt}{2.409pt}}
\put(588.0,716.0){\rule[-0.200pt]{0.400pt}{2.409pt}}
\put(606.0,174.0){\rule[-0.200pt]{0.400pt}{132.977pt}}
\put(606.0,174.0){\rule[-0.200pt]{0.400pt}{4.818pt}}
\put(606,116){\makebox(0,0){ 10}}
\put(606.0,706.0){\rule[-0.200pt]{0.400pt}{4.818pt}}
\put(728.0,174.0){\rule[-0.200pt]{0.400pt}{2.409pt}}
\put(728.0,716.0){\rule[-0.200pt]{0.400pt}{2.409pt}}
\put(799.0,174.0){\rule[-0.200pt]{0.400pt}{2.409pt}}
\put(799.0,716.0){\rule[-0.200pt]{0.400pt}{2.409pt}}
\put(849.0,174.0){\rule[-0.200pt]{0.400pt}{2.409pt}}
\put(849.0,716.0){\rule[-0.200pt]{0.400pt}{2.409pt}}
\put(888.0,174.0){\rule[-0.200pt]{0.400pt}{2.409pt}}
\put(888.0,716.0){\rule[-0.200pt]{0.400pt}{2.409pt}}
\put(920.0,174.0){\rule[-0.200pt]{0.400pt}{2.409pt}}
\put(920.0,716.0){\rule[-0.200pt]{0.400pt}{2.409pt}}
\put(947.0,174.0){\rule[-0.200pt]{0.400pt}{2.409pt}}
\put(947.0,716.0){\rule[-0.200pt]{0.400pt}{2.409pt}}
\put(971.0,174.0){\rule[-0.200pt]{0.400pt}{2.409pt}}
\put(971.0,716.0){\rule[-0.200pt]{0.400pt}{2.409pt}}
\put(991.0,174.0){\rule[-0.200pt]{0.400pt}{2.409pt}}
\put(991.0,716.0){\rule[-0.200pt]{0.400pt}{2.409pt}}
\put(1010.0,174.0){\rule[-0.200pt]{0.400pt}{132.977pt}}
\put(1010.0,174.0){\rule[-0.200pt]{0.400pt}{4.818pt}}
\put(1010,116){\makebox(0,0){ 100}}
\put(1010.0,706.0){\rule[-0.200pt]{0.400pt}{4.818pt}}
\put(1131.0,174.0){\rule[-0.200pt]{0.400pt}{2.409pt}}
\put(1131.0,716.0){\rule[-0.200pt]{0.400pt}{2.409pt}}
\put(1202.0,174.0){\rule[-0.200pt]{0.400pt}{2.409pt}}
\put(1202.0,716.0){\rule[-0.200pt]{0.400pt}{2.409pt}}
\put(1252.0,174.0){\rule[-0.200pt]{0.400pt}{2.409pt}}
\put(1252.0,716.0){\rule[-0.200pt]{0.400pt}{2.409pt}}
\put(1292.0,174.0){\rule[-0.200pt]{0.400pt}{2.409pt}}
\put(1292.0,716.0){\rule[-0.200pt]{0.400pt}{2.409pt}}
\put(1324.0,174.0){\rule[-0.200pt]{0.400pt}{2.409pt}}
\put(1324.0,716.0){\rule[-0.200pt]{0.400pt}{2.409pt}}
\put(1351.0,174.0){\rule[-0.200pt]{0.400pt}{2.409pt}}
\put(1351.0,716.0){\rule[-0.200pt]{0.400pt}{2.409pt}}
\put(1374.0,174.0){\rule[-0.200pt]{0.400pt}{2.409pt}}
\put(1374.0,716.0){\rule[-0.200pt]{0.400pt}{2.409pt}}
\put(1395.0,174.0){\rule[-0.200pt]{0.400pt}{2.409pt}}
\put(1395.0,716.0){\rule[-0.200pt]{0.400pt}{2.409pt}}
\put(1413.0,174.0){\rule[-0.200pt]{0.400pt}{132.977pt}}
\put(1413.0,174.0){\rule[-0.200pt]{0.400pt}{4.818pt}}
\put(1413,116){\makebox(0,0){ 1000}}
\put(1413.0,706.0){\rule[-0.200pt]{0.400pt}{4.818pt}}
\put(203.0,174.0){\rule[-0.200pt]{0.400pt}{132.977pt}}
\put(203.0,174.0){\rule[-0.200pt]{291.489pt}{0.400pt}}
\put(1413.0,174.0){\rule[-0.200pt]{0.400pt}{132.977pt}}
\put(203.0,726.0){\rule[-0.200pt]{291.489pt}{0.400pt}}
\put(808,29){\makebox(0,0){x = (predicted alg time)}}
\put(808,813){\makebox(0,0){y = $\frac{\mbox{(Simplex time / alg time)}}{\mbox{(predicted Simplex time / predicted alg time)}}$}}
\put(300,343){\raisebox{-.8pt}{\makebox(0,0){$\Diamond$}}}
\put(543,384){\raisebox{-.8pt}{\makebox(0,0){$\Diamond$}}}
\put(786,394){\raisebox{-.8pt}{\makebox(0,0){$\Diamond$}}}
\put(438,276){\raisebox{-.8pt}{\makebox(0,0){$\Diamond$}}}
\put(681,290){\raisebox{-.8pt}{\makebox(0,0){$\Diamond$}}}
\put(924,302){\raisebox{-.8pt}{\makebox(0,0){$\Diamond$}}}
\put(373,326){\raisebox{-.8pt}{\makebox(0,0){$\Diamond$}}}
\put(616,339){\raisebox{-.8pt}{\makebox(0,0){$\Diamond$}}}
\put(858,349){\raisebox{-.8pt}{\makebox(0,0){$\Diamond$}}}
\put(328,350){\raisebox{-.8pt}{\makebox(0,0){$\Diamond$}}}
\put(571,370){\raisebox{-.8pt}{\makebox(0,0){$\Diamond$}}}
\put(814,379){\raisebox{-.8pt}{\makebox(0,0){$\Diamond$}}}
\put(390,371){\raisebox{-.8pt}{\makebox(0,0){$\Diamond$}}}
\put(633,390){\raisebox{-.8pt}{\makebox(0,0){$\Diamond$}}}
\put(876,398){\raisebox{-.8pt}{\makebox(0,0){$\Diamond$}}}
\put(370,449){\raisebox{-.8pt}{\makebox(0,0){$\Diamond$}}}
\put(613,464){\raisebox{-.8pt}{\makebox(0,0){$\Diamond$}}}
\put(856,467){\raisebox{-.8pt}{\makebox(0,0){$\Diamond$}}}
\put(477,322){\raisebox{-.8pt}{\makebox(0,0){$\Diamond$}}}
\put(720,341){\raisebox{-.8pt}{\makebox(0,0){$\Diamond$}}}
\put(963,350){\raisebox{-.8pt}{\makebox(0,0){$\Diamond$}}}
\put(424,348){\raisebox{-.8pt}{\makebox(0,0){$\Diamond$}}}
\put(667,370){\raisebox{-.8pt}{\makebox(0,0){$\Diamond$}}}
\put(910,381){\raisebox{-.8pt}{\makebox(0,0){$\Diamond$}}}
\put(390,401){\raisebox{-.8pt}{\makebox(0,0){$\Diamond$}}}
\put(633,419){\raisebox{-.8pt}{\makebox(0,0){$\Diamond$}}}
\put(876,437){\raisebox{-.8pt}{\makebox(0,0){$\Diamond$}}}
\put(370,399){\raisebox{-.8pt}{\makebox(0,0){$\Diamond$}}}
\put(613,440){\raisebox{-.8pt}{\makebox(0,0){$\Diamond$}}}
\put(856,451){\raisebox{-.8pt}{\makebox(0,0){$\Diamond$}}}
\put(477,345){\raisebox{-.8pt}{\makebox(0,0){$\Diamond$}}}
\put(720,355){\raisebox{-.8pt}{\makebox(0,0){$\Diamond$}}}
\put(963,362){\raisebox{-.8pt}{\makebox(0,0){$\Diamond$}}}
\put(359,449){\raisebox{-.8pt}{\makebox(0,0){$\Diamond$}}}
\put(602,488){\raisebox{-.8pt}{\makebox(0,0){$\Diamond$}}}
\put(845,497){\raisebox{-.8pt}{\makebox(0,0){$\Diamond$}}}
\put(424,384){\raisebox{-.8pt}{\makebox(0,0){$\Diamond$}}}
\put(667,393){\raisebox{-.8pt}{\makebox(0,0){$\Diamond$}}}
\put(910,407){\raisebox{-.8pt}{\makebox(0,0){$\Diamond$}}}
\put(392,391){\raisebox{-.8pt}{\makebox(0,0){$\Diamond$}}}
\put(635,419){\raisebox{-.8pt}{\makebox(0,0){$\Diamond$}}}
\put(877,432){\raisebox{-.8pt}{\makebox(0,0){$\Diamond$}}}
\put(553,286){\raisebox{-.8pt}{\makebox(0,0){$\Diamond$}}}
\put(795,295){\raisebox{-.8pt}{\makebox(0,0){$\Diamond$}}}
\put(1038,299){\raisebox{-.8pt}{\makebox(0,0){$\Diamond$}}}
\put(479,334){\raisebox{-.8pt}{\makebox(0,0){$\Diamond$}}}
\put(722,349){\raisebox{-.8pt}{\makebox(0,0){$\Diamond$}}}
\put(965,361){\raisebox{-.8pt}{\makebox(0,0){$\Diamond$}}}
\put(426,378){\raisebox{-.8pt}{\makebox(0,0){$\Diamond$}}}
\put(669,396){\raisebox{-.8pt}{\makebox(0,0){$\Diamond$}}}
\put(912,405){\raisebox{-.8pt}{\makebox(0,0){$\Diamond$}}}
\put(437,466){\raisebox{-.8pt}{\makebox(0,0){$\Diamond$}}}
\put(680,515){\raisebox{-.8pt}{\makebox(0,0){$\Diamond$}}}
\put(922,540){\raisebox{-.8pt}{\makebox(0,0){$\Diamond$}}}
\put(484,450){\raisebox{-.8pt}{\makebox(0,0){$\Diamond$}}}
\put(726,456){\raisebox{-.8pt}{\makebox(0,0){$\Diamond$}}}
\put(969,460){\raisebox{-.8pt}{\makebox(0,0){$\Diamond$}}}
\put(458,464){\raisebox{-.8pt}{\makebox(0,0){$\Diamond$}}}
\put(701,482){\raisebox{-.8pt}{\makebox(0,0){$\Diamond$}}}
\put(944,492){\raisebox{-.8pt}{\makebox(0,0){$\Diamond$}}}
\put(587,337){\raisebox{-.8pt}{\makebox(0,0){$\Diamond$}}}
\put(829,338){\raisebox{-.8pt}{\makebox(0,0){$\Diamond$}}}
\put(1072,342){\raisebox{-.8pt}{\makebox(0,0){$\Diamond$}}}
\put(444,448){\raisebox{-.8pt}{\makebox(0,0){$\Diamond$}}}
\put(687,490){\raisebox{-.8pt}{\makebox(0,0){$\Diamond$}}}
\put(930,511){\raisebox{-.8pt}{\makebox(0,0){$\Diamond$}}}
\put(525,417){\raisebox{-.8pt}{\makebox(0,0){$\Diamond$}}}
\put(768,421){\raisebox{-.8pt}{\makebox(0,0){$\Diamond$}}}
\put(1011,425){\raisebox{-.8pt}{\makebox(0,0){$\Diamond$}}}
\put(484,412){\raisebox{-.8pt}{\makebox(0,0){$\Diamond$}}}
\put(726,426){\raisebox{-.8pt}{\makebox(0,0){$\Diamond$}}}
\put(969,429){\raisebox{-.8pt}{\makebox(0,0){$\Diamond$}}}
\put(458,412){\raisebox{-.8pt}{\makebox(0,0){$\Diamond$}}}
\put(701,431){\raisebox{-.8pt}{\makebox(0,0){$\Diamond$}}}
\put(944,445){\raisebox{-.8pt}{\makebox(0,0){$\Diamond$}}}
\put(587,347){\raisebox{-.8pt}{\makebox(0,0){$\Diamond$}}}
\put(829,358){\raisebox{-.8pt}{\makebox(0,0){$\Diamond$}}}
\put(1072,364){\raisebox{-.8pt}{\makebox(0,0){$\Diamond$}}}
\put(444,468){\raisebox{-.8pt}{\makebox(0,0){$\Diamond$}}}
\put(687,479){\raisebox{-.8pt}{\makebox(0,0){$\Diamond$}}}
\put(930,478){\raisebox{-.8pt}{\makebox(0,0){$\Diamond$}}}
\put(525,397){\raisebox{-.8pt}{\makebox(0,0){$\Diamond$}}}
\put(768,411){\raisebox{-.8pt}{\makebox(0,0){$\Diamond$}}}
\put(1011,420){\raisebox{-.8pt}{\makebox(0,0){$\Diamond$}}}
\put(485,428){\raisebox{-.8pt}{\makebox(0,0){$\Diamond$}}}
\put(728,441){\raisebox{-.8pt}{\makebox(0,0){$\Diamond$}}}
\put(971,449){\raisebox{-.8pt}{\makebox(0,0){$\Diamond$}}}
\put(460,427){\raisebox{-.8pt}{\makebox(0,0){$\Diamond$}}}
\put(703,453){\raisebox{-.8pt}{\makebox(0,0){$\Diamond$}}}
\put(945,463){\raisebox{-.8pt}{\makebox(0,0){$\Diamond$}}}
\put(588,352){\raisebox{-.8pt}{\makebox(0,0){$\Diamond$}}}
\put(831,367){\raisebox{-.8pt}{\makebox(0,0){$\Diamond$}}}
\put(1074,372){\raisebox{-.8pt}{\makebox(0,0){$\Diamond$}}}
\put(446,428){\raisebox{-.8pt}{\makebox(0,0){$\Diamond$}}}
\put(688,472){\raisebox{-.8pt}{\makebox(0,0){$\Diamond$}}}
\put(931,485){\raisebox{-.8pt}{\makebox(0,0){$\Diamond$}}}
\put(526,400){\raisebox{-.8pt}{\makebox(0,0){$\Diamond$}}}
\put(769,415){\raisebox{-.8pt}{\makebox(0,0){$\Diamond$}}}
\put(1012,422){\raisebox{-.8pt}{\makebox(0,0){$\Diamond$}}}
\put(520,467){\raisebox{-.8pt}{\makebox(0,0){$\Diamond$}}}
\put(763,518){\raisebox{-.8pt}{\makebox(0,0){$\Diamond$}}}
\put(1005,545){\raisebox{-.8pt}{\makebox(0,0){$\Diamond$}}}
\put(579,482){\raisebox{-.8pt}{\makebox(0,0){$\Diamond$}}}
\put(822,482){\raisebox{-.8pt}{\makebox(0,0){$\Diamond$}}}
\put(1064,486){\raisebox{-.8pt}{\makebox(0,0){$\Diamond$}}}
\put(515,474){\raisebox{-.8pt}{\makebox(0,0){$\Diamond$}}}
\put(757,545){\raisebox{-.8pt}{\makebox(0,0){$\Diamond$}}}
\put(1000,570){\raisebox{-.8pt}{\makebox(0,0){$\Diamond$}}}
\put(548,496){\raisebox{-.8pt}{\makebox(0,0){$\Diamond$}}}
\put(790,508){\raisebox{-.8pt}{\makebox(0,0){$\Diamond$}}}
\put(1033,508){\raisebox{-.8pt}{\makebox(0,0){$\Diamond$}}}
\put(529,494){\raisebox{-.8pt}{\makebox(0,0){$\Diamond$}}}
\put(772,528){\raisebox{-.8pt}{\makebox(0,0){$\Diamond$}}}
\put(1015,548){\raisebox{-.8pt}{\makebox(0,0){$\Diamond$}}}
\put(628,428){\raisebox{-.8pt}{\makebox(0,0){$\Diamond$}}}
\put(871,434){\raisebox{-.8pt}{\makebox(0,0){$\Diamond$}}}
\put(1114,435){\raisebox{-.8pt}{\makebox(0,0){$\Diamond$}}}
\put(579,441){\raisebox{-.8pt}{\makebox(0,0){$\Diamond$}}}
\put(822,451){\raisebox{-.8pt}{\makebox(0,0){$\Diamond$}}}
\put(1064,455){\raisebox{-.8pt}{\makebox(0,0){$\Diamond$}}}
\put(548,440){\raisebox{-.8pt}{\makebox(0,0){$\Diamond$}}}
\put(790,451){\raisebox{-.8pt}{\makebox(0,0){$\Diamond$}}}
\put(1033,458){\raisebox{-.8pt}{\makebox(0,0){$\Diamond$}}}
\put(529,456){\raisebox{-.8pt}{\makebox(0,0){$\Diamond$}}}
\put(772,477){\raisebox{-.8pt}{\makebox(0,0){$\Diamond$}}}
\put(1015,476){\raisebox{-.8pt}{\makebox(0,0){$\Diamond$}}}
\put(628,401){\raisebox{-.8pt}{\makebox(0,0){$\Diamond$}}}
\put(871,416){\raisebox{-.8pt}{\makebox(0,0){$\Diamond$}}}
\put(1114,424){\raisebox{-.8pt}{\makebox(0,0){$\Diamond$}}}
\put(521,459){\raisebox{-.8pt}{\makebox(0,0){$\Diamond$}}}
\put(764,526){\raisebox{-.8pt}{\makebox(0,0){$\Diamond$}}}
\put(1007,534){\raisebox{-.8pt}{\makebox(0,0){$\Diamond$}}}
\put(580,457){\raisebox{-.8pt}{\makebox(0,0){$\Diamond$}}}
\put(823,464){\raisebox{-.8pt}{\makebox(0,0){$\Diamond$}}}
\put(1066,467){\raisebox{-.8pt}{\makebox(0,0){$\Diamond$}}}
\put(549,465){\raisebox{-.8pt}{\makebox(0,0){$\Diamond$}}}
\put(792,480){\raisebox{-.8pt}{\makebox(0,0){$\Diamond$}}}
\put(1035,489){\raisebox{-.8pt}{\makebox(0,0){$\Diamond$}}}
\put(700,339){\raisebox{-.8pt}{\makebox(0,0){$\Diamond$}}}
\put(943,351){\raisebox{-.8pt}{\makebox(0,0){$\Diamond$}}}
\put(1186,360){\raisebox{-.8pt}{\makebox(0,0){$\Diamond$}}}
\put(531,466){\raisebox{-.8pt}{\makebox(0,0){$\Diamond$}}}
\put(774,489){\raisebox{-.8pt}{\makebox(0,0){$\Diamond$}}}
\put(1016,511){\raisebox{-.8pt}{\makebox(0,0){$\Diamond$}}}
\put(677,505){\raisebox{-.8pt}{\makebox(0,0){$\Diamond$}}}
\put(920,508){\raisebox{-.8pt}{\makebox(0,0){$\Diamond$}}}
\put(1163,506){\raisebox{-.8pt}{\makebox(0,0){$\Diamond$}}}
\put(735,435){\raisebox{-.8pt}{\makebox(0,0){$\Diamond$}}}
\put(978,438){\raisebox{-.8pt}{\makebox(0,0){$\Diamond$}}}
\put(1221,439){\raisebox{-.8pt}{\makebox(0,0){$\Diamond$}}}
\put(603,510){\raisebox{-.8pt}{\makebox(0,0){$\Diamond$}}}
\put(846,521){\raisebox{-.8pt}{\makebox(0,0){$\Diamond$}}}
\put(1089,527){\raisebox{-.8pt}{\makebox(0,0){$\Diamond$}}}
\put(677,464){\raisebox{-.8pt}{\makebox(0,0){$\Diamond$}}}
\put(920,467){\raisebox{-.8pt}{\makebox(0,0){$\Diamond$}}}
\put(1163,473){\raisebox{-.8pt}{\makebox(0,0){$\Diamond$}}}
\put(639,481){\raisebox{-.8pt}{\makebox(0,0){$\Diamond$}}}
\put(881,484){\raisebox{-.8pt}{\makebox(0,0){$\Diamond$}}}
\put(1124,489){\raisebox{-.8pt}{\makebox(0,0){$\Diamond$}}}
\put(616,492){\raisebox{-.8pt}{\makebox(0,0){$\Diamond$}}}
\put(859,502){\raisebox{-.8pt}{\makebox(0,0){$\Diamond$}}}
\put(1101,504){\raisebox{-.8pt}{\makebox(0,0){$\Diamond$}}}
\put(736,448){\raisebox{-.8pt}{\makebox(0,0){$\Diamond$}}}
\put(979,455){\raisebox{-.8pt}{\makebox(0,0){$\Diamond$}}}
\put(1222,456){\raisebox{-.8pt}{\makebox(0,0){$\Diamond$}}}
\put(678,484){\raisebox{-.8pt}{\makebox(0,0){$\Diamond$}}}
\put(921,491){\raisebox{-.8pt}{\makebox(0,0){$\Diamond$}}}
\put(1164,495){\raisebox{-.8pt}{\makebox(0,0){$\Diamond$}}}
\put(203.0,174.0){\rule[-0.200pt]{0.400pt}{132.977pt}}
\put(203.0,174.0){\rule[-0.200pt]{291.489pt}{0.400pt}}
\put(1413.0,174.0){\rule[-0.200pt]{0.400pt}{132.977pt}}
\put(203.0,726.0){\rule[-0.200pt]{291.489pt}{0.400pt}}
\end{picture}

}
\end{center}


The speedup is typically at least as predicted in~(\ref{eqn:predspeedup}), and often more.

To make this more concrete,
consider the case when $r\approx c$ and $\eps=0.01$.
Then the estimate simplifies to about $(r/310)^2/\ln r$.
For $r\ge 900$ or so, the algorithm here 
should be faster than Simplex, and for each factor-10 increase in $r$,
the speedup should increase by a factor of almost 100.


\subsection{Implementation issues}
The primary implementation issue is implementing the random sampling efficiently and precisely.
The data structures in
\cite{matias2003dgd,hagerup1993oag},
have two practical drawbacks.
The constant factors in the running times are moderately large,
and they implicitly or explicitly require that the probabilities being sampled
remain in a polynomially bounded range
(in the algorithm here, this can be accomplished 
by rescaling the data structure periodically).
However, the algorithm here uses these data structures in a restricted way.
Using the underlying ideas,
we built a data structure from scratch with very fast entry-update time
and moderately fast sample time.
We focused more on reducing the update time than the sampling time,
because we expect more update operations than sampling operations.
Full details are beyond the scope of this paper.
An open-source implementation is at \cite{fastpc2013}.

\subsection{Data} 

The following table tabulates the details of the experimental results
described earlier:
``t-alg'' is the time for the algorithm here in seconds;
``t-sim'' is the time for Simplex to find a $(1-\eps)$-optimal soln;
``t-sim\%'' is that time divided by the time for Simplex to complete;
``alg/sim'' is t-alg/t-sim.
{

\scriptsize

\bigskip
\begin{tabular}[t]{@{}|c@{~~}c@{~~}c@{~}c@{~}r@{~}r@{~}c@{~}c|} \hline
$r$	& $c$	& $k$	& $100\eps$	& t-alg	& t-sim	& t-sim\% & alg/sim \\ \hline
739 & 739 & 2 & 2.0 & 1 & 3 & 0.31 & 0.519
\\ 739 & 739 & 2 & 1.0 & 7 & 6 & 0.51 & 1.251
\\ 739 & 739 & 2 & 0.5 & 33 & 7 & 0.64 & 4.387
\\ 739 & 739 & 5 & 2.0 & 3 & 1 & 0.51 & 2.656
\\ 739 & 739 & 5 & 1.0 & 15 & 1 & 0.63 & 8.840
\\ 739 & 739 & 5 & 0.5 & 63 & 2 & 0.76 & 30.733
\\ 739 & 739 & 4 & 2.0 & 2 & 2 & 0.51 & 0.970
\\ 739 & 739 & 4 & 1.0 & 11 & 3 & 0.64 & 3.317
\\ 739 & 739 & 4 & 0.5 & 46 & 4 & 0.76 & 11.634
\\ 739 & 739 & 3 & 2.0 & 2 & 3 & 0.43 & 0.561
\\ 739 & 739 & 3 & 1.0 & 9 & 5 & 0.60 & 1.745
\\ 739 & 739 & 3 & 0.5 & 38 & 6 & 0.72 & 6.197
\\ 1480 & 740 & 3 & 2.0 & 2 & 9 & 0.37 & 0.304
\\ 1480 & 740 & 3 & 1.0 & 13 & 13 & 0.53 & 0.959
\\ 1480 & 740 & 3 & 0.5 & 57 & 16 & 0.64 & 3.478
\\ 1480 & 740 & 2 & 2.0 & 2 & 24 & 0.44 & 0.102
\\ 1480 & 740 & 2 & 1.0 & 11 & 33 & 0.60 & 0.342
\\ 1480 & 740 & 2 & 0.5 & 51 & 39 & 0.71 & 1.313
\\ 1480 & 740 & 5 & 2.0 & 4 & 4 & 0.41 & 0.928
\\ 1480 & 740 & 5 & 1.0 & 18 & 6 & 0.56 & 2.930
\\ 1480 & 740 & 5 & 0.5 & 77 & 7 & 0.66 & 10.447
\\ 1480 & 740 & 4 & 2.0 & 3 & 6 & 0.34 & 0.495
\\ 1480 & 740 & 4 & 1.0 & 15 & 10 & 0.49 & 1.496
\\ 1480 & 740 & 4 & 0.5 & 64 & 12 & 0.60 & 5.239
\\ 740 & 1480 & 3 & 2.0 & 3 & 14 & 0.35 & 0.211
\\ 740 & 1480 & 3 & 1.0 & 14 & 21 & 0.51 & 0.667
\\ 740 & 1480 & 3 & 0.5 & 63 & 29 & 0.71 & 2.139
\\ 740 & 1480 & 2 & 2.0 & 2 & 13 & 0.27 & 0.192
\\ 740 & 1480 & 2 & 1.0 & 11 & 25 & 0.51 & 0.462
\\ 740 & 1480 & 2 & 0.5 & 54 & 34 & 0.68 & 1.597
\\ 740 & 1480 & 5 & 2.0 & 5 & 7 & 0.59 & 0.699
\\ 740 & 1480 & 5 & 1.0 & 22 & 9 & 0.72 & 2.460
\\ 740 & 1480 & 5 & 0.5 & 94 & 10 & 0.82 & 9.054
\\ 740 & 1480 & 1 & 2.0 & 2 & 23 & 0.24 & 0.097
\\ 740 & 1480 & 1 & 1.0 & 9 & 41 & 0.44 & 0.237
\\ 740 & 1480 & 1 & 0.5 & 47 & 55 & 0.59 & 0.848
\\ 740 & 1480 & 4 & 2.0 & 3 & 12 & 0.47 & 0.313
\\ 740 & 1480 & 4 & 1.0 & 17 & 15 & 0.61 & 1.130
\\ 740 & 1480 & 4 & 0.5 & 73 & 19 & 0.75 & 3.803
\\ \hline
\end{tabular}
\begin{tabular}[t]{@{}|c@{~~}c@{~~}c@{~}c@{~}r@{~}r@{~}c@{~}c|} \hline
$r$	& $c$	& $k$	& $100\eps$	& t-alg	& t-sim	& t-sim\% & alg/sim \\ \hline
 1110 & 1110 & 3 & 2.0 & 3 & 21 & 0.30 & 0.142
\\ 1110 & 1110 & 3 & 1.0 & 13 & 33 & 0.48 & 0.399
\\ 1110 & 1110 & 3 & 0.5 & 58 & 43 & 0.62 & 1.354
\\ 1110 & 1110 & 6 & 2.0 & 6 & 5 & 0.64 & 1.327
\\ 1110 & 1110 & 6 & 1.0 & 29 & 6 & 0.76 & 4.763
\\ 1110 & 1110 & 6 & 0.5 & 121 & 6 & 0.83 & 17.903
\\ 1110 & 1110 & 5 & 2.0 & 4 & 9 & 0.48 & 0.480
\\ 1110 & 1110 & 5 & 1.0 & 20 & 13 & 0.64 & 1.575
\\ 1110 & 1110 & 5 & 0.5 & 86 & 15 & 0.77 & 5.439
\\ 1110 & 1110 & 4 & 2.0 & 3 & 17 & 0.43 & 0.203
\\ 1110 & 1110 & 4 & 1.0 & 16 & 24 & 0.60 & 0.649
\\ 1110 & 1110 & 4 & 0.5 & 68 & 29 & 0.71 & 2.325
\\ 1111 & 2222 & 1 & 2.0 & 3 & 94 & 0.15 & 0.036
\\ 1111 & 2222 & 1 & 1.0 & 15 & 198 & 0.30 & 0.077
\\ 1111 & 2222 & 1 & 0.5 & 78 & 344 & 0.53 & 0.227
\\ 1111 & 2222 & 4 & 2.0 & 5 & 94 & 0.49 & 0.057
\\ 1111 & 2222 & 4 & 1.0 & 26 & 123 & 0.64 & 0.212
\\ 1111 & 2222 & 4 & 0.5 & 119 & 148 & 0.77 & 0.803
\\ 1111 & 2222 & 3 & 2.0 & 4 & 109 & 0.35 & 0.042
\\ 1111 & 2222 & 3 & 1.0 & 21 & 163 & 0.52 & 0.134
\\ 1111 & 2222 & 3 & 0.5 & 104 & 222 & 0.71 & 0.467
\\ 1111 & 2222 & 6 & 2.0 & 9 & 23 & 0.66 & 0.426
\\ 1111 & 2222 & 6 & 1.0 & 44 & 26 & 0.76 & 1.664
\\ 1111 & 2222 & 6 & 0.5 & 187 & 29 & 0.84 & 6.346
\\ 1111 & 2222 & 2 & 2.0 & 3 & 83 & 0.18 & 0.047
\\ 1111 & 2222 & 2 & 0.5 & 91 & 269 & 0.57 & 0.339
\\ 1111 & 2222 & 5 & 2.0 & 6 & 63 & 0.57 & 0.110
\\ 1111 & 2222 & 5 & 1.0 & 32 & 77 & 0.69 & 0.415
\\ 1111 & 2222 & 5 & 0.5 & 140 & 88 & 0.79 & 1.594
\\ 2222 & 1111 & 4 & 2.0 & 4 & 53 & 0.38 & 0.092
\\ 2222 & 1111 & 4 & 1.0 & 23 & 75 & 0.54 & 0.311
\\ 2222 & 1111 & 4 & 0.5 & 107 & 91 & 0.65 & 1.185
\\ 2222 & 1111 & 3 & 2.0 & 4 & 53 & 0.29 & 0.080
\\ 2222 & 1111 & 3 & 1.0 & 21 & 84 & 0.46 & 0.253
\\ 2222 & 1111 & 3 & 0.5 & 97 & 115 & 0.63 & 0.848
\\ 2222 & 1111 & 6 & 2.0 & 7 & 21 & 0.49 & 0.373
\\ 2222 & 1111 & 6 & 1.0 & 34 & 26 & 0.61 & 1.297
\\ 2222 & 1111 & 6 & 0.5 & 148 & 30 & 0.71 & 4.816
\\ 2222 & 1111 & 2 & 2.0 & 3 & 102 & 0.36 & 0.037
\\ 2222 & 1111 & 2 & 1.0 & 17 & 139 & 0.49 & 0.127
\\ 2222 & 1111 & 2 & 0.5 & 88 & 173 & 0.61 & 0.513
\\ 2222 & 1111 & 5 & 2.0 & 5 & 42 & 0.41 & 0.141
\\ 2222 & 1111 & 5 & 1.0 & 27 & 57 & 0.56 & 0.472
\\ 2222 & 1111 & 5 & 0.5 & 120 & 70 & 0.68 & 1.696
\\ \hline
\end{tabular}

\begin{tabular}[t]{@{}|c@{~~}c@{~~}c@{~}c@{~}r@{~}r@{~}c@{~}c|} \hline
$r$	& $c$	& $k$	& $100\eps$	& t-alg	& t-sim	& t-sim\% & alg/sim \\ \hline
 1666 & 1666 & 4 & 2.0 & 5 & 117 & 0.40 & 0.045
\\ 1666 & 1666 & 4 & 1.0 & 24 & 163 & 0.56 & 0.153
\\ 1666 & 1666 & 4 & 0.5 & 111 & 201 & 0.69 & 0.554
\\ 1666 & 1666 & 3 & 2.0 & 4 & 112 & 0.29 & 0.040
\\ 1666 & 1666 & 3 & 1.0 & 21 & 185 & 0.48 & 0.114
\\ 1666 & 1666 & 3 & 0.5 & 98 & 245 & 0.64 & 0.400
\\ 1666 & 1666 & 6 & 2.0 & 8 & 42 & 0.51 & 0.210
\\ 1666 & 1666 & 6 & 1.0 & 38 & 55 & 0.66 & 0.697
\\ 1666 & 1666 & 6 & 0.5 & 165 & 63 & 0.76 & 2.612
\\ 1666 & 1666 & 2 & 2.0 & 3 & 109 & 0.20 & 0.036
\\ 1666 & 1666 & 2 & 1.0 & 18 & 221 & 0.41 & 0.083
\\ 1666 & 1666 & 2 & 0.5 & 88 & 313 & 0.58 & 0.282
\\ 1666 & 1666 & 5 & 2.0 & 6 & 82 & 0.44 & 0.080
\\ 1666 & 1666 & 5 & 1.0 & 29 & 109 & 0.58 & 0.269
\\ 1666 & 1666 & 5 & 0.5 & 130 & 133 & 0.71 & 0.981
\\ 1666 & 3332 & 2 & 2.0 & 5 & 354 & 0.12 & 0.017
\\ 1666 & 3332 & 2 & 1.0 & 30 & 857 & 0.29 & 0.036
\\ 1666 & 3332 & 2 & 0.5 & 162 & 1594 & 0.54 & 0.102
\\ 1666 & 3332 & 5 & 2.0 & 9 & 509 & 0.51 & 0.020
\\ 1666 & 3332 & 5 & 1.0 & 51 & 654 & 0.65 & 0.078
\\ 1666 & 3332 & 5 & 0.5 & 227 & 762 & 0.76 & 0.299
\\ 1666 & 3332 & 1 & 2.0 & 5 & 350 & 0.09 & 0.015
\\ 1666 & 3332 & 1 & 1.0 & 24 & 1003 & 0.25 & 0.025
\\ 1666 & 3332 & 1 & 0.5 & 135 & 1881 & 0.46 & 0.072
\\ 1666 & 3332 & 4 & 2.0 & 7 & 578 & 0.38 & 0.014
\\ 1666 & 3332 & 4 & 1.0 & 42 & 899 & 0.58 & 0.047
\\ 1666 & 3332 & 4 & 0.5 & 204 & 1087 & 0.71 & 0.188
\\ 1666 & 3332 & 3 & 2.0 & 6 & 533 & 0.20 & 0.013
\\ 1666 & 3332 & 3 & 1.0 & 36 & 1095 & 0.41 & 0.033
\\ 1666 & 3332 & 3 & 0.5 & 180 & 1741 & 0.65 & 0.104
\\ 1666 & 3332 & 6 & 2.0 & 13 & 255 & 0.56 & 0.051
\\ 1666 & 3332 & 6 & 1.0 & 60 & 319 & 0.70 & 0.190
\\ 1666 & 3332 & 6 & 0.5 & 271 & 361 & 0.79 & 0.752
\\ 3332 & 1666 & 5 & 2.0 & 9 & 275 & 0.38 & 0.033
\\ 3332 & 1666 & 5 & 1.0 & 45 & 392 & 0.54 & 0.115
\\ 3332 & 1666 & 5 & 0.5 & 213 & 482 & 0.66 & 0.441
\\ 3332 & 1666 & 4 & 2.0 & 7 & 274 & 0.30 & 0.028
\\ 3332 & 1666 & 4 & 1.0 & 40 & 414 & 0.45 & 0.097
\\ 3332 & 1666 & 4 & 0.5 & 195 & 556 & 0.60 & 0.352
\\ 3332 & 1666 & 3 & 2.0 & 6 & 316 & 0.24 & 0.020
\\ 3332 & 1666 & 3 & 1.0 & 34 & 544 & 0.41 & 0.063
\\ 3332 & 1666 & 3 & 0.5 & 178 & 703 & 0.53 & 0.254
\\ 3332 & 1666 & 6 & 2.0 & 11 & 154 & 0.39 & 0.071
\\ 3332 & 1666 & 6 & 1.0 & 52 & 218 & 0.56 & 0.238
\\ 3332 & 1666 & 6 & 0.5 & 233 & 273 & 0.70 & 0.854
\\ \hline
\end{tabular}
\begin{tabular}[t]{@{}|c@{~~}c@{~~}c@{~}c@{~}r@{~}r@{~}c@{~}c|} \hline
$r$	& $c$	& $k$	& $100\eps$	& t-alg	& t-sim	& t-sim\% & alg/sim \\ \hline
 2499 & 2499 & 2 & 2.0 & 5 & 530 & 0.13 & 0.011
\\ 2499 & 2499 & 2 & 1.0 & 29 & 1556 & 0.40 & 0.019
\\ 2499 & 2499 & 2 & 0.5 & 159 & 2275 & 0.58 & 0.070
\\ 2499 & 2499 & 5 & 2.0 & 9 & 580 & 0.42 & 0.016
\\ 2499 & 2499 & 5 & 1.0 & 46 & 793 & 0.58 & 0.059
\\ 2499 & 2499 & 5 & 0.5 & 217 & 960 & 0.70 & 0.227
\\ 2499 & 2499 & 4 & 2.0 & 8 & 662 & 0.31 & 0.012
\\ 2499 & 2499 & 4 & 1.0 & 42 & 1064 & 0.50 & 0.040
\\ 2499 & 2499 & 4 & 0.5 & 195 & 1369 & 0.64 & 0.143
\\ 2499 & 2499 & 7 & 2.0 & 17 & 125 & 0.50 & 0.139
\\ 2499 & 2499 & 7 & 1.0 & 76 & 162 & 0.65 & 0.475
\\ 2499 & 2499 & 7 & 0.5 & 327 & 190 & 0.77 & 1.715
\\ 2499 & 2499 & 3 & 2.0 & 6 & 618 & 0.18 & 0.011
\\ 2499 & 2499 & 3 & 1.0 & 35 & 1079 & 0.32 & 0.032
\\ 2499 & 2499 & 3 & 0.5 & 174 & 1774 & 0.53 & 0.099
\\ 2500 & 5000 & 6 & 2.0 & 19 & 2525 & 0.52 & 0.008
\\ 2500 & 5000 & 6 & 1.0 & 98 & 3337 & 0.69 & 0.029
\\ 2500 & 5000 & 6 & 0.5 & 458 & 3828 & 0.79 & 0.120
\\ 2500 & 5000 & 7 & 2.0 & 26 & 1042 & 0.60 & 0.026
\\ 2500 & 5000 & 7 & 1.0 & 124 & 1272 & 0.73 & 0.098
\\ 2500 & 5000 & 7 & 0.5 & 556 & 1427 & 0.82 & 0.390
\\ 5000 & 2500 & 3 & 2.0 & 10 & 2165 & 0.23 & 0.005
\\ 5000 & 2500 & 3 & 1.0 & 62 & 3828 & 0.40 & 0.016
\\ 5000 & 2500 & 3 & 0.5 & 338 & 5586 & 0.58 & 0.061
\\ 5000 & 2500 & 6 & 2.0 & 17 & 1352 & 0.39 & 0.013
\\ 5000 & 2500 & 6 & 1.0 & 90 & 1832 & 0.53 & 0.049
\\ 5000 & 2500 & 6 & 0.5 & 418 & 2297 & 0.66 & 0.182
\\ 5000 & 2500 & 5 & 2.0 & 14 & 1752 & 0.33 & 0.008
\\ 5000 & 2500 & 5 & 1.0 & 82 & 2592 & 0.49 & 0.032
\\ 5000 & 2500 & 5 & 0.5 & 397 & 3330 & 0.63 & 0.119
\\ 5000 & 2500 & 4 & 2.0 & 12 & 1916 & 0.26 & 0.006
\\ 5000 & 2500 & 4 & 1.0 & 70 & 3177 & 0.44 & 0.022
\\ 5000 & 2500 & 4 & 0.5 & 367 & 4197 & 0.58 & 0.087
\\ 3750 & 3750 & 7 & 2.0 & 23 & 1828 & 0.50 & 0.013
\\ 3750 & 3750 & 7 & 1.0 & 111 & 2343 & 0.64 & 0.047
\\ 3750 & 3750 & 7 & 0.5 & 506 & 2712 & 0.74 & 0.187
\\ 3750 & 3750 & 6 & 2.0 & 18 & 3061 & 0.40 & 0.006
\\ 3750 & 3750 & 6 & 1.0 & 91 & 4263 & 0.55 & 0.022
\\ 3750 & 3750 & 6 & 0.5 & 432 & 5279 & 0.68 & 0.082
\\ \hline
\end{tabular}

}

\section{Future directions}\label{sec:conclusion}
Can one extend the coupling technique to {\em mixed} packing and covering problems?
What about the special case of $\exists x \ge 0; Ax \approx b$ (important for computer tomography).
What about covering with ``box'' constraints (upper bounds on individual variables)?
Perhaps most importantly, 
what about general (not explicitly given)
packing and covering, e.g.~to maximum multicommodity flow (where $P$ is the polytope whose vertices correspond to all $s_i\rightarrow t_i$ paths)?  
In all of these cases, correctness of a natural algorithm is easy to establish, 
but the running time is problematic.
This seems to be because the coupling approach requires 
that fast primal {\em and} dual algorithms of a particular kind must both exist.
Such algorithms are known for each of the above-mentioned problems,
but the natural algorithm for each dual problems is slow.

The algorithm seems a natural candidate for solving {\em dynamic} problems, 
or sequences of closely related problems (e.g. each problem comes from the previous one by a small change in the constraint matrix).
Adapting the algorithm to start with a given primal/dual pair seems straightforward
and may be useful in practice.

Can one use coupling to improve {\em parallel and distributed} algorithms for packing and covering (e.g.~\cite{Luby93Parallel,Young01Sequential}), perhaps reducing the dependence on $\eps$ from $1/\eps^4$ to $1/\eps^3$?
(In this case, instead of incrementing a {\em randomly} chosen variable in each of the primal and dual solutions,
one would increment {\em all} primal and dual variables deterministically 
in each iteration:
increment the primal vector $\xp$ by $\alpha \pd$
and the dual vector $\xd$ by $\alpha\pp$ for the maximal $\alpha$
so that the correctness proof goes through.
Can one bound the number of iterations, assuming the matrix is appropriately preprocessed?)

\section*{Acknowledgments}
Thanks to two anonymous referees for helpful suggestions.
The first author would like to thank the Greek State Scholarship Foundation (IKY).
The second author's research was partially supported by
NSF grants 0626912, 0729071, and 1117954.

\bibliographystyle{spmpsci}
\bibliography{bib,bib2}

\section{Appendix: Utility Lemmas}

\newcommand{\epsoverdelta}{{\textstyle\frac{\eps}{\delta}}}
\renewcommand{\epsoverdelta}{\eps}

The first is a one-sided variant of Wald's equation:
\begin{lemma}\cite[lemma 4.1]{Young00KMedians}\label{lemma:walds}
Let $K$ be any finite number.
Let $x_0,x_1,\ldots,x_T$ be a sequence of random variables, 
where $T$ is a random stopping time with finite expectation. 

If $\E[x_{t} - x_{t-1} \giv x_{t-1}] \le \mu$ 
and (in every outcome)
$x_{t}-x_{t-1}\le K$ for $t\le T$, 
then
$\E[x_T-x_0] \le\mu\,\E[T]$.
\end{lemma}

The second is the Azuma-like inequality tailored for random stopping times.
\begin{lemma}
\label{lemma:chst}\label{lemma:chernoff}
Let $X=\sum_{t=1}^T x_t$ and $Y=\sum_{t=1}^T y_t$ 
be sums of non-negative random variables, 
where $T$ is a random stopping time with finite expectation,
and, for all $t$,
$|x_t-y_t|\le 1$ and
\[\textstyle
 \E\big[\,x_{t} - y_{t} \,\giv\, \sum_{s< t} x_s, \sum_{s< t} y_s\,\big]
~\le~
 0.\]
Let $\eps\in[0,1]$ and $A\in\R$.  
Then
\[\Pr\big[\,(1-\eps) X \,\ge\, Y + A\, \big] ~\le~ \exp({-\epsoverdelta}A).\]
\end{lemma}
\begin{Proof}
Fix $\lambda>0$.
Consider the sequence $\pi_0,\pi_1,\ldots,\pi_T$ where
$\pi_t = 0$ for $t> \lambda\E[T]$ and otherwise
\[
\pi_t
~\doteq~
\prod_{s\le t} (1+\eps)^{x_s}(1-\eps)^{y_s}
 ~=~
\pi_{t-1}(1+\eps)^{x_t}(1-\eps)^{y_t}
 ~\le~
\pi_{t-1}(1+\eps x_t -\eps y_t)
\]
(using $(1+\eps)^x(1-\eps)^y\le (1+\eps x-\eps y)$ when $|x-y|\le 1$).

From $\E[x_{t} -y_{t} \giv \pi_{t-1}]\le 0$, 
it follows that $\E[\pi_{t} \giv \pi_{t-1}] \le \pi_{t-1}$.

Note that, from the use of $\lambda$,
$\sum_{s\le t} x_s-y_s$ and (therefore) $\pi_t-\pi_{t-1}$ are bounded.
Thus Wald's (Lemma~\ref{lemma:walds}), implies
$\E[\pi_T] \le \pi_0 = 1$.

Applying the Markov bound,
\[
\Pr[ \pi_T \ge \exp(\eps A)]
~\le~
\exp({-\eps} A).
\]

So assume $\pi_T < \exp(\eps A)$.
Taking logs, if $T\le \lambda\E[T]$,
\[
X\ln(1+\eps) -Y\ln(1/(1-\eps)) 
 ~=~ 
\ln \pi_T
~<~ 
\eps A.\]

Dividing by $\ln(1/(1-\eps))$
and applying the inequalities $\ln(1+\eps)/\ln(1/(1-\eps))\ge 1-\eps$ and
$\eps/\ln(1/(1-\eps))\le 1$,  gives
$(1-\eps) X < Y + A$.
Thus,
\[
\Pr[(1-\eps)X\ge Y+A]
~\le~
\Pr[T\ge \lambda \E[T]]
+
\Pr[\pi_T \ge \exp(\eps A)]
~\le~
1/\lambda + \exp(-\eps A).
\]
Since $\lambda$ can be arbitrarily large, the lemma follows.
\end{Proof}

\end{document}